\documentclass[bibyear]{aa}  
\usepackage{mathtools}
\usepackage{natbib}
\usepackage{multirow}
\usepackage{siunitx}
\usepackage{comment}
\usepackage{booktabs}
\usepackage{graphicx}
\usepackage{menukeys}
\usepackage{algorithm}
\usepackage{ulem}
\usepackage[noend]{algpseudocode}
\usepackage{placeins}
\graphicspath{{Fig/}}
\usepackage[colorlinks=true, allcolors=blue]{hyperref}
\usepackage{txfonts}

\usepackage{cuted}
\makeatletter
\renewcommand*\aa@pageof{, page \thepage{} of \pageref*{LastPage}}
\makeatother
\begin{document} 

   \title{Revisiting rotationally excited CH at radio wavelengths: A case study towards W51}

   \author{Arshia M. Jacob\inst{1,2}
          \and
          Meera Nandakumar\inst{3}
          \and
          Nirupam Roy\inst{3}
          \and
          Karl M. Menten\inst{2}
          \and
          David A. Neufeld\inst{1}
          \and 
          Alexandre Faure\inst{4}
          \and 
          Maitraiyee Tiwari\inst{2} 
          \and 
          Thushara G. S. Pillai\inst{5, 6} 
          \and
          Timothy Robishaw\inst{7}
          \and 
          Carlos A. Dur\'{a}n\inst{8}
          }

   \institute{William H. Miller III Department of Physics \& Astronomy, Johns Hopkins University, 3400 North Charles Street, Baltimore, MD 21218, USA \email{ajacob@mpifr-bonn.mpg.de}
   \and
   Max-Planck-Institut f\"{u}r Radioastronomie, Auf dem H\"{u}gel 69, 53121 Bonn, Germany
   \and
   Department of Physics, Indian Institute of Science, Bangalore 560012, India
   \and 
Univ. Grenoble Alpes, CNRS, IPAG, F-38000 Grenoble, France
   \and 
   Haystack Observatory, Massachusetts Institute of Technology, 99 Millstone Rd., Westford, MA 01886, USA
   \and
   Institute for Astrophysical Research, Boston University, 725 Commonwealth Avenue, Boston, MA 02215, USA
      \and 
   Dominion Radio Astrophysical Observatory, Herzberg Astronomy and Astrophysics Research Centre, National Research Council
Canada, PO Box 248, Penticton, BC V2A 6J9, Canada
\and 
Instituto de Radioastronom\'{i}a Milim\'{e}trica-- IRAM, Avenida Divina Pastora 7, Local 20, E-18012, Granada, Spain\\
   }

   \date{Received 13 February, 2024 / Accepted 8 November, 2024}
  \titlerunning{Revisiting rotationally excited CH}
   \authorrunning{A. M. Jacob et al.}
 \abstract{Ever since they were first detected in the interstellar medium, the radio wavelength (3.3~GHz) hyperfine-structure splitting transitions in the rotational ground state of CH have been observed to show anomalous excitation. Astonishingly, this behaviour has been uniformly observed towards a variety of different sources probing a wide range of physical conditions. While the observed level inversion can be explained globally by a pumping scheme involving collisions, a description of the extent of `over-excitation' observed in individual sources requires the inclusion of radiative processes, involving transitions at higher rotational levels. Therefore, a complete description of the excitation mechanism in the CH ground
state, observed towards individual sources entails observational constraints from the
rotationally excited levels of CH and in particular that of its first rotationally excited state ($^{2}\Pi_{3/2}, N=1, J = 3/2$).}
 {Given the limited detections of these lines, the objective of this work is to characterise the physical and excitation properties of the rotationally excited lines of CH between the $\Lambda$-doublet levels of its $^{2}\Pi_{3/2}, N=1, J=3/2$ state near 700~MHz, and investigate their influence on the pumping mechanisms of the ground-state lines of CH. }
 {This work presents the first interferometric search for the rotationally excited lines of CH between the $\Lambda$-doublet levels of its $^{2}\Pi_{3/2}, N=1, J=3/2$ state near 700~MHz carried out using the upgraded Giant Metrewave Radio Telescope (uGMRT) array towards six star-forming regions, W51\,E, Sgr\,B2\,(M), M8, M17, W43 and DR21~Main. 
 }
 {We detect the two main hyperfine structure lines within the first rotationally excited state of CH, in absorption towards W51\,E. 
 To jointly model the physical and excitation conditions traced by lines from both the ground and first rotationally excited states of CH, we performed non-local thermodynamic equilibrium (LTE) radiative transfer calculations using the code MOLPOP-CEP. These models account for the effects of line overlap and are aided by column density constraints from the far-infrared (FIR) wavelength rotational transitions of CH that connect to the ground state and use collisional rate coefficients for collisions of CH with H, H$_2$ and electrons (the latter are computed in this work using cross-sections estimated within the Born approximation). 
 }{The non-LTE analysis revealed that physical properties typical of diffuse and translucent clouds best reproduce the higher rates of level inversion seen in the ground-state lines at 3.3~GHz, observed at velocities near 66~km~s$^{-1}$ along the sightline towards W51\,E. In contrast, the excited lines near 700~MHz were only excited in much denser environments with ${n_{\rm H}\sim10^{5}}~$cm$^{-3}$ towards which the anomalous excitation in two of the three ground state lines is quenched, but not in the 3.264~GHz line. This is in alignment with our observations and suggests that while FIR pumping and line overlap effects are essential for exciting and producing line inversion in the ground state, excitation to the first rotational level is dominated by collisional excitation from the ground state. For the rotationally excited state of CH, the models indicate low excitation temperatures and column densities of $2\times10^{14}~$cm$^{-2}$. Furthermore, modelling these lines helps us understand the complexities of the spectral features observed in the 532/536~GHz rotational transitions of CH. These transitions, connecting sub-levels of the first rotationally excited state to the ground state, play a crucial role in trapping FIR radiation and enhancing the degree of inversion seen in the ground state lines. Based on the physical conditions constrained, we predict the potential of detecting hyperfine-splitting transitions arising from higher rotationally excited transitions of CH in the context of their current non-detections.
 
 }{}{} 
 

   \maketitle
   
%

\section{Introduction} \label{sec:intro}
First identified by \citet{Swings1937}, via one of the transitions between its A$^{2}\Delta$-X$^2\Pi$ electronic states near 4300.2~\AA\ in optical absorption spectra observed towards several early-type stars by \citet{Dunham1937} \citep[and later][]{Adams1941}, the methylidyne radical, CH, has the unique distinction of being the first molecule to be detected in the interstellar medium (ISM). An exciting new discovery, the presence of interstellar CH was soon corroborated by the subsequent identification of other CH lines near 3880~\AA\ corresponding to its B$^{2}\Sigma$-X$^2\Pi$ electronic transitions by \citet{McKellar1940}. The successful detection and identification of CH in the ISM motivated laboratory investigations of other unidentified absorption lines and led to the birth of molecular astrophysics. To date, CH has been observed across the electromagnetic spectrum from far-UV \citep{Watson2001} to long radio wavelengths \citep{Rydbeck1973, Turner1974} and maintains an important role as a fundamental building block in gas-phase interstellar chemistry and as a  diagnostic probe of diffuse and CO-dark molecular gas \citep{Federman1982, Sheffer2008, Gerin2010, Weselak2010}.

The lines resulting from the hyperfine structure (HFS) splitting of the $\Lambda$-doublet levels of the rotational ground state of CH at 3.3~GHz (9~cm) have been of particular interest owing to the ease with which they can be observed in the midst of the microwave range. However, the interpretation of these observations proved challenging, since the line intensities of these 
lines were found to be inconsistent with predictions made under conditions of local thermodynamic equilibrium (LTE). In particular, these lines are almost always seen in emission, even along lines of sight towards bright background continuum sources: The observed deviations were interpreted as being caused by the amplification of radiation by stimulated emission \citep[for example,][]{Zuckerman1975, Rydbeck1976, Genzel1979}. The anomalous excitation 
of these (weak) maser lines could in principle be qualitatively understood by assuming a pumping cycle that involves collisional excitation to rotational states (via collisions with atomic and molecular hydrogen), followed by radiative decay back to the ground state \citep{Bertojo1976, Elitzur1977}. However, the relative intensities of CH's ground state HFS lines and in particular the dominance of its lowest frequency satellite line at 3.264~GHz was not well understood and this greatly limited the use of CH as a probe of molecular gas properties at radio wavelengths.

\begin{figure*}
\centering
    \includegraphics[width=0.95\textwidth]{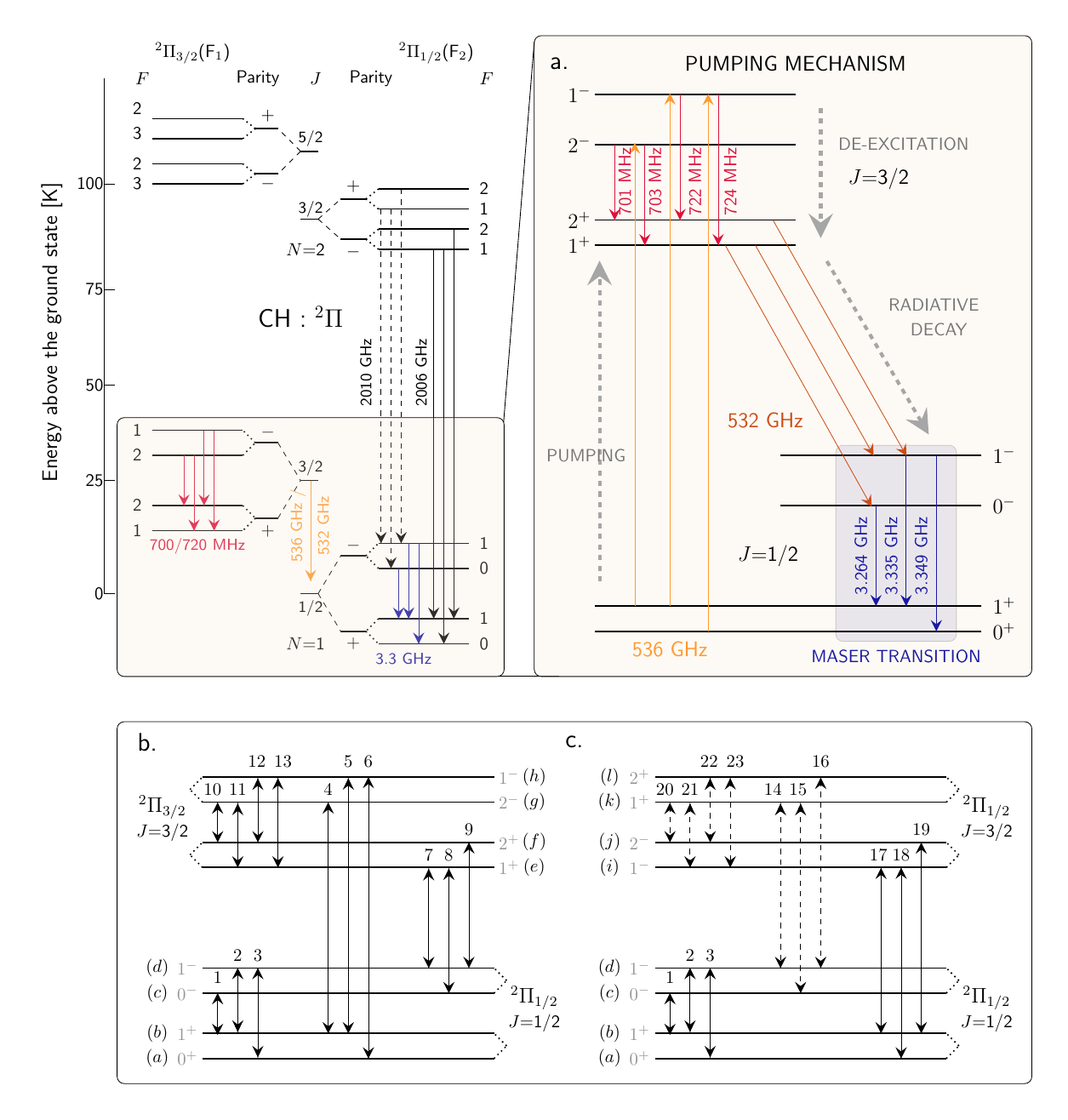}
    \caption{Lowest rotational energy levels of CH, where the rotational transitions relevant to this work are labelled and marked using arrows. Note that the $\Lambda$-doublet and HFS splitting level separations are not drawn to scale and that the transitions labelled by dashed arrows are not observed. a.) The inset zooms in on the ground and first rotationally excited levels, to show the sub-mm transitions connecting the two levels and highlights the general pumping mechanism. b.) All allowed transitions and HFS levels for lines lying within and connecting the $^{2}\Pi_{3/2}, J=3/2$ and $^{2}\Pi_{1/2}, J=1/2$ states are marked and labelled, respectively, to illustrate the effect of line overlap as referenced in the text. c.) Same as b.) but for the $^{2}\Pi_{1/2}, J=3/2$ and $^{2}\Pi_{1/2}, J=1/2$ states.} 
    \label{fig:energy_level}
\end{figure*}

\citet{Bujarrabal1984} demonstrated that, while the universal inversion observed in the 3.3~GHz CH lines is a consequence of parity discrimination induced by collisions, radiative processes have to be considered to explain the `over-excitation' observed in the lowest frequency satellite line. Recently, by including the effects of pumping from the radiative trapping of photons caused by the overlap of lines from higher rotational states at far-infrared (FIR) wavelengths, \citet{Jacob2021} (henceforth Paper I) were able to model enhancements in the intensities of the lower satellite line and further study the physical and excitation conditions traced. The non-LTE radiative transfer models carried out in Paper I were facilitated by the latest HFS resolved collisional rate coefficients for collisions with atomic and molecular hydrogen as well as helium, computed by \citet{Dagdigian2018} and \citet{Marinakis2019}, respectively. Furthermore, the column densities of CH in the models presented in Paper I were constrained using values determined from the high spectral resolution observations of the $^{2}\Pi_{1/2}, N, J = 2, 3/2 \rightarrow 1, 1/2$ FIR rotational transition of CH near 2006~GHz (${\sim \! 149~\mu}$m), which share a common lower energy level with the HFS lines of the rotational ground state (see Fig.~\ref{fig:energy_level}). The latter was observed using the upGREAT\footnote{German REciever for Astronomy at Terahertz frequencies.} instrument \citep{Risacher2016} on the Stratospheric Observatory for Infrared Astronomy \citep[SOFIA;][]{Young2012} and was seen in widespread and typically optically thin absorption, which makes it a reliable tool for measuring column densities \citep{Wiesemeyer2018, Jacob2019}. Given their importance in exciting the ground state lines, it is imperative to observe transitions of CH arising from higher rotational levels to obtain a comprehensive picture of the excitation of CH in different astrophysical systems.\\

As discussed above, the observed enhancements in the intensity of the lowest frequency HFS satellite line of the CH ground state is primarily due to line overlap effects from higher rotational states. The rotational ground state of CH is radiatively connected to only two rotationally excited states namely, the ${^{2}\Pi_{3/2}, N=1, J =3/2}$, and ${^{2}\Pi_{1/2}, N=2, J = 3/2}$ (see transitions highlighted in Fig.~\ref{fig:energy_level}). The ${^{2}\Pi_{3/2}, N=1, J = 3/2}$ level corresponds to the first rotationally excited level lying at an upper-level energy of only 25.7~K above the ground state. This transition is particularly important, given that its energy is low enough to selectively affect the relative populations of the ground state $\Lambda$-doublet levels \citep{Rydbeck1973}. The HFS lines of the ${N, J = 1, 3/2 \rightarrow 1, 1/2}$ transition near 532/536~GHz (560~$\mu$m) connect the rotational ground 
state to the first excited state (lying at 25.7~K) and the corresponding spectra were observed using
the Heterodyne Instrument for the Far Infrared (HIFI) on board the Herschel Space Observatory \citep{Gerin2010}. 
In addition to the 532/536~GHz rotational lines of CH, HFS transitions between the $\Lambda$-doublet levels of the lowest rotationally excited state, $^{2}\Pi_{3/2}, N=1, J =3/2$, near 700~MHz (42~cm) were first successfully detected by \citet{Ziurys1985}. They observed  the two main HFS transitions  ($F = 2^{-}\rightarrow2^{+}$ and $F = 1^{-}\rightarrow1^{+}$) towards the star-forming region W51\,A using the Arecibo 305~m telescope. While focusing on the detection towards the W51 system, \citet{Ziurys1985} also mentioned the detection of the same transitions towards three other Galactic sight lines, namely, W3, W43 and Orion B, but do not provide details. Shortly thereafter \citet{Turner1988} detected the two satellite lines ($F = 2^{-}\rightarrow1^{+}$ and $F = 1^{-}\rightarrow2^{+}$) of the rotationally excited CH system towards W51\,A, once again using the Arecibo 305~m telescope.

Apart from contributing to our understanding of the masing action of the ground state HFS lines of CH, the detection of the 700~MHz transitions of CH can also serve as a potentially highly interesting probe of magnetic fields as they exhibit Zeeman splitting \citep{Truppe2013, Truppe2014}. While there have been recent attempts to observe these lines \citep[for example,][]{Tremblay2020}, observations of the low frequency HFS transitions corresponding to the $N, J = 1, 3/2$ $\Lambda$-doublet system are nowadays challenging due to man-made radio frequency interference (RFI). As a consequence, to date the only unambiguous detections of the 700~MHz lines of CH remain those made by \citet{Ziurys1985} and \citet{Turner1988}. In addition, efforts to detect other rotationally excited lines of CH within the 4.8~GHz to 24~GHz range, such as those reported by \citet{Matthews1986} and more recently by \citet{Tan2020} towards Galactic sources have been unsuccessful.   \\
With the goal to further our understanding of the anomalous excitation of 
the ground state $\Lambda-$doublet HFS lines of CH, this work revisits the radical's rotationally excited lines at 700~MHz. A combined analysis of the CH transitions within the ground state and first excited levels at radio wavelengths alongside the sub-millimetre (sub-mm)/FIR rotational lines of CH 
promises to help form a comprehensive picture of excitation mechanisms relevant for the ground state lines of this species. In Sect.~\ref{sec:excitation_theory} we recount in more detail the excitation mechanism of the CH ground state lines. Section~\ref{sec:observations} presents the interferometric observations of the HFS lines lying within the 700~MHz $^{2}\Pi_{3/2}, N, J = 1, 3/2$ $\Lambda$-doublet level of CH carried out using the upgraded Giant Metrewave Radio Telescope \citep[uGMRT;][]{Gupta2017}. The observational results are presented alongside a description of the non-LTE radiative transfer models used, in Section~\ref{sec:results}. The resultant physical and excitation conditions are discussed in Sect.~\ref{sec:discussion}, with the main conclusions of this work summarised in Sect.~\ref{sec:conclusions}.

\section{Spectroscopy of CH and excitation of its ground state}\label{sec:excitation_theory}

This section describes the spectroscopy of CH and details the excitation mechanism of the CH ground state with particular emphasis on the role of the first rotationally excited level.\\

The CH $X^{2}\Pi$ ground electronic state  conforms to Hund's case (b) coupling such that each principal quantum level, $N$, splits into two spin-orbit manifolds -- $^{2}\Pi_{1/2}$ and $^{2}\Pi_{3/2}$, due to spin-orbit interactions (see Fig.~\ref{fig:energy_level}). As a result, the absolute ground state of this species is the $^{2}\Pi_{1/2}, J=1/2$ level where $\boldsymbol{J}$ is the rotational quantum number. The orientation of the valence electron's orbital momentum axis, with respect to the rotational axis of the molecule, further  splits the rotational levels into $\Lambda$-doublet levels, which are distinguished based on their parity (either + or $-$). In addition, owing to the non-zero nuclear spin of the hydrogen atom ($I_{\rm H} = 1/2$), each rotational level is further split into HFS levels, $\boldsymbol{F}$. We refer the reader to \citet{Truppe2014} for more information pertaining to the electronic ground state of CH and in particular for details regarding the peculiar ordering of the HFS levels between the $^{2}\Pi_{1/2}$ and $^{2}\Pi_{3/2}$ ladders. \\                          

As discussed briefly in Sect.~\ref{sec:intro}, the fact that the ground state lines of CH portray anomalous excitation across a wide range of physical conditions indicates that there is a universal excitation scheme for CH. While the populations of all three HFS components within the ground state are inverted, observations reveal that the excitation in the lower satellite line ($F=0^- \rightarrow 1^+$) is particularly enhanced. Therefore, the excitation of the CH ground state must involve a combination of radiative and collisional processes, with variations in the relative strengths observed towards different sources dictated by their ambient physical properties. 
In the first step, the underlying pumping cycle must involve collisional excitation to the first rotational level $^{2}\Pi_{3/2}, N=1, J=3/2$. However, given that the rate of radiative decay between rotational levels is faster than that via collisions, collisional excitation to higher rotational levels may be assumed to almost instantaneously decay\footnote{This includes transitions satisfying the selection rule, $N\geq1, \Delta J = 0, \pm 1$ which for the ground state of CH at $N=1, J=1/2$ includes only two rotational levels, at $N=1, J=3/2$ and $N=2, J=3/2$.} back to the ground rotational level. 

The population inversion that results from this decay arises due to parity selection rules of the collisional excitation rates between the rotational ground state and excited levels. The parity selection or parity discrimination rules favour the excitation of the lower half of the $\Lambda$-doublet, such that the rate of collisional excitation between $+\leftrightarrow +$ levels exceeds that between $-\leftrightarrow -$ levels \citep{Bujarrabal1984, Bouloy1984}. In practise, this means that collisional excitation from the (+) parity $\Lambda$-doublet levels of the $N=1, J=1/2$ ground state to the (+) parity doublet levels of the $N=1, J=3/2$ and $N=2, J=3/2$ excited states is more probable than collisional excitations between their corresponding ($-$) parity $\Lambda$-doublet levels. As detailed in Paper I, this is the case because of differences in the state-to-state scattering cross-sections between the upper and lower levels of the CH $\Lambda$-doublets. Consequently, this results in the overpopulation of the lower half (+) of the $^{2}\Pi_{3/2}, N=1, J=3/2$ doublet, which in turn preferentially overpopulates the upper half ($-$) of the ground state $\Lambda$-doublet through subsequent radiative decay. The radiative decay channels are illustrated in the inset of Fig.~\ref{fig:energy_level}a. \\

In addition to collisions, radiation may also pump the ground state maser. Analogous to the case of OH, discussed in depth by \citet{Litvak1969}, population inversion in the ground state of CH can also be caused by external or internal sources of FIR radiation or via resonance line fluxes. This radiation can be absorbed by the sub-mm/FIR rotational transitions of CH, resulting in a net transfer of population between the HFS-split levels of the $\Lambda$-doublet states thereby reinforcing the masing action in the ground state. Such a pumping scheme induced by FIR radiation or FIR pumping has been invoked to understand the enhanced emission observed in the lower satellite line of CH (at 3.264~GHz) toward the sub-mm/FIR bright continua emitted from the
dusty envelopes of H{\small II} regions \citep{Bujarrabal1984, Stacey1987}.

 The pumping cycle can be visualised as follows for transitions between the $^{2}\Pi_{3/2}, J=3/2$ and $^{2}\Pi_{1/2}, J=3/2$ states, which we refer to as the the 560~$\mu$m pump: consider the absorption of FIR photons through line 7 (see Fig.~\ref{fig:energy_level}b.) which transfers molecules from energy level $d \rightarrow e$, this is followed by radiative decay from energy level $e$ to both the energy levels in $^{2}\Pi_{1/2}(+)$, i.e., $d$ and $c$. Among these, decays via line 8 between levels $e \rightarrow c$ are more probable due to its higher Einstein A-coefficients (see Table~\ref{tab:spec_properties}). This results in a net population transfer from level $d$ to $c$, leading to level inversion in line 1 (3.264~GHz). Conversely, for transitions excited via line 9 ($d \rightarrow f$), an increase in the photon trapping has no effect on the excitation of the ground state because the $f \rightarrow d$ rotational transition serves as the only permitted de-excitation route. As a consequence, for CH, line overlap via the 560~$\mu$m pump favours the overpopulation of energy level $c$ or the $0^-$ level of the ground state. Another, although less likely overlap would occur if molecules were transferred from energy levels $b$ to $a$ through the absorption of FIR photons in line 5, resulting in the under-population of energy level $b$ or the $1^+$ level of the ground state. Similarly, the relative roles played by the rotational lines connecting the $^{2}\Pi_{1/2}, J=3/2$ state to the ground state or the 160~$\mu$m pump can be visualised using the schematic presented in Fig.~\ref{fig:energy_level}c.

\begin{figure}
    \centering
\includegraphics[width=0.45\textwidth]{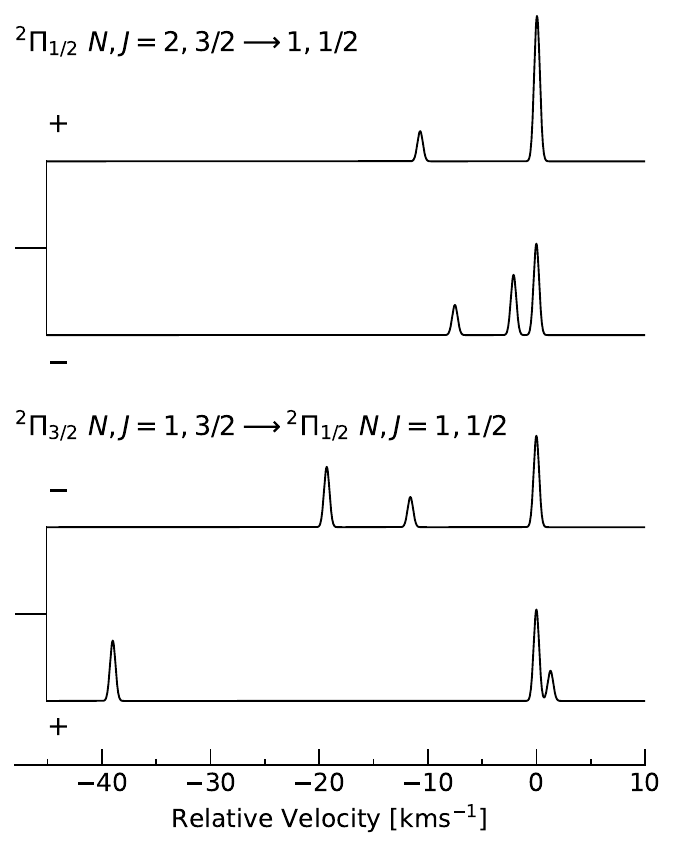}
    \caption{Normalised spectrum of the HFS splitting of the sub-mm and FIR transitions of CH connecting the $^{2}\Pi_{1/2}, N=2, J=3/2$ (top) and $^{2}\Pi_{3/2}, N=1, J=3/2$ (bottom) rotational levels to the ground state. The line intensities of the HFS triplets are computed assuming optically thin conditions at LTE for a Doppler line width at a gas temperature of 50~K. The velocity scale is displayed relative to the strongest HFS component.}
    \label{fig:HFS-structure}
\end{figure}

Additionally, the radiative transfer rates and, therefore, the FIR pumping cycle can be significantly altered by the overlapping of two lines as a result of their Doppler width (line broadening) or
Doppler shifts caused by bulk velocity gradients in the gas. 
This is because line overlap affects photon trapping, allowing for a photon emitted in one transition to be (partially) absorbed in a different one. As a consequence, this modifies the degree to which the pumping radiation in overlapping lines are absorbed. Furthermore, the process of line overlapping in the sub-mm/FIR lines is specific to the HFS splitting. As shown in Fig.~\ref{fig:HFS-structure} the (+) parity halves of the FIR lines in both the $^{2}\Pi_{1/2}, J=3/2$ and $^{2}\Pi_{3/2}, J=1/2$ orbital manifolds show larger HFS splits. Such uneven splits between $\Lambda$-doublet levels creates an asymmetry in the overlapping of the pumping lines that are coupled to the ground state across a broad range of column densities, thus enhancing the potential for a stronger maser effect \citep[see also][]{Elitzur1977, Bujarrabal1984}. However, while both the 560~$\mu$m and 160~$\mu$m pumps contribute toward ground state inversion through effects of line overlap, the 560~$\mu$m pump is more efficient due to greater asymmetries in the HFS splittings of the involved lines, particularly for line widths $\geq2~$km~s$^{-1}$ (see Fig.~\ref{fig:HFS-structure}).

\section{Observations} \label{sec:observations}

\begin{table*}
    \centering
    \caption{Spectroscopic properties of the CH ground-state HFS transitions.}
    \begin{tabular}{llrlcc}
    \hline \hline 
       \multicolumn{2}{c}{Transition}  &   Frequency  &  \multicolumn{1}{c}{$A_{\text{ul}}$\tablefootmark{a}} & Energy & Telescope \\
       $N, J$ & $F^{\prime} - F^{\prime\prime}$ & \multicolumn{1}{c}{[MHz]} & \multicolumn{1}{c}{[s$^{-1}$]} & \multicolumn{1}{c}{[K]}\\ 
         \hline 
         $1, 1/2$ &
        $0^{-}$ -- $1^{+}$ & 3263.793 & 2.876($-$10) & 0.15 & VLA \\
        & $1^{-}$ -- $1^{+}$ & 3335.479 & 2.045($-$10) \\
        & $1^{-}$ -- $0^{+}$ & 3349.192 & 1.036($-$10) \\
         \hline 
         $1, 3/2$ & $2^{-}$ -- $2^{+}$ &  701.677 & 2.067($-$12) & 25.76 & uGMRT\\
         & $2^{-}$ -- $1^{+}$ &  703.978 & 0.231($-$12) \\
         & $1^{-}$ -- $2^{+}$ &  722.487 & 0.417($-$12) \\
         & $1^{-}$ -- $1^{+}$ &  724.788 & 2.109($-$12) \\
         \hline 
         & & \multicolumn{1}{c}{[GHz]} & \\
         \hline 
          $1, 3/2$ -- $1, 1/2$ & $1^{+}$ -- $ 1^{-}$ & 532.72159 &  2.069($-$4) & 25.76 & Herschel/HIFI
     \\ 
                      & $ 2^{+}$ -- $1^{-}$\tablefootmark{b} & 532.72389 &  6.207($-$4)
                      \\
                       & $1^{+}$ -- $0^{-}$ & 532.79327 &  4.139($-$4)
                       \\
      & $2^{-}$ -- $1^{+}$\tablefootmark{b} & 536.76105 &  6.378($-$4)
     \\
                       & $1^{-}$ -- $1^{+}$ & 536.78185 &  2.126($-$4)
                       \\
                       & $1^{-}$ -- $0^{+}$ & 536.79557 &  4.251($-$4)
                       \\
    \hline
    $2, 3/2$ -- $1, 1/2$ & $1^{-}$ -- $1^{+}$ & 2006.749 & 1.117($-$2) & 96.31 & SOFIA/upGREAT\\
                    & $1^{-}$ -- $0^{+}$ & 2006.763 & 2.234($-$2)   \\
					& $2^{-}$ -- $1^{+}$\tablefootmark{b} & 2006.799 & 3.350($-$2) \\
     & $1^{+}$ -- $1^{-}$\tablefootmark{c} & 2010.739 & 1.128($-$2) & \\
                    & $1^{+}$ -- $0^{-}$\tablefootmark{c} & 2010.811 & 2.257($-$2) &  \\
					& $2^{+}$ -- $1^{-}$\tablefootmark{c} & 2010.812 & 3.385($-$2) &  \\
     \hline
    \end{tabular}
    \tablefoot{The columns are (from left to right): the transition as described by the hyperfine quantum number ($\boldsymbol{F}$), the frequency of the transition, the Einstein A coefficient ($A_{\rm ul}$), the upper level energies and the telescope (receiver) used for carrying out these observations. The frequencies of the HFS lines near 3.3~GHz and 700~MHz were measured in the laboratory by \citet{Truppe2013} with uncertainties of 3~Hz and $\approx\!20$~Hz, respectively, while those of the 532/536~GHz rotational lines where measured by \citet{Truppe2014} with 0.6~kHz uncertainty. The frequencies of the 2006/2010~GHz rotational lines are taken from \citet{Davidson2001} and have uncertainties of $\approx\!0.15$~MHz. \tablefootmark{a}{The values in the brackets represent the exponent of the Einstein A coefficient.}\tablefootmark{b}{Indicates the HFS transition that was used to set the velocity scale in the analysis.}\tablefootmark{c}{The CH HFS transitions at $2010\,$GHz were not observed because of strong contamination from atmospheric ozone features at $149.1558\, \mu$m and $149.7208\, \mu$m.}}
    \label{tab:spec_properties}
\end{table*}

Observations of the HFS lines between the $\Lambda$-doublet levels of the first excited state of CH near 700~MHz were carried out on 2019 September 7, 10, 12, 16, 17, and 23 using band\,--\,4 (550\,--\,850~MHz) receivers of the uGMRT (Project id: 36\_012, PI: Jacob) for a total observing time (including overheads) of 4.6~hours per source. The frequencies and spectroscopic properties of the different CH transitions studied in this work are tabulated in Table~\ref{tab:spec_properties}. In this pilot study, a total of six well-known H~{\small\!II} regions with associated photodissociation regions (PDRs) were observed based on their continuum fluxes at 700~MHz. The target list includes W51\,E\, a source within the W51\,A region that contains young stellar objects, towards which all four of the rotationally excited HFS lines of CH have been detected before \citep{Ziurys1985, Turner1988}. These observations in particular form the test bed for assessing the amount of single dish flux that is recovered by the interferometer. In addition to this high-mass star forming region in W51\,A, our sample consists of W43 (one of the regions towards which \citealt{Ziurys1985} also report a detection of the 700$\,$MHz CH lines), Sgr~B2~(M), DR21~Main, M8 and M17. All of these sources are prominent star forming regions that contain multiple high mass young stellar objects that excite H~{\small II} regions and PDRs. It is worth mentioning that previous attempts to search for these CH transitions towards Sgr~B2~(M) \citep{Ziurys1983} were unsuccessful. However, re-observing them now holds significance due to the lack of accurate frequency measurements from the laboratory in those early attempts. Apart from being extensively studied regions, the subset of sources selected for this study are all characterised by strong mm and FIR continuum emission towards which complementary observations of the other relevant CH (radio and sub-mm/FIR) transitions have been made. Information on the sources discussed in our study and observational parameters are summarised in Table~\ref{tab:source_parameters}. \\

\begin{table*}
\centering
  \caption{Summary of source parameters. }
    \begin{tabular}{lll c c c cr cc}
    \hline \hline 
         \multicolumn{1}{c}{Source} & \multicolumn{2}{c}{Coordinates (J2000)}  & $d$ & $\upsilon_{\text{LSR}}$ & \multicolumn{1}{c}{Frequency} & \multicolumn{1}{c}{$\theta_{\rm maj} \times \theta_{\rm min}$} & P.A. & rms  & $S_{\rm cont. peak}$\\ 
           &   $\alpha$~[hh:mm:ss] & $\delta$~[dd:mm:ss]  
           & [kpc] & [km~s$^{-1}$] & \multicolumn{1}{c}{[GHz]} & \multicolumn{1}{c}{[$^{\prime\prime}\times{}^{\prime\prime}$]} & [$^{\circ}$] & [mJy/beam] &  [Jy]\\
         \hline  
         Sgr~B2~(M) & 17:47:20.50 & $-$28:23:06.00 &
         8.2  & 64.0 & 701 & $27.6 \times 27.4$&  $71.4$ & 5.8 &  0.69 \\
         & & & & & 703 &$25.7\times 22.5$ & $-2.8$& 4.2 & 0.67  \\
        & & & & & 722 & $30.2\times 25.0$ & 6.2 & 5.7 & 0.70 \\
        & & & & & 724 &  $32.3 \times 26.8$&  $-6.4$ &6.4 & 0.81 \\
         M8 & 18:03:37.00 & $-$24:23:12.00 & 1.3 & 10.0 & 701 & $38.0 \times 22.6$&  $18.0$ & 26.4 & 1.80 \\
         & & & & & 703 &$33.7\times 22.8$ & 23.8 & 25.7 & 1.81   \\
        & & & & & 722 & $41.2\times 26.5$ &18.0 & 21.8 & 1.64 \\
         & & & &  & 724 &  $43.6 \times 27.9$&  $14.5$ & 21.9 & 1.65  \\
         M17 & 18:20:47.00 &  $-$16:10:18.00 & 1.8 &  22.0& 701 & $46.6 \times 40.2$&  $36.2$ & 9.3 & 2.60 \\
         & & & & & 703 & $28.7 \times 21.4$& 32.6& 16.3 & 2.44   \\
        & & & & & 722 & $33.8 \times 23.9$ & 33.1& 17.2 & 2.31 \\
         & & & &  & 724 &  $33.4 \times 24.4$&  $33.3$ & 13.3 & 2.39 \\
         W43 & 18:47:32.40 & $-$01:56:31.00 & 3.1 &  97.8 & 701 & $20.8 \times 15.2$&  $43.7$ & 2.0 & 1.44 \\
         & & & & & 703 & $25.5 \times 18.2$ & 57.1 & 0.8 & 1.85  \\
        & & & & & 722 & $33.1 \times 21.2 $& 50.5 & 1.2 & 1.79 \\
         & & & &  & 724 &  $31.6 \times 22.1$&  $44.2$ & 2.5 & 1.40  \\
         W51\,E & 19:23:43.90 & +14:30:31.00 & 5.4 &  57.0 & 701 & $35.5 \times 26.2$&  $33.2$ & 8.4 & 3.11 \\
         & & & & & 703 & $35.5 \times 30.1$ & 27.0 & 4.7 & 2.34   \\
        & & & & & 722 & $38.6 \times 26.9$ & 39.8 & 6.8 & 3.06 \\
          & & & & & 724 & $42.1 \times 31.2$&  $32.9$ & 24.9 & 2.36 \\
         DR21~Main & 20:39:01.59 &  +42:19:37.80 & 1.5 &  $-$4.0 & 701 &  $10.2 \times 6.4$&  $36.2$ & 2.4 & 2.33 \\
         & & & & & 703 & $ 13.9 \times 8.4$ & 44.4 & 2.1 & 2.16   \\
        & & & & & 722 & $13.9 \times 8.3$& 44.4 & 1.6 & 2.23 \\
         & & & &  & 724 &  $10.2 \times 6.5$&  $43.5$ & 4.4 & 2.39 \\
         \hline 
         \end{tabular}
         \tablefoot{The columns are (from left to right): the source designation, the equatorial source coordinates, the heliocentric distances, systemic velocity of the source, FWHM of the synthesised beams ($\theta_{\rm maj} \times \theta_{\rm min}$), position angles (P.A.) and the root-mean-square (rms) noise levels of the CH lines and the continuum peak near 700~MHz, respectively. The rms noise levels are quoted for a spectral channel width of $1.3$~km~s$^{-1}$. }
  \tablebib{For the heliocentric distances: Sgr~B2~(M), W43:~\citet{Reid2019}; M8:~\citet{Damiani2019}; M17:~\citet{GaiaCollab2018}; W51\,E:~\citet{Sato2010}; DR21~Main:~\citet{Rygl2012}.}
 
    \label{tab:source_parameters}
\end{table*}

Tuned to 713~MHz, the GMRT wideband backend (GWB) permitted simultaneous observations of all four of the HFS lines arising from the CH $\Lambda$-doublet in the $N, J=1,3/2$ level with a bandwidth of 50~MHz across 16384~channels, yielding a spectral resolution of 3.05~kHz (${\sim\!0.13}$~km~s$^{-1}$). Furthermore, this setup ensures sufficient velocity as well as continuum coverage since these lines are expected to appear in absorption. The quasars 3C48 and 3C286 were used as both the bandpass and flux calibrators, while J1925+211, J1822$-$096, J2052+365, J1833$-$210, 1751-253, and J1822$-$096 were used as phase calibrators for the different epochs in which W51\,E, W43, DR21~Main, Sgr~B2~(M), M8 and M17 were observed, respectively. In addition to the CH lines, the spectral window setup used also covered five hydrogen-, helium-, and carbon- radio recombination lines (RRLs) with principal quantum numbers, $n$, between 207 and 211. Although they are excellent tracers of PDRs and can probe magnetic fields through Zeeman splitting similar to the H\,{\small I} 21~cm line \citep{Troland1977,Greve1980}, the RRLs will not be discussed here further as this work focuses on the rotationally excited lines of CH.

The data were calibrated and imaged using the Common Astronomy Software Applications \citep[CASA;][]{CASA2022} and the Astronomical Image Processing System \citep[AIPS;][]{AIPS1985}. The primary calibration and `flagging' were carried out manually, following the standard calibration procedure. Apart from a few short baseline antenna pairs that were corrupted by RFI, the majority of the data was free from RFI. 
 Continuum subtraction was performed on the visibility data using the task {\tt uvsub} in CASA, for which a model image of the continuum emission was made by including only channels that were free from any line emission or absorption. The final continuum-subtracted visibilities were CLEANed and imaged with a pixel size of $3^{\prime\prime}$ and an image size of $1\rlap{.}^{\circ}7\times 1\rlap{.}^{\circ}7$ using {\tt IMAGR} in AIPS.

For consistency in the analysis that follows, the intensity scales were converted from specific brightness, $S$, in units of Jy per beam to brightness temperature, $T_{\rm B}$, in Kelvin, using $T_{\rm B}/S$ = {$1.22\times10^{6}/\left(\nu^{2}({\rm GHz})\,\theta_{\rm min}(^{\prime\prime})\times \theta_{\rm max}(^{\prime\prime})\right)$~K/Jy} based on the Rayleigh-Jeans relation. For the ease of comparison, the spectra of the 700~MHz CH lines were extracted from a region encompassing the same area as that by the full-width at half maximum (FWHM) beam width of the Herschel/HIFI band 1a receiver, of $40\rlap{.}^{\prime\prime}4$ and centred at the same position towards which both the sub-mm and FIR observations (where available) were carried out. These pointing positions correspond to the map's centres, with their respective positions tabulated in Table~\ref{tab:source_parameters} for each source. The spectra were further post-processed and analysed using Python packages including: SciPy \citep{2020SciPy-NMeth}, Astropy \citep{Astropy2013, Astropy2018} and APLpy \citep{AplPy2012}.

\section{Results} \label{sec:results}
Our search for the elusive 700~MHz lines of CH has resulted in the successful detection of this line only towards W51\,E, which verifies the original detection by \citet{Ziurys1985} and \citet{Turner1988}. With a complete census of CH lines obtained only towards W51\,E, the analysis and subsequent modelling are only performed towards this sight line. The non-detection of the 700~MHz lines toward the other observed targets is summarised in Table~\ref{tab:source_parameters}, with a discussion on the probable reasons for these non-detections provided in Sect.~\ref{subsec:nondetection}.

\subsection{Spectral line profiles}\label{subsec:line_profile}
The W51\,A cloud complex is located in the Sagittarius spiral arm at a distance of 5.41$^{+0.31}_{-0.28}$~kpc \citep{Sato2010} and is one of the most active and luminous sites of massive star formation in the Milky Way. The chemical richness of this region may be attributed to the presence of strong feedback amidst on-going star formation. 

In addition to the extensively studied massive protocluster W51 Main (M) or IRS1, the W51\,A region also harbours two other luminous condensations of high-mass young stellar objects -- W51~North (N) or IRS2 and W51 East (E), both of which hosts a number of hyper-, and ultra-compact {H\,{\small II}} regions \citep[see][and references therein]{Sato2010, Ginsburg2017}. While both sources are comparably bright at radio and sub-mm wavelengths, the radio emission from W51\,E, arising from a cluster of H~{\small\!II} regions unresolved by our uGMRT beam was found to be more compact than that from W51\,N \citep{Ginsburg2017a}. 

\begin{figure*}
    \centering
    \includegraphics[width=0.9\textwidth]{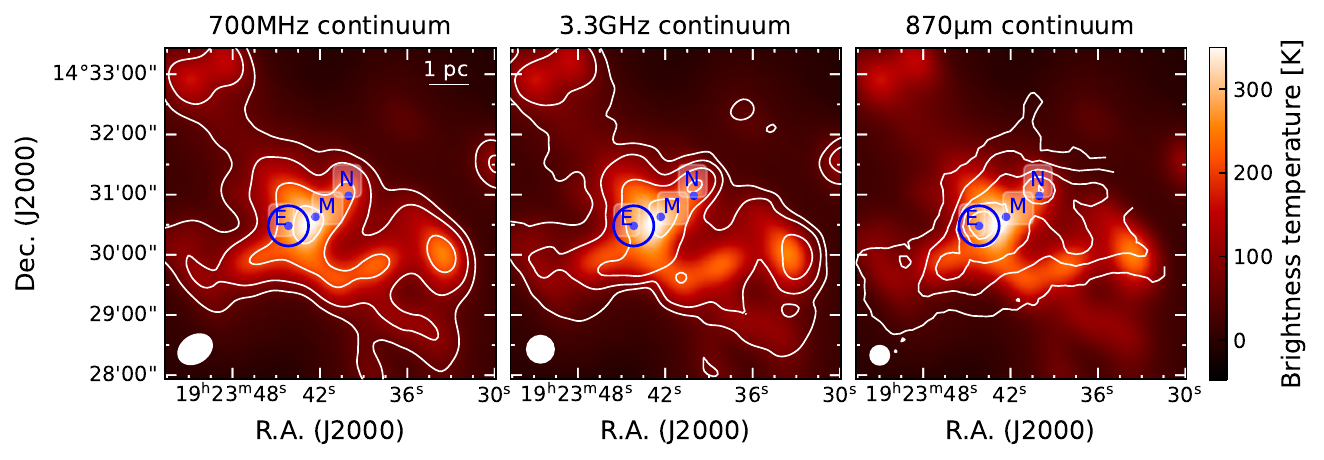}
    \caption{Overview of the radio continuum data. From left-to-right the panels display contours of the background continuum emission at 700~MHz, 3.3~GHz, and 870~$\mu$m (in white), respectively, overlaid atop the 700~MHz continuum emission. Labelled and marked (in blue) are the positions of the main young cluster W51 M and the two young stellar objects W51 N and E, which harbour several high mass young stellar objects at different (early) evolutionary stages. The position from which the CH spectra are extracted is marked by the beam of the Herschel/HIFI band 1a in blue. The filled white ellipse at the bottom left-hand corner of each plot displays, from left-to-right the (synthesised) beams of the uGMRT, VLA and APEX/LABOCA observations, respectively. }
    \label{fig:continuum_map}
\end{figure*}

Figure~\ref{fig:continuum_map} presents an overview of the radio continuum emission at 700~MHz, towards the W51\,A complex. A comparison with the 3.3~GHz data presented in Paper I reveals that the continuum emission has a similar morphology at both frequencies.  
Notably, the distribution of the radio continuum emissions differs from that of the dust emission seen at 870~$\mu$m in the APEX Telescope Large Survey of the Galaxy \citep[ATLASGAL;][]{Schuller2009}, such that the peak of the dust emission is also offset from W51\,M (see Fig.~\ref{fig:continuum_map}). This is unsurprising, given that the 870~$\mu$m dust emission traces the densest parts of the H~{\small II} region where the strong dust continuum peak coincides with the position towards which \citet{Ziurys1985} first detected the CH 700~MHz lines albeit with the 7$\rlap{.}^{\prime}$7 beam of the Arecibo 305~m telescope. Therefore, the observations discussed here are carried out towards a position along the molecular ridge of W51\,E as marked by the blue circle in Fig.~\ref{fig:continuum_map}. This is the position towards which \citet{Gerin2010} observed the 532/536~GHz sub-millimetre CH lines with HIFI on Herschel and 
it coincides with the strong dust continuum peak observed at 870~$\mu$m with ATLASGAL.

\begin{figure*}
\begin{center}
    \includegraphics[width=0.422\textwidth]{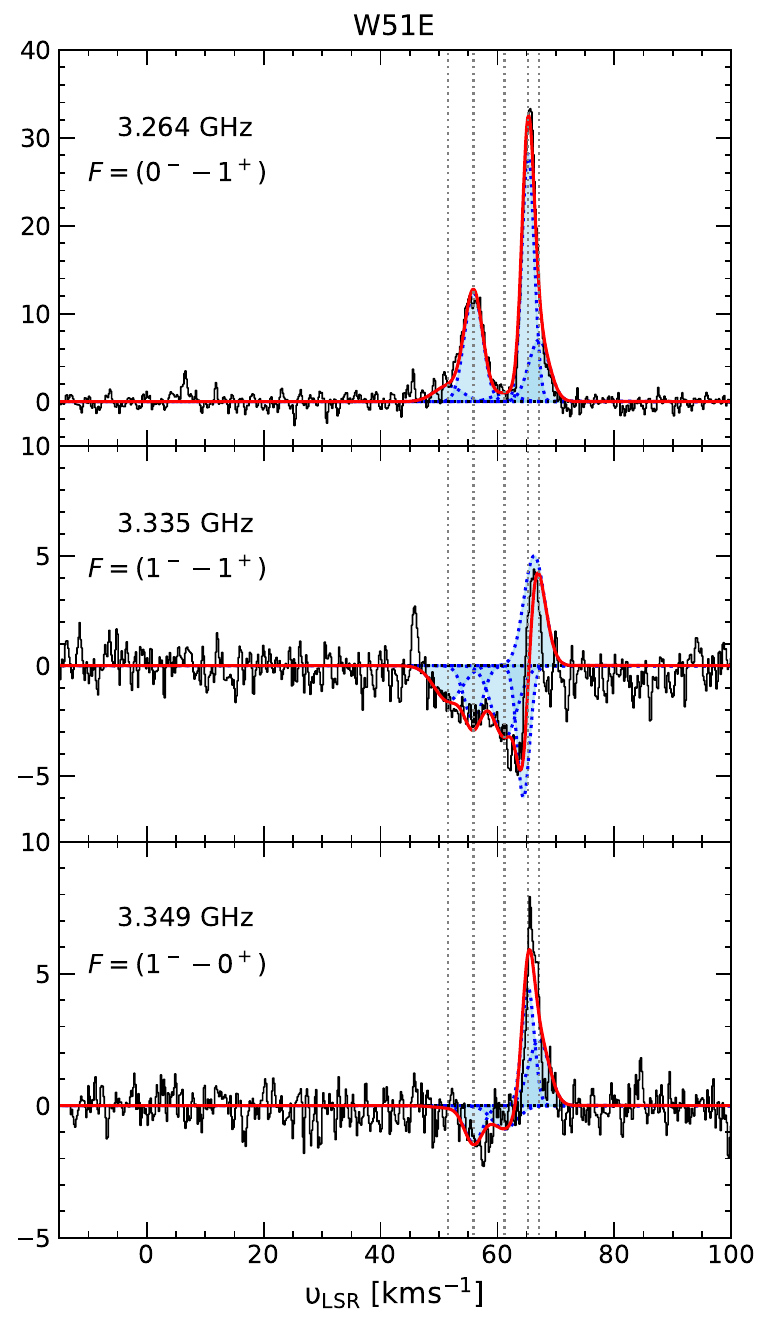} \quad \includegraphics[width=0.42\textwidth]{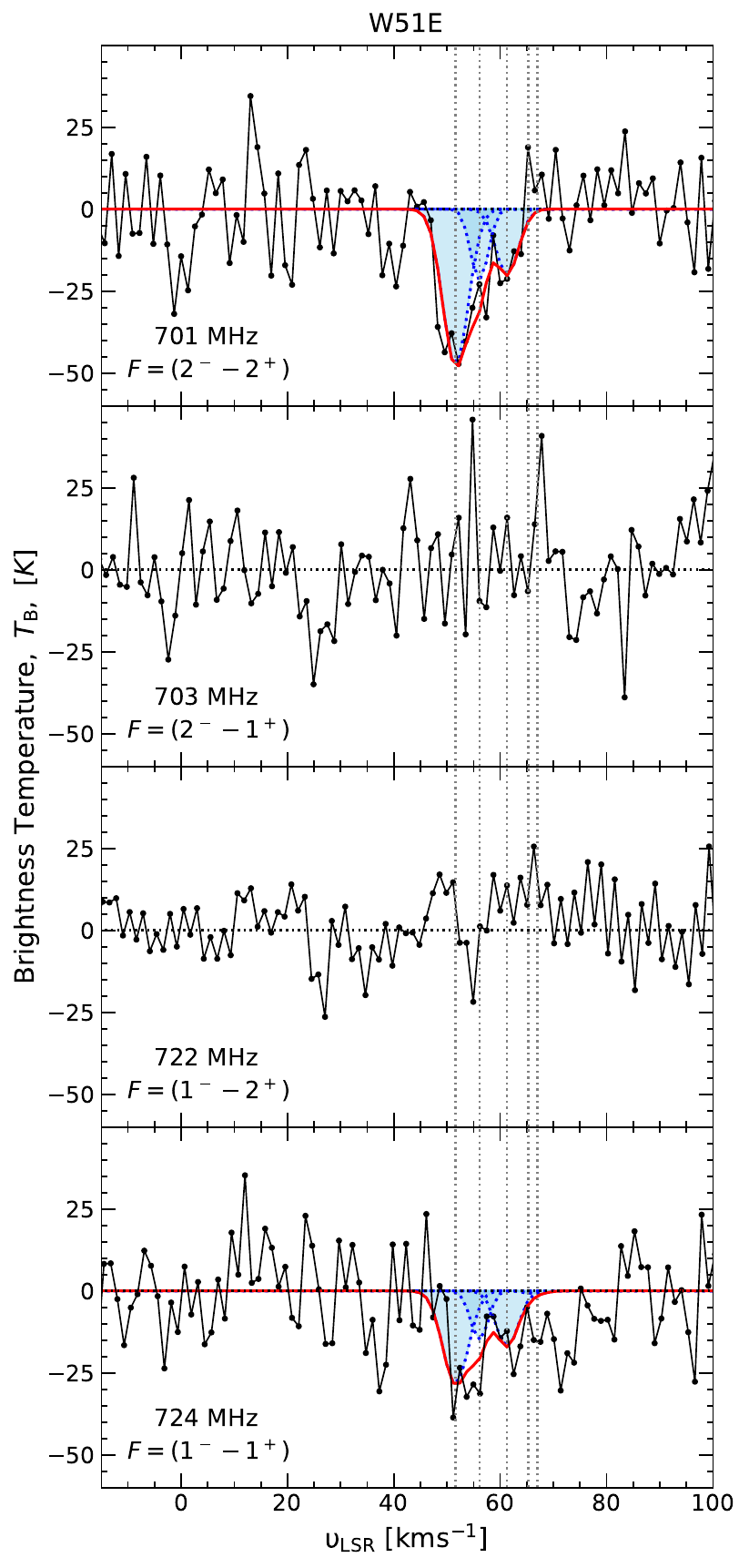}
    \caption{Baseline-subtracted spectra of the ground state (left), and first rotationally excited state (right) HFS lines of CH towards W51\,E. The individual fits to different velocity components are displayed by dotted blue curves and highlighted by the blue shaded regions, while the solid red curves display the combined fits. The dotted grey lines mark the centroids of the different fitted Gaussian components. }
    \label{fig:spectra-700}
        \end{center}
\end{figure*}

The calibrated and baseline-subtracted spectra of the HFS lines of the first rotationally excited level of CH towards W51\,E are presented in Fig.~\ref{fig:spectra-700}. The spectra are extracted from an area equivalent to a FWHM beam width of 40$\rlap{.}^{\prime\prime}$4 corresponding to the Hershel/HIFI beam at 530~GHz (see Sect.~\ref{sec:observations}). The main lines, at 701~MHz and 724~MHz
are seen in absorption and detected at $3.1\sigma$ and $2.6\sigma$ levels, respectively, while absorption in the satellite lines is not discernible at the current rms noise level (see Table~\ref{tab:source_parameters}). This is expected under conditions of LTE since the relative line strengths follow a ratio of ${I_{\rm 701~MHZ}:I_{\rm 703~MHz}:I_{\rm 722~MHz}:I_{\rm 724~MHz} \sim 1.20:0.13:0.13:0.67}$. 
Overall, the observed line shapes and widths are consistent with those presented in \citet{Turner1988}, but the absorption depths seen in the spectra presented here lie closer to 53~km~s$^{-1}$ than 60~km~s$^{-1}$ as noted by \citet{Turner1988}. These differences can be attributed to differences in beam sizes used in both sets of observations, where the 7$\rlap{.}^{\prime}$7 beam of the Arecibo 305~m telescope used by \citet{Ziurys1985} and \citet{Turner1988} contains contributions from not just the W51\,E ridge but from the entire W51\,A giant molecular cloud. For comparison, Fig.~\ref{fig:spectra-700} also displays the three ground-state HFS transitions of CH near 3.3~GHz towards the same position observed using the Karl G. Jansky Very Large Array (VLA).  
While the VLA data were previously presented in Paper I, in order to carry out a fair comparison the spectra presented here were extracted from the same region, enclosed within a 40$\rlap{.}^{\prime\prime}$4 beam and converted to brightness temperature scales using a conversion factor of 146~K/Jy. 

\begin{figure}
\centering 
    \includegraphics[width=0.41\textwidth]{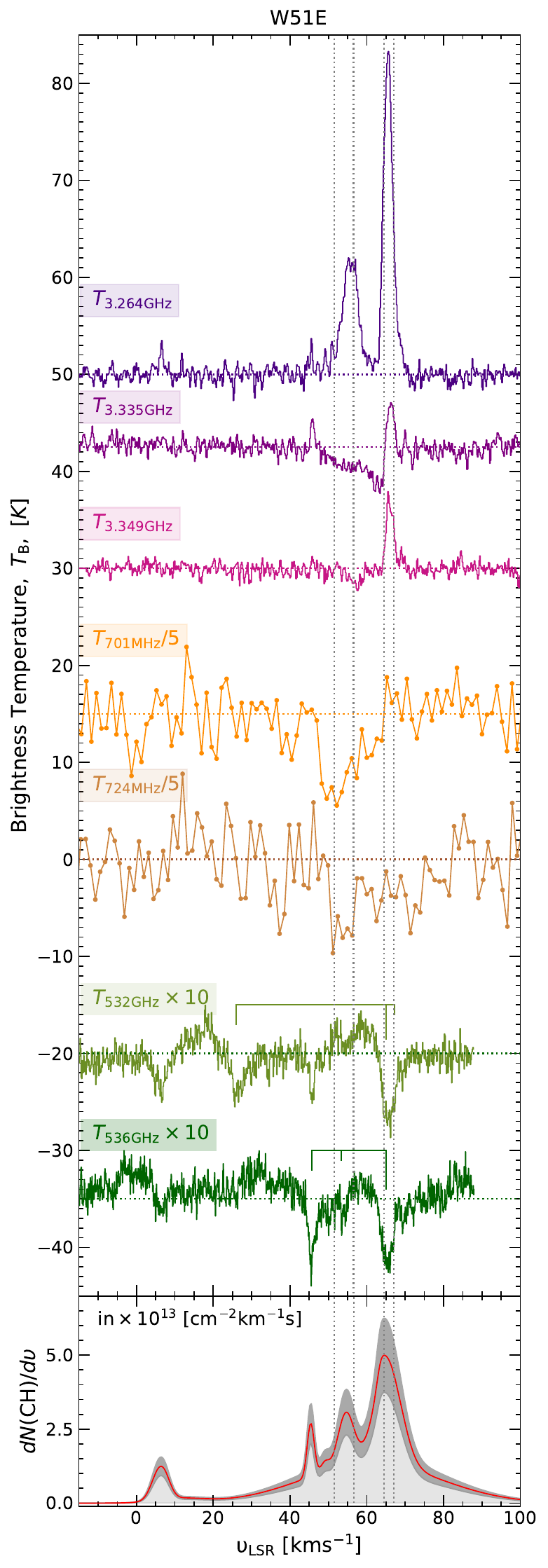}
    \caption{Top panel: Baseline-subtracted spectra of the CH transitions discussed in this work, displayed on brightness temperature scales. The spectra are offset and scaled for ease of viewing. The relative intensities and positions of the HFS splitting lines corresponding to the $N=1, J=3/2\rightarrow1/2$ rotational transitions near 532~GHz and 536~GHz are indicated above the spectra and aligned to a velocity of 57~km~s$^{-1}$ corresponding to the LSR velocity of the strongest component. Bottom panel: HFS deconvolved column density distribution of CH obtained from the $N, J=2, 3/2\rightarrow1,1/2$ transition near 2007~GHz. The dotted grey lines mark the centroid velocities of the different Gaussian components that were fitted in Fig.~\ref{fig:spectra-700}.}
    \label{fig:spectrum_panelplot}
\end{figure}

The bulk of the emission and absorption observed towards W51\,E lies in the velocity range between 51\,km~s$^{-1}$ and 70~km~s$^{-1}$. However, due to its tangential position in the Sagittarius spiral arm, the line of sight towards W51\,E traces very little molecular cloud material across the Galactic plane besides local gas near 7\,km~s$^{-1}$. 
The dominant absorption  
observed in the spectra of the excited CH lines at 700~MHz lies close to 55~km~s$^{-1}$, which is within the local standard of rest (LSR) velocity range (53\,--\,60~km~s$^{-1}$) determined for multiple molecular lines by \citet{Kalenskii2022} toward the position of W51 e1/e2, which is within $5^{\prime\prime}$ of our analysed position. These lines were observed with an angular resolution ($\approx\!40^{\prime\prime}$), close to our synthesised beam FWHM. 
Furthermore, it is consistent with the velocity of the emission peak seen in the C$^+$ 158~$\mu$m line towards W51\,E by \citet{Gerin2015}. Similarly, low-lying rotational transitions of H$_2$CO and CH$_3$OH also display deep absorption line features centred at ${\sim\!56~}$km~s$^{-1}$ \citep{Ginsburg2017}. 
In contrast, the most prominent features observed in the spectra of the other CH transitions discussed in this work, towards W51\,E, whether in absorption or emission, cover velocities from 62~km~s$^{-1}$ to 72~km~s$^{-1}$ and are centred around 66~km~s$^{-1}$ (see panel plot in Fig.~\ref{fig:spectrum_panelplot}). This velocity range is associated with a more extended foreground gas component at 68~km~s$^{-1}$ which was first identified in early H{\small\,I} surveys \citep{Burton1970} and dubbed the high-velocity streamer. 
CH at lower velocities than the last two components (down to $\upsilon_{\rm LSR} = 45$~km~s$^{-1}$) traces gas associated with the far-side of this cloud which is likely interacting with the Sagittarius spiral arm. While it is unclear whether the different cloud components discussed above are physically related or if they represent distinct structures, they are believed to be kinematically connected \citep[see][for more details on the complex geometry of this sightline]{Ginsburg2015}. 

Multiple Gaussian profiles are used to fit the individual velocity components for both the ground- and excited-state lines, simultaneously. The Gaussian model decomposition was carried out using the Python package lmfit. This package employs the Levenberg-Marquardt algorithm to carry out a non-linear least squares fit, with the positions and linewidths for a given Gaussian component set to the same value between the different lines modelled. The best-fit values are determined by minimising the residual sum of squares or the difference between the observations and predicted fit parameter values. The selection of individual velocity components toward the background continuum of W51~E was motivated by the detection of distinct features in the CH spectra as well as those identified when using other species, as discussed above. Furthermore, given the relatively low signal-to-noise ratios of the observed 700~MHz lines in particular, we assessed the robustness of the best-fit parameters by estimating the posterior probability density distribution using a Markov Chain Monte-Carlo sampler \citep[MCMC;][]{EMCEE2013}.  

The results of the MCMC simulations are visualised through probability density distributions in Appendix~\ref{appendix:fit_checks} with the best-fit spectral line properties summarised in Table~\ref{tab:spectral_fit_prop}. These parameters serve as crucial inputs for the subsequent non-LTE modelling. We attribute minor differences between the fit parameters derived in this work and those listed in Table 3 of Paper I for the 65~km~s$^{-1}$ and 67~km~s$^{-1}$ velocity components of the ground-state lines to differences in the spectral line shapes. These differences arise due to the larger beam width of $40\rlap{.}^{\prime\prime}4$, used to extract the 3.3~GHz spectra presented in this work in comparison to the 13$\rlap{.}^{\prime\prime}$5 beam used in Paper I. In particular, we observe an additional feature near 60~km~s$^{-1}$ in the 3.335~GHz line of CH. The multi-component Gaussian fitting routine includes a component at this velocity and its contributions are included in the subsequent analyses even though this feature is not clearly detected in either of the 3.3~GHz satellite lines. Furthermore, a comparison with the spectral line profiles of other species, like CO toward this sight line, by \citet{Fujita2021} also only report weak emission arising from the 60~km~s$^{-1}$ cloud component.      \\

Unlike the other CH transitions discussed here, both the low- and high-velocity extended gas components do not appear in the 700~MHz lines of CH. 
Qualitatively this suggests that the physical conditions traced by the 66~km~s$^{-1}$ high-velocity streamer, which is itself composed of multiple components \citep[see decomposition presented in][]{Jacob2021}, traces low density gas in comparison to the main cloud component at 55~km~s$^{-1}$. This is also corroborated by gas density maps derived from H$_2$CO absorption maps at 2~cm and 6~cm by \citet{Ginsburg2015}, who derive gas densities, $n_{\rm H}$, of ${>\!10^{5}}$~cm$^{-3}$ towards W51\,E and $10^{3}~{\rm cm}^{-3} \leq n_{\rm H} \leq 3\times10^{4}~$cm$^{-3}$ in the high-velocity filament. Moreover, this is supported by the emission features seen in the 532~GHz and 536~GHz CH lines at 55~km~s$^{-1}$, while foreground absorption is seen from the low-density component at 67~km~s$^{-1}$.

\begin{table}[]
\small
\centering
    \caption{Summary of the line fit parameters for the CH HFS lines detected within the $^{2}\Pi_{1/2}, N=1, J=1/2$ ground state and $^{2}\Pi_{3/2}, N=1, J=3/2$ rotationally excited state towards W51\,E.}
    \begin{tabular}{lllrr}
    \hline \hline
         Frequency  &   \multicolumn{1}{c}{$\upsilon_{\rm LSR}$} & \multicolumn{1}{c}{$\Delta\upsilon$}  & \multicolumn{1}{c}{Peak  $T_{\rm B}$} &  \multicolumn{1}{c}{$\int T_{\rm B}\Delta\upsilon$}\\
          \![MHz]&  [km~s$^{-1}$] & [km~s$^{-1}$] & \multicolumn{1}{c}{[K]} & \multicolumn{1}{c}{[K~km~s$^{-1}$]} \\
         \hline 
         3264 & $51.6\pm3.1$ & $5.7 \pm 0.3$ & $1.8 \pm 0.1$ & $10.1\pm0.8$\\
         & $56.1\pm3.2$ & $3.8 \pm 0.2$ & $12.0 \pm 0.7$ & $45.6\pm3.6$\\
         & $61.3\pm2.6$ & $5.0\pm0.2$& $0.8\pm0.03$ & $4.1\pm0.2$\\
         & $65.3\pm3.2$ & $2.5 \pm 0.1$ & $28.4 \pm 1.5$ & $71.0\pm4.7$\\
         & $67.0\pm2.4$ & $4.1 \pm 0.2$ & $7.0 \pm 0.3$ & $28.7 \pm1.8$\\
         
         3335 & $51.7\pm3.0$ & $5.7 \pm 0.3$ & $-1.6\pm 0.1$ & $-9.1 \pm0.7$\\
         & $56.1\pm3.2$ & $3.9 \pm 0.2$  & $-2.5\pm 0.1$ & $-9.8\pm0.6$\\
         & $61.3\pm3.6$ & $5.0\pm0.3$ & $-3.3\pm0.2$ & $-16.5\pm1.4$\\
         & $65.3\pm2.7$ & $2.5 \pm 0.1$ & $-6.0\pm0.3$ & $-15.0\pm1.0$\\
& $67.0\pm3.0$ & $4.1 \pm 0.2$ & $5.0\pm0.2$ & $20.5\pm1.3$\\
         3349 & $51.6\pm3.0$ & $5.6 \pm 0.3$ & $-0.07 \pm 0.01$ & $-0.4\pm0.06$\\
         & $56.1 \pm 3.2$ & $3.9 \pm 0.2$ & $-1.4 \pm 0.1$ & $-5.5\pm0.5$\\
         & $61.3\pm3.5$ & $5.0\pm0.3$ & $-0.8\pm0.04$ & $-3.8\pm0.3$\\ 
         & $65.3 \pm 3.6$ & $2.5 \pm 0.2$ & $3.9 \pm 0.2$ & $9.8\pm1.0$\\
         & $67.0 \pm 3.3$ & $4.1 \pm 0.2$ & $2.5 \pm 0.1$ & $10.3\pm0.6$\\
         701  & $51.5 \pm 6.5$ & $5.7 \pm 0.7$ & $-47.7 \pm 5.8$ & $-267.1\pm47.0$\\
         & $56.2 \pm 6.3$ & $3.9 \pm 0.5$ & $-22.0 \pm 2.5$ & $-85.8\pm14.7$ \\
         & $61.4 \pm 7.1$ & $5.0\pm0.6$ & $-20.2\pm2.2$ & $-101.0\pm 16.3$\\
         724 & $51.7\pm8.0$ & $5.7\pm0.8$ & $-28.5\pm4.0$ & $162.5\pm32.2$\\
         & $56.3\pm7.2$ & $3.9\pm0.5$ & $-15.0\pm2.0$ & $-58.5\pm10.8$\\
         & $61.4\pm8.0$ & $5.0\pm0.7$ & $-17.0\pm2.2$ & $-85.0\pm16.0$ \\
         \hline
    \end{tabular}
    \label{tab:spectral_fit_prop}
\end{table}

\subsection{Non-LTE radiative transfer modelling}\label{subsec:nonLTE_modelling}
Following the analysis presented in Paper I, we perform non-LTE radiative transfer calculations using the radiative transfer code MOLPOP-CEP \citep{Asensio2018}, which takes into account the effects of line overlap for a single-zone plane-parallel slab geometry. The code provides solutions to the radiative transfer equation for multi-level systems based on the (coupled) escape probability formalism presented in \citet{Elitzur2006}. The models are run for each of the different velocity components identified in Sect.~\ref{subsec:line_profile}, with their line widths (a key quantity for accounting for the effects of line overlap) being fixed to their intrinsic values, which is tabulated in Table~\ref{tab:spectral_fit_prop}.

Each slab is illuminated by both internal and external radiation fields, where the former is dominated by emission from warm dust grains, while radiative excitation by the latter is due to the cosmic microwave background and  interstellar radiation field. The internal radiation field of dust is set by an uniform temperature as computed as $B_\nu(T)(1-e^{-\tau_{\nu}})$, where $\tau_{\nu}$ is the dust optical depth at a frequency, $\nu$ assuming standard ISM dust properties. The dust temperature used was derived from spectral energy distributions (SEDs) generated from mid-infrared to sub-millimetre wavelengths (8\,--\,870~$\mu$m) dust continuum images and taken from \citet{Konig2017}. Moreover, because the code assumes uniform physical conditions across the slab, the models are run in multiple iterations to examine a broad range of physical conditions. Each model iteration is run for a distinct  
set of physical conditions that are represented by a 
density-temperature grid of size $93 \times 93$ with values of gas density, $n_{\rm H}$, ranging from 100 -- 10$^{8}$~cm$^{-3}$ with a logarithmic step of log[$n$(H)~/cm$^{-3}$] =0.065 and gas temperatures, $T_{\rm kin}$, between 20 and 204~K in steps 2~K. 
In the analysis that follows, we use a constant value for the CH abundance of 3.5$^{+2.1}_{-1.4}\times10^{-8}$, as estimated by \citet{Sheffer2008} and CH column densities estimated from the 2006~GHz FIR lines (between $8\times10^{13}$ and $2\times10^{14}~$cm$^{-2}$). The radiative transfer calculations are carried out using HFS resolved collisional rate coefficients computed by \citet{Dagdigian2018} for collisions between CH and atomic hydrogen and ortho-H$_2$, while the collisions for CH and para-H$_2$ were scaled (by the reduced mass) from those computed for collisions with He by \citet{Marinakis2019}. These rate coefficients were computed for all transitions between the lowest 32 HFS levels (up to upper-level energies of 300~K) for the lowest vibrational state of the electronic ground state of CH. As noted in Paper I, only this specific combination of collisional rate coefficients is capable of producing level inversion in ground-state lines of CH. The inability of the directly computed collisional rates in producing inversion may likely be due to the unusually high collisional cross-sections computed for collisions of CH with para-H$_2$ by \citet{Dagdigian2018}, with values comparable to that for collisions between CH and ortho-H$_2$. The scattering calculations presented by \citet{Dagdigian2018} incorporate inelastic collisions through both direct and indirect pathways, where the latter involve an intermediary collision complex. Notably, this author finds a higher proportion of transitions to be induced by indirect collisions with para-H$_{2}$ compared to ortho-H$_{2}$, except for CH--para-H$_2$ collisions involving the $F^{\prime}-F^{\prime\prime}=1\,$--$\,0$ transition of the ground state at gas temperatures above 100~K (see their Figs.~2 and 3). These differences, along with other HFS--specific anomalies for collisions involving higher rotational levels, will influence the excitation mechanisms of the CH ground state line and requires further examination. 

We refer the reader to Section~3.1 of \citet{Dagdigian2018} and Section~4.3 for a detailed discussion pertaining to the statistical description employed for computing indirect collisions alongside Appendix~A of Paper I for more details on the specific combination of collisional rate coefficients used. Furthermore, contributions from the ortho- and para-H$_2$ collision partners is weighted based on the temperature-dependent ortho-to-para ratio, which nears 3 above 200~K \citep{Sternberg1999}.

In addition to atomic and molecular hydrogen, collisional excitation by electrons has long been suggested to play a critical role in exciting the ground-state transitions of CH, particularly in regions with high electron fractions, $x_{\rm e} = n_{\rm e}/n_{\rm H}>10^{-5}$--10$^{-4}$, such as in PDRs associated with H~{\small II} regions. \citet{Bouloy1984} illustrated that while collisions with electrons do not play a role in inverting the ground-state lines of CH, they are crucial for thermalising these lines. In the models presented in Paper I, contributions from collisions with electrons in the excitation of CH had not been considered because accurate collisional rate coefficients were unavailable (to the best of our knowledge). 

However, for a comprehensive treatment of the radiative and collisional (de-)excitation of the CH ground state, it is essential to account for collisions with all relevant collision partners, including electrons. To address this, in the following section and Appendix~\ref{subsec:collisions_with_electrons}, we present scattering calculations for CH-electron collisions using the dipole Born approximation.

Furthermore, contributions from the different collision partners are weighted according to the column averaged molecular fractions \citep[$f^{N}_{\rm H_{2}} = 2N({\rm H}_{2})/(N({\rm H})+2N({\rm H}_{2}))$;][]{winkel2017hydrogen} and electron fractions, $x_{\rm e}$, traced for collisions with atomic hydrogen, molecular hydrogen, and electrons, respectively. The values for $N({\rm H_2})$ are estimated from $N({\rm CH})$ following the CH--H$_2$ relationship presented in \citet{Sheffer2008} and those for $N(\rm {H})$ are taken from \citet{winkel2017hydrogen}. The molecular fractions used for the different cloud components lie between 0.65 and 0.78 \citep[see also Fig.~8 of][]{winkel2017hydrogen}. With line parameters and spatial distributions indicating an origin in the neutral gas near the C$^+$/C/CO transition layer, coincident with CH, we use carbon radio recombination line (CRRL) data as proxies for the physical conditions traced by electrons in these regions. Lacking PDR models of CRRL emission toward W51~E, we use properties computed for W49 -- a source with similar physical properties and comparable CRRL line strengths and widths, as discussed in Appendix~C of \citet{Jacob2021CH2}. We use a value of 3000~cm$^{-3}$ as an upper limit for the electron density, $n_{\rm e}$, which was computed from PDR models of CRRL emission by \citet{Roshi2006} toward W49~N using C75$\alpha$ and C91$\alpha$ lines.  
In addition, we assumed the upper limit on the electron fraction to be set by the C$^+$ abundance of $1.5\times10^{-4}$ \citep{sofia2004interstellar} up until $n_{\rm H} = 3\times10^{6}~$cm$^{-3}$. This results in values of $x_{\rm e}$ between $1.5\times10^{-4}$ and $3.0\times10^{-5}$. The results obtained from such MOLPOP-CEP calculations have already been benchmarked in Paper I, in which CH observations towards the widely studied Taurus Molecular Cloud-1 (TMC-1) were modelled.

\subsection{Rate coefficients for collisions of CH with electrons}\label{subsec:collisions_with_electrons}
In this section, we present scattering calculations for electron-CH collisions.  
Since long-range forces dominate the collisional cross-sections for electron excitation, these cross-sections and the subsequently derived collisional rate coefficients scale with the square of the electric dipole moment, $\mu_{\rm e}$. For 
CH, which possesses a relatively strong dipole moment of 1.46~D, this results in large cross-sections for dipole-allowed rotational transitions 
for which $\Delta J =1$. 
A detailed formalism, deriving the collisional (de-) excitation rate coefficient is presented in Appendix~\ref{appendix:collisions-e}, where we recover the expression derived by \citet{Bouloy1984} in the limit that $k_{\rm B} T >> \nu_{ij}$.\\

The collisional rate coefficients are computed for all transitions involved in the lowest 32 levels of CH for gas temperatures ranging from 10~K to 300~K in intervals of 10~K. Their corresponding frequencies and Einstein A coefficients were taken from the Cologne Database for Molecular Spectroscopy \citep[CDMS;][]{muller2005cologne}. As discussed in \citet{Loreau2022} for pure rotationally excited levels the derived rate coefficients should be accurate while for ro-vibrational excitation more accurate rate coefficients require $R$-matrix based calculations. Furthermore, collisional excitations by electrons make negligible contributions to the non-dipole permitted transitions. 

\begin{figure*}
    \centering
    \includegraphics[width=0.31\textwidth]{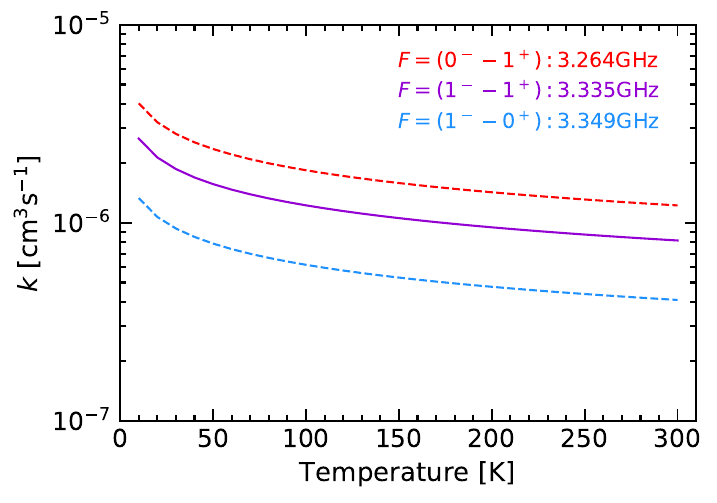} \quad
    \includegraphics[width=0.31\textwidth]{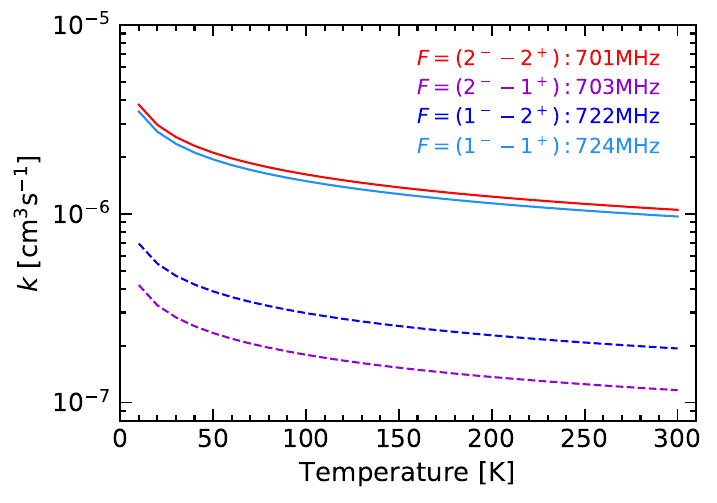} \quad
    \includegraphics[width=0.31\textwidth]{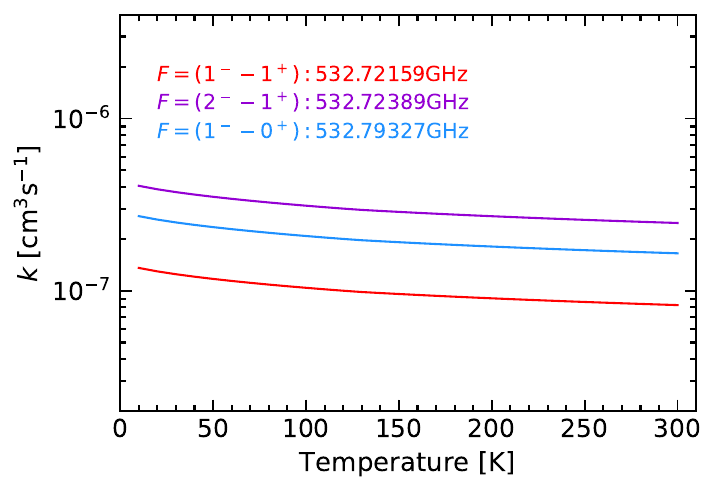} 
    \includegraphics[width=0.3\textwidth]{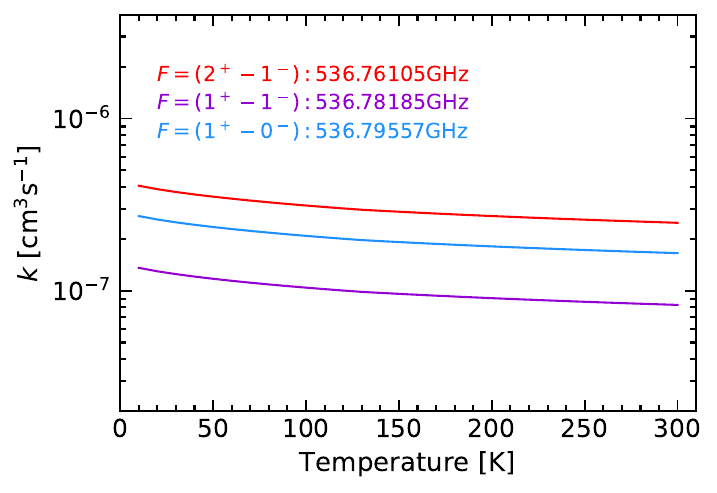} \quad
    \includegraphics[width=0.3\textwidth]{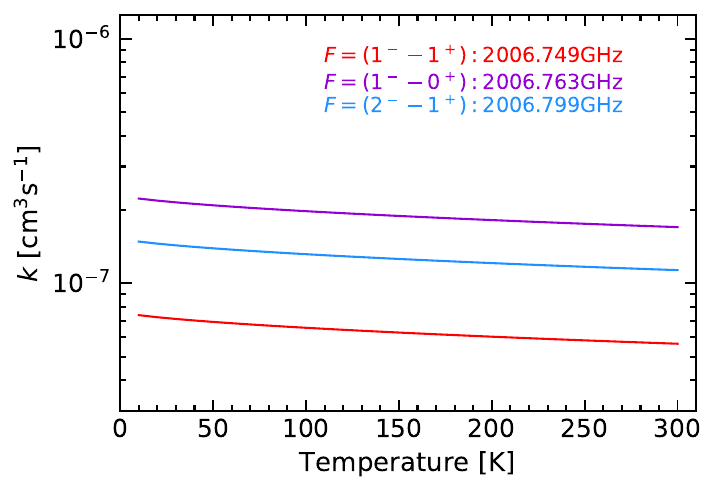} 
    \caption{Top row: Temperature-dependent rate coefficients for de-excitation transitions induced by collisions with electrons within $\Lambda$-doublet levels of the CH ground  $^{2}\Pi_{1/2}, N=1, J=1/2$ and rotationally excited $^{2}\Pi_{3/2}, N=1, J=3/2$ states. The solid and dashed curves mark the main and satellite lines within the two states, respectively. Middle and bottom rows: Same as the top row but for the rotational transitions connecting the $^{2}\Pi_{3/2}, N=1, J=3/2 \rightarrow ^{2}\Pi_{1/2}, N=1, J=1/2$ and the $^{2}\Pi_{1/2}, N=2, J=3/2 \rightarrow ^{2}\Pi_{1/2}, N=1, J=1/2$ levels, respectively.}
    \label{fig:rate_coeffs}
\end{figure*}

Figure~\ref{fig:rate_coeffs} presents the temperature-dependent rate coefficients computed using the formalism discussed above, for HFS transitions within the $\Lambda$-doublet levels of the ground state and the first rotationally excited state ($^{2}\Pi_{3/2}, N=1, J=3/2$) of CH for collisions with electrons. The collisional rate coefficients derived for the main lines between both sets of $\Lambda$-doublet levels are comparable to one another, while the satellite lines show consistently lower rates across the temperature range studied, except for the lower frequency satellite transition within the ground state 
at 3.264~GHz. In contrast, the collisional (de-)excitation rates computed for these transitions for collisions with atomic and molecular hydrogen are lower than those computed here for collisions of CH with electrons by factors between $10^{-6}$ and $10^{-4}$, across the temperature range. This dataset
will be made available through the Excitation of Molecules and
Atoms for Astrophysics (EMAA) database\footnote{\href{https://emaa.osug.fr/details/CH}{https://emaa.osug.fr/} and \href{https://dx.doi.org/10.17178/EMAA}{https://dx.doi.org/10.17178/EMAA}}.

A thorough analysis of the impact that electron collisions have on the excitation of these levels will be discussed further in the following sections. 

\section{Discussion} \label{sec:discussion}

Facilitated by the non-LTE radiative transfer models described above, the following sections provide detailed insights into the physical and excitation properties that render the 55~km~s$^{-1}$ clouds more conducive to exciting the first rotationally excited level of CH. 
\subsection{Physical conditions traced by the ground and first rotationally excited states of CH}\label{subsec:phyconditions}

\begin{figure}
	\centering 
	\includegraphics[width=0.45\textwidth]{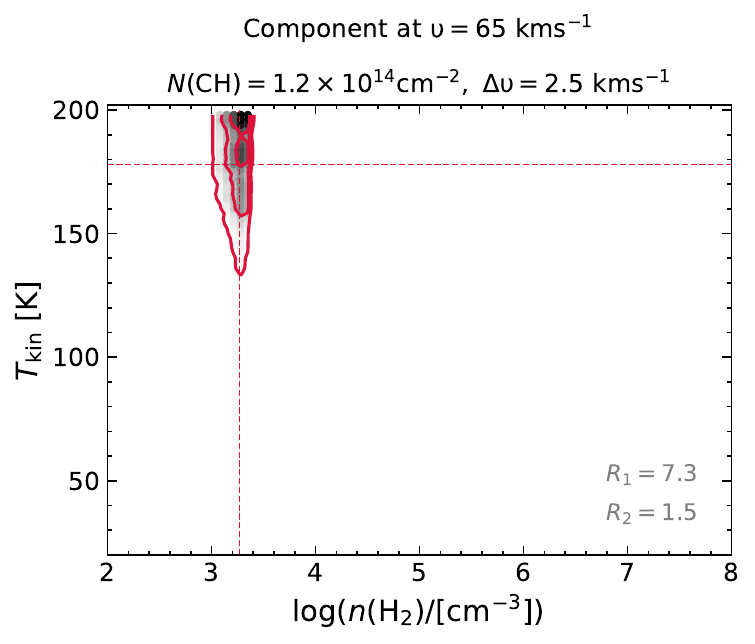}\quad
	\includegraphics[width=0.45\textwidth]{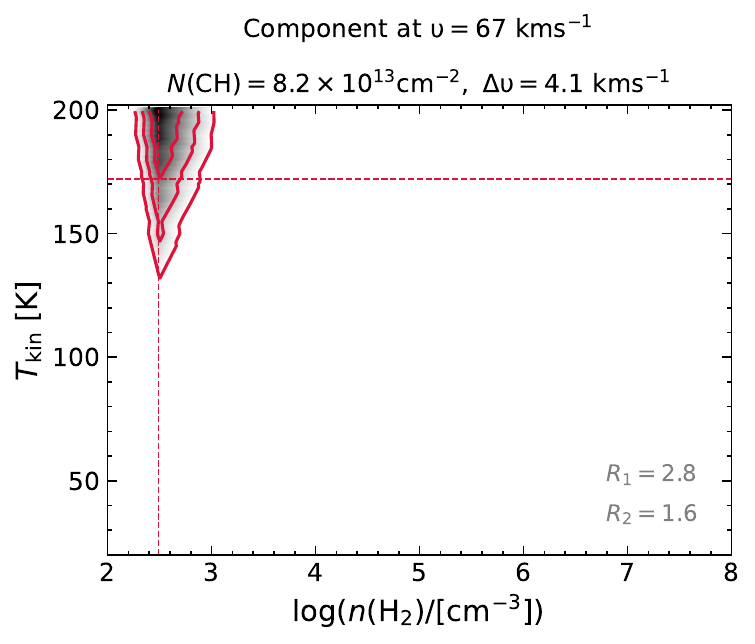}
	\caption{MOLPOP-CEP non-LTE radiative transfer modelling results for the W51\,E cloud components at $\upsilon_{\rm LSR} = 65~{\rm km~s}^{-1}$ (top) and 67~km~s$^{-1}$ (bottom), respectively. The red contours display the 1, 2, and 3~$\sigma$ levels of the $\chi^2$ distributions of the modelled line ratios that best reproduce the observed line ratios between the 3.264~GHz,
		and 3.349~GHz lines ($R_1$) and those between the 3.335~GHz, and 3.349~GHz lines ($R_2$) across the $n_{\rm H_{2}}$--$T_{\rm kin}$ parameter space probed for the CH column densities and line widths specified above. The dashed red lines mark the minimised values and $1~\sigma$ limits.}
	\label{fig:Models_comp_65_67}
\end{figure}

\begin{figure}
	\centering 	\includegraphics[width=0.45\textwidth]{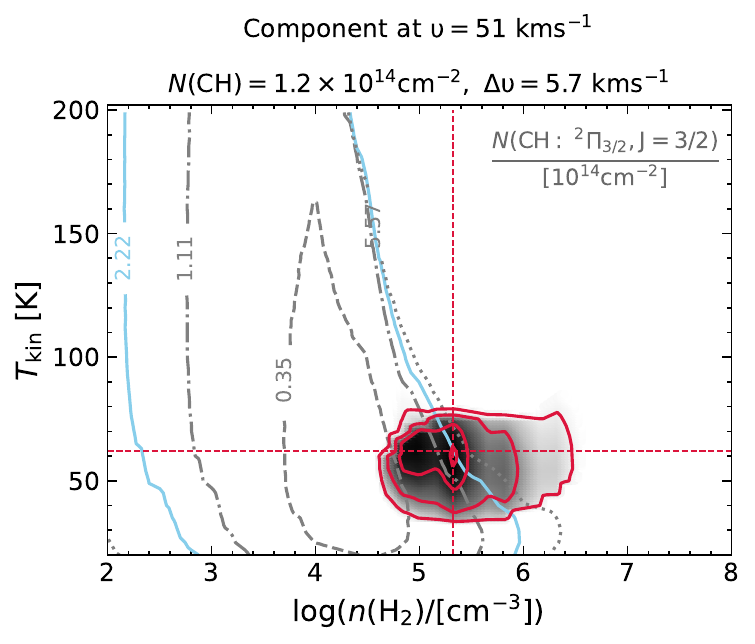}\quad	\includegraphics[width=0.45\textwidth]{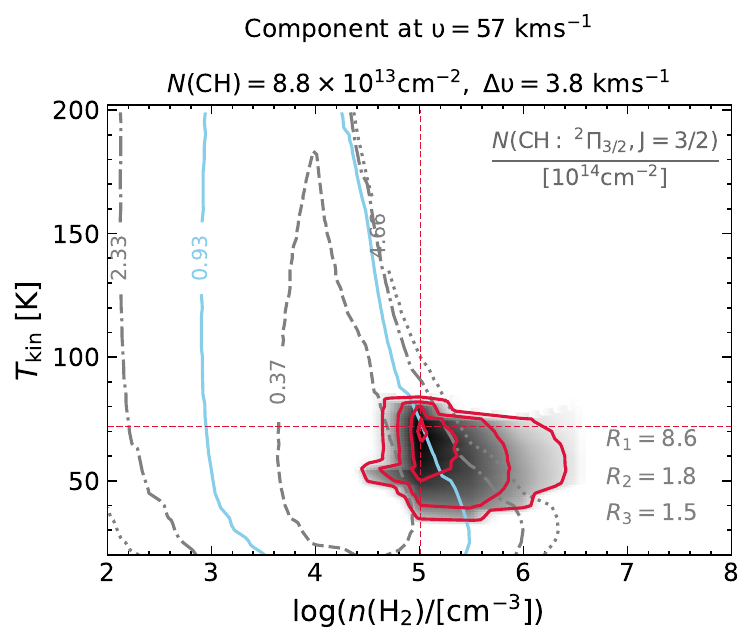}\quad
 	\centering 	\includegraphics[width=0.45\textwidth]{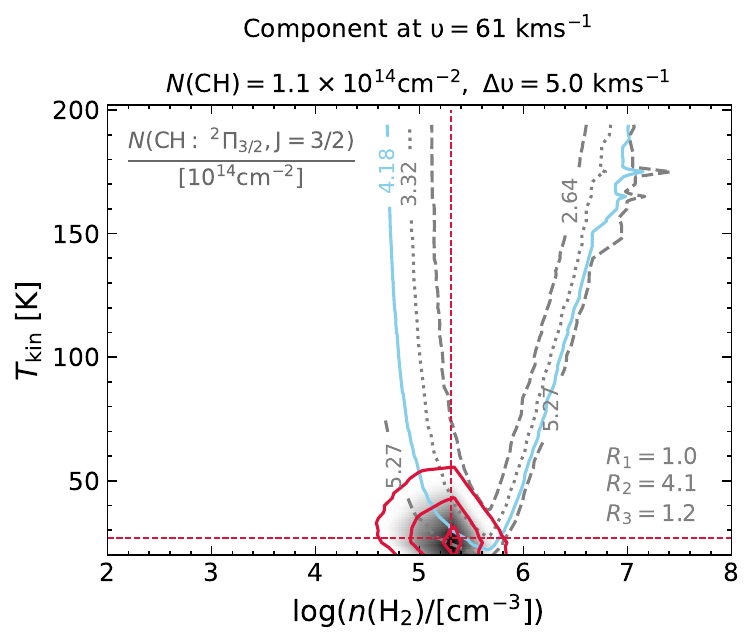}\quad	
	\caption{Same as Fig.~\ref{fig:Models_comp_65_67} but for cloud components at $\upsilon_{\rm LSR} =51~{\rm km~s}^{-1}$ (top), 57~km~s$^{-1}$ (centre) and 61~km~s$^{-1}$ (bottom), respectively. While the colour map and contours in red displays the 1, 2, and 3$\sigma$ levels of the $\chi^2$ distributions of the modelled line ratios between the 3.264~GHz,
		and 3.349~GHz lines ($R_1$), the 3.335~GHz, and 3.349~GHz lines ($R_2$), and those between the 701~MHz and 724~MHz ($R_3$), the contours in grey with varying line styles mark the observed line ratios of $R_3$ for fixed values of $N$(CH) (as labelled) corresponding to the population in the excited level. The solid light blue curve marks the $N$(CH) value of the excited state for which the model is best constrained. }
	\label{fig:Models_comp_52_57}
\end{figure}

As a first step, we reproduced the physical conditions obtained in Paper I for the W51\,E cloud components that have velocities of 65~km~s$^{-1}$ and 67~km~s$^{-1}$. As demonstrated in Paper I, the physical conditions of the regions probed are constrained by minimising the $\chi^2$ values of the observed line ratios across the modelled density-temperature parameter space. Specifically, the use of line ratios has the advantage that it removes the dependence on the beam filling factor or the fraction of the spatial resolution of the measurement filled by the emitting region. The independent line ratios used in this analysis are $R_1 = T_{\rm 3.264~GHz}/T_{\rm 3.349~GHz}$ and $R_2 = T_{\rm 3.335~GHz}/T_{\rm 3.349~GHz}$. The physical conditions thus inferred for the two velocity components are displayed in Fig~\ref{fig:Models_comp_65_67}. Overall, these results are consistent with those presented in Paper I for the cloud components at 65~km~s$^{-1}$ and 67~km~s$^{-1}$, respectively. While the addition of collisions between CH and electrons to the non-LTE radiative transfer analysis, has not significantly altered the derived gas densities, the effect on gas temperature is uncertain because the modelled gas temperatures for these velocity components are not well constrained. 
Minor variations in the derived gas densities and temperatures (with those in Paper I) may also be attributed to differences in the modelled line ratios that arise primarily from the multi-component fit presented in this work (see Sect.~\ref{subsec:line_profile}). Despite reduced $\chi^2$ values that are comparable to a value of 1, there are multiple local minima across the parameter space modelled. The presence of multiple minima suggests a range of parameter sets that locally optimise the $\chi^2$ minimisation, resulting in only 1~$\sigma$ lower limits for the gas temperatures modelled towards the 65~km~s$^{-1}$ and 67~km~s$^{-1}$ clouds. The gas densities and temperatures derived towards the different cloud components identified are summarised in Table~\ref{tab:modelled_physicalparams} alongside their observed line intensity ratios. 

Having established the pumping mechanisms affecting the ground-state lines of CH, in the combined analysis that follows, we make use of this knowledge to provide clues about the excitation observed in higher states. Therefore, in a manner similar to that described above, we also constrained the physical conditions traced by the cloud components observed at 51~km~s$^{-1}$, 57~km~s$^{-1}$ and 61~km~s$^{-1}$. Given the clear detection of the main lines of the CH transitions near 700~MHz, towards these clouds, these models are further constrained by using the line ratios observed between the 701~MHz and 724~MHz lines, $R_3 = T_{\rm 701~MHz}/T_{\rm 724~MHz}$ (see Table~\ref{tab:modelled_physicalparams}). 

Although the column density can be derived from the apparent optical depths, deduced from the line-to-continuum ratio this requires additional assumptions or knowledge of the beam-filling factors. We ignore beam filling effects when computing column densities from the 2006~GHz CH lines, given that its beam is contained within the 3.3~GHz emission which is resolved in our VLA observations. Therefore, the modelled outputs for the $R_3$ line ratio are presented for a range of CH column densities since the population and consequently the column densities of the HFS lines in the rotationally excited state are unknown. The combined analysis presented in Fig.~\ref{fig:Models_comp_52_57} jointly minimises the $\chi^2$ of all three line ratios yielding not only the physical conditions traced by these cloud components but also places constraints on the population in the first rotationally excited state. The population in the excited state in terms of CH column densities hence deduced are a few $\times10^{14}$~cm$^{-2}$. The CH column densities hence determined for these velocity components, for the first excited state are comparable if not slightly greater than the values estimated using the absorption spectrum of the 2006~GHz lines of CH which connect to its ground state. This results in gas densities of a few times $\times10^{5}$~cm$^{-3}$ towards these cloud components, 
values that are consistent with the gas densities of the clumps that harbour star formation sites derived by \citet{Ginsburg2017} using observations of H$_2$CO. Toward both the 51~km~s$^{-1}$ and 57~km~s$^{-1}$ cloud components, the models reproduce gas temperatures of $\sim\!60\,$--$\,70$~K which is slightly warmer than those typically found in dense molecular clouds. 

  Gas temperatures up to 153~K have previously been derived toward these clouds, using non-LTE radiative transfer models by \citet{Remijan2004} as traced by CH$_3$CN and also large velocity gradient models of $^{13}$CO emission by \citet{Fujita2021}. The latter have attributed the higher gas temperatures in these clouds as arising due to gas heating from nearby massive stars in this region. However, for the 61~km~s$^{-1}$ cloud, we derive similar gas densities of a few $\times 10^{5}$~cm$^{-3}$ while $T_{\rm kin}\!\sim\!27~$K, assuming that the excited level has a population of $N({\rm CH})\!\sim\!4.2\times10^{14}$~cm$^{-2}$. For this cloud component, the validity of such high CH column densities in the first excited level is questionable, as indicated by reduced $\chi^2$ values less than 1 likely because of underestimations in the uncertainties of the fitted components.  \\

\begin{figure}
    \centering
    \includegraphics[width=0.5\textwidth]{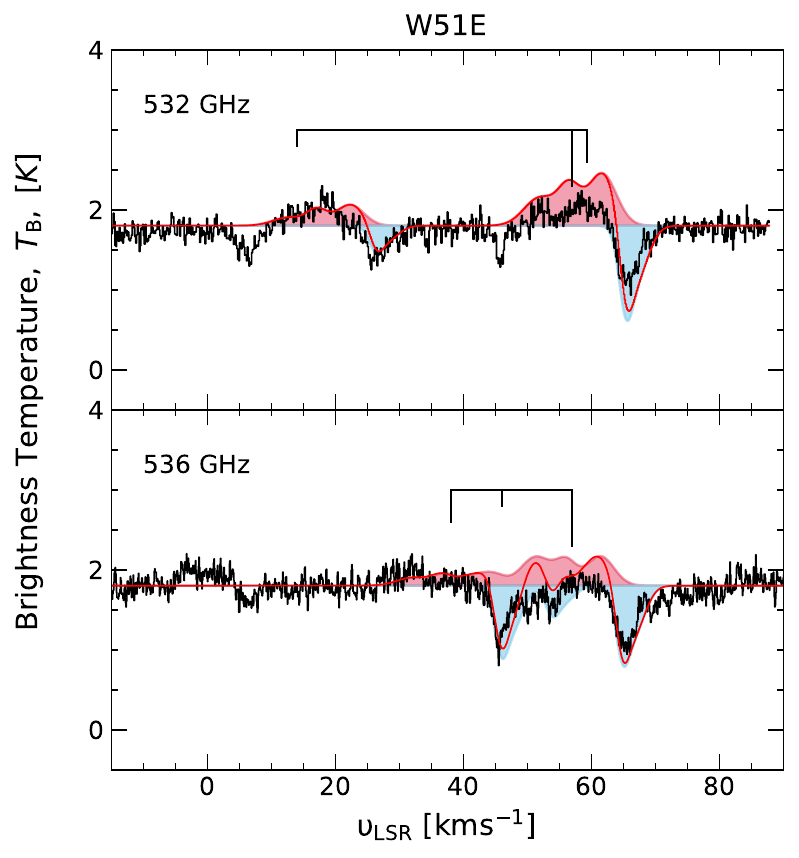}
    \caption{MOLPOP-CEP model constrained brightness temperature (in red) to the emission and absorption components observed in the $N, J = 1,3/2\rightarrow1,1/2$ transitions of CH near 532~GHz (upper panel) and 536~GHz (lower panel), respectively, towards W51\,E. The pink and blue shaded regions indicate the fits to only the emission and only the absorption components, respectively. }
    \label{fig:fits-532-536}
\end{figure}

The quantitative analysis presented here is in agreement with the qualitative assessment made in Sect~\ref{subsec:line_profile} whereby cloud components with LSR velocities near 55~km~s$^{-1}$ are denser than those closer to 66~km~s$^{-1}$. Furthermore, the higher densities derived towards the 55~km~s$^{-1}$ cloud components correspond to the velocities at which we observe emission in the 532/536~GHz rotational lines of CH (see Fig.~\ref{fig:spectrum_panelplot}). Due to the complex nature of their line profiles, consisting of a combination of emission and absorption features, contributions from the different components were difficult to gauge and hence to fit. However, aided by the physical conditions derived by modelling the 700~MHz lines, we are now able to simultaneously model and constrain the observed emission features in the 532/536~GHz lines. The resulting fits to the emission components observed in the 532/536~GHz lines of CH towards W51\,E are presented in Fig.~\ref{fig:fits-532-536}. This aids in modelling and subsequently removing contributions from the emission features seen in the spectra of the 532/536~GHz lines resulting in pure absorption spectra. This simplifies the analysis of the absorption features observed towards velocity components arising in the material surrounding the background source. The derived column densities hence are in agreement with the values derived from the 2006~GHz lines discussed in Sect.~\ref{subsec:line_profile}, within the uncertainties. This exercise illustrates the important role played by the $N, J = 1,3/2$ 700~MHz lines in analysing the rotational lines at 532/536~GHz that connect this level to the ground state. 

\begin{table}[]
    \centering
     \caption{Physical conditions modelled using MOLPOP-CEP non-LTE radiative transfer analysis towards different velocity components in W51E. }
    \begin{tabular}{cccc}
    \hline \hline 
    Velocity component &  $\left|R_{i}\right|$ & log$_{\rm 10}$($n_{\rm H}/[{\rm cm^{-3}}]$) & $T_{\rm kin}$\\
~[km~s$^{-1}$] & & & [K] \\
    \hline
    51 & 25.2  & $5.32^{+0.18}_{-0.44}$ & $63^{+10.0}_{-12.0}$\\
       & 22.8  & \\
       & 1.8   & \\
    57 & 8.6   & $5.02^{+0.26}_{-0.20}$ & $ 72^{+8.0}_{-19.0}$\\
       & 1.8   & \\
       & 1.5   & \\
       61 & 1.0 & $5.3^{+0.10}_{-0.14}$& $27^{+4.0}_{-5.0}$ \\
       & 4.1 \\
       & 1.2 \\
    65 & 7.3   & $3.25^{+0.08}_{-0.12}$ & $\geq 178$ \\
       & 1.5   & \\
       & --    & \\
    67 & 2.8   & $2.49^{+0.22}_{-0.17}$ & $\geq172$ \\
       & 1.6   & \\
       & --   & \\
         \hline
    \end{tabular}
   
    \label{tab:modelled_physicalparams}
    \tablefoot{The columns are from left to right: the velocity component modelled, the observed line ratios ($R_i$) where each row corresponds to $R_1 = T_{\rm 3.264~GHz}/T_{\rm 3.349~GHz}$, $R_2 = T_{\rm 3.335~GHz}/T_{\rm 3.349~GHz}$, and $R_3 = T_{\rm 701~MHz}/T_{\rm 724~MHz}$ (where detected), the modelled gas densities and temperatures alongside their 2$\sigma$ errors.}
\end{table}

\subsection{Excitation scheme}\label{subsec:excitation}

\begin{figure*}
	\centering
 \includegraphics[width=0.32\textwidth]{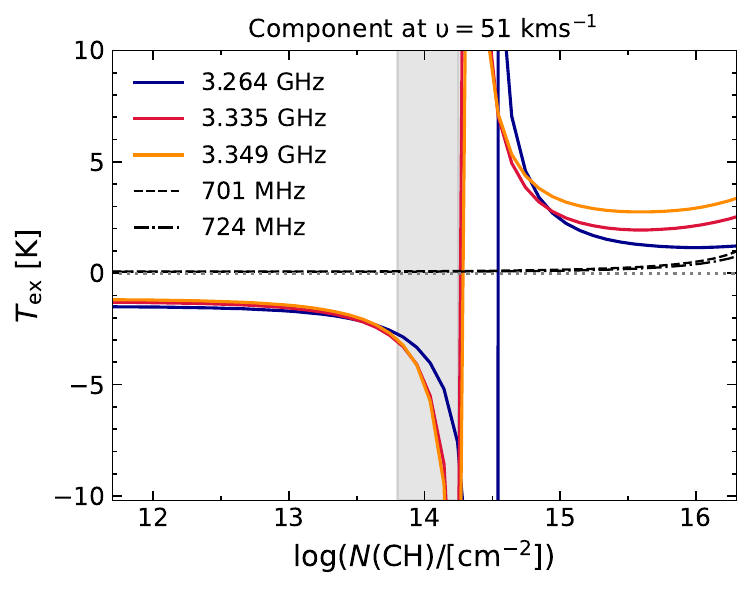}\quad
	\includegraphics[width=0.32\textwidth]{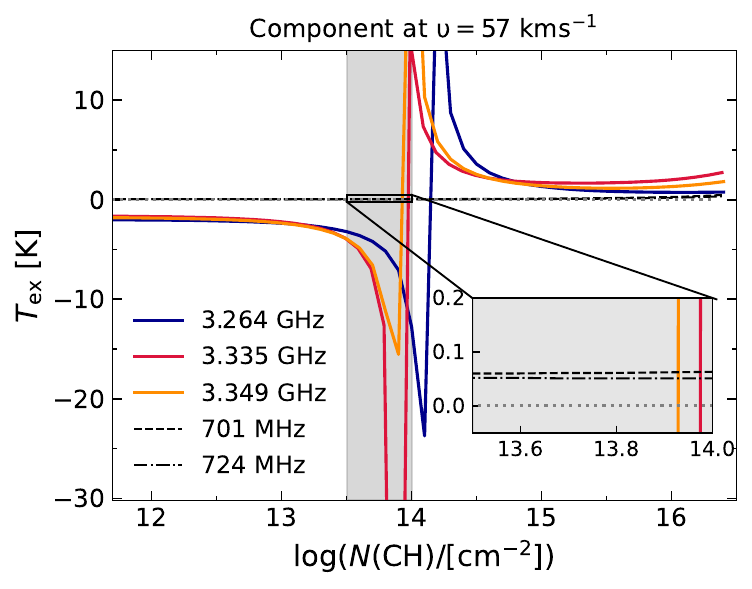}\quad
	\includegraphics[width=0.32\textwidth]{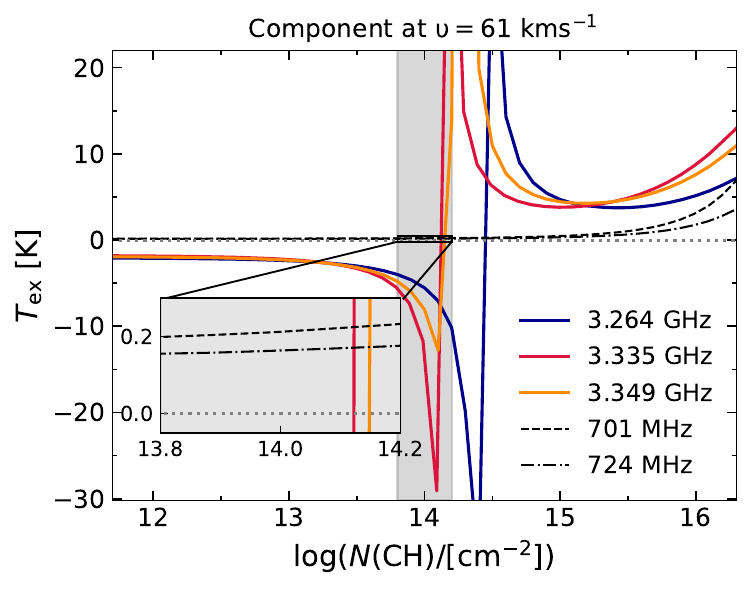}\\
		\includegraphics[width=0.32\textwidth]{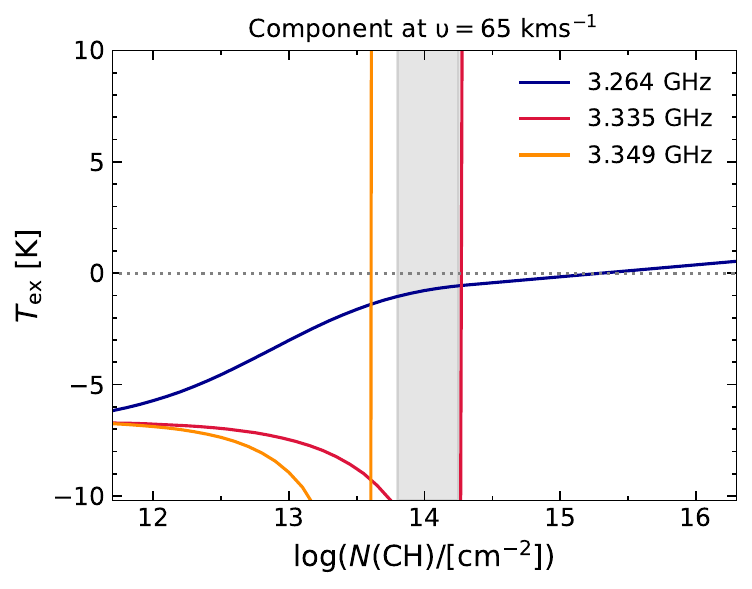}\quad
		\includegraphics[width=0.32\textwidth]{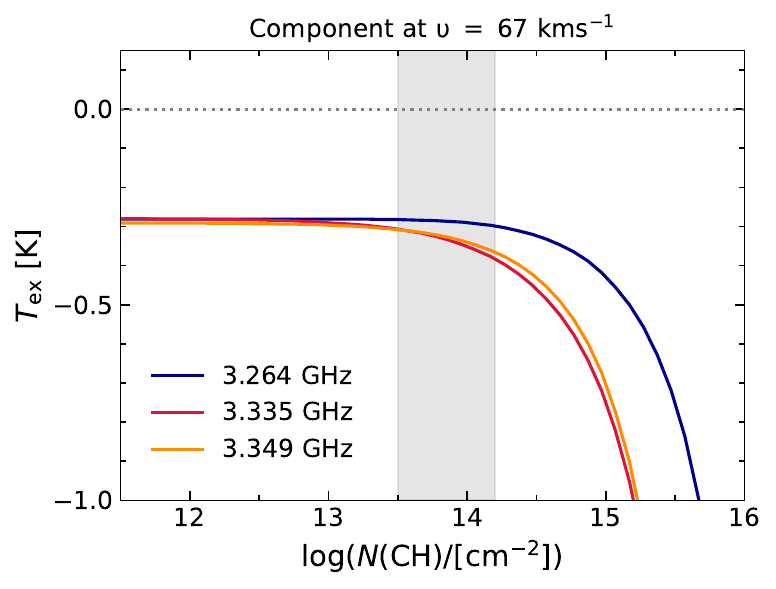}
		\caption{Clockwise from the top-left: Modelled excitation temperatures of the 3.264~GHz (dark blue), 3.349~GHz (dark orange), and 3.335~GHz (red) ground-state lines of CH, alongside that of the 701~MHz (dashed black) and 724~MHz (dashed-dotted black) lines of the first excited level of CH (where detected) as a function of CH column densities for the physical conditions derived for the  51, 57, 65, 67 and 61~km~s$^{-1}$ cloud components of W51\,E. The inset panels expand on the $T_{\rm ex}$ values corresponding to CH column densities between $7\times 10^{13}$ and $3\times10^{14}$~cm~$^{-2}$. The region shaded in grey highlights the range of $N({\rm CH})$ values derived toward W51~E for each velocity component.}
\label{fig:excitation_coldens_models}
\end{figure*}

In this section we examine the excitation mechanisms prevalent in both the ground and excited states of CH. The excitation conditions for the pertinent CH HFS lines are derived from the slab models that most effectively constrain the physical conditions for each cloud component, as detailed in Sect.~\ref{subsec:phyconditions}. The resulting excitation curves are presented in Fig.~\ref{fig:excitation_coldens_models}.

Akin to Paper~1, the modelled excitation temperatures derived for the 65~km~s$^{-1}$ and 67~km~s$^{-1}$ velocity components are low but negative (presented in the lower panels of Fig.~13 in Paper~1 and Fig.~\ref{fig:excitation_coldens_models} in this work, respectively). For the 67~km~s$^{-1}$ cloud component, all three CH ground-state lines are characterised by a similar excitation temperature of ${\sim\!-0.3}$~K with marginal deviations seen in all but the lower satellite line, across the range of CH column densities most relevant for our analysis (as highlighted in the figure). This difference is reflective of the slightly higher degrees of inversion observed in the main line. When compared to the results presented in Paper~1, the modelled excitation temperatures show more deviations for the 65~km~s$^{-1}$ cloud component. By modelling the excitation conditions assuming a kinetic temperature of 178~K, we find that as the CH column density approaches values of ${\sim \! 10^{14}~}$cm$^{-2}$, the excitation temperatures of both the main and upper satellite lines of CH exhibit a steady decline, followed by a sharp increase in the value. This turning point characterises the quenching of the inversion in these lines, which is in agreement with our observations, that show both lines in absorption for this velocity component or a low positive value of $T_{\rm ex}$. In the classical picture, quenching in a simple two-level system takes place when the gas volume density nears or exceeds the critical density of the rotational line connecting the two levels. For the ground-state lines of CH, this is governed, in particular by the $N, J = 1, 3/2 \rightarrow 1, 1/2$ rotational transitions near 532/536~GHz. Therefore, at gas densities above ${\sim 4\times10^{6}}$~cm$^{-3}$, which corresponds to typical translucent gas conditions, collisional processes compete with radiative ones to de-excite the level, as the critical density for the 532/536~GHz transitions is exceeded. However, this cannot be the case for the 65~km~s$^{-1}$ cloud since the modelled gas density towards the absorbed region is only a few $\times 10^{3}$~cm$^{-3}$. Because the gas densities determined for this component are notably lower than the critical density, the shift in excitation temperatures (from negative to positive values) for the main and upper satellite lines, along with the enhanced inversion in the lower satellite line, occurs as a consequence of radiative processes. This includes effects such as FIR pumping and optical depth effects (see Sect.~\ref{sec:excitation_theory}), processes that preferentially enhances masing observed in the lower satellite line via radiative trapping and line overlap. As collisions do not sufficiently compete, the inversion in the ground state is driven by the radiative decay of the lower-half of the first excited state $\Lambda$-doublet, back to the ground state. As a result, the effects of line overlap are more pronounced in regions where the sub-mm/FIR radiation field is strong and gas temperatures are relatively high. 

Similarly, for the 51~km~s$^{-1}$,  57~km~s$^{-1}$ and 61~km~s$^{-1}$ clouds, we observe that the level inversion in both the main and upper satellite lines of the ground state are quenched within the range of CH column densities modelled. Owing to the higher gas densities of these cloud components this suggests that in these velocity components the excitation of these lines must arise from collisional excitation competing with radiative decay. In the HFS lines within the first rotationally excited level, produced by dipole-allowed collisions, the excitation temperatures are found to have low values, typically a few hundred mK towards these clouds, a value that is fortuitously in agreement with the excitation condition of ${T_{\rm ex}< 2~}$K derived by \citet{Turner1988}. The trends in the modelled excitation temperatures remain consistent across CH column densities, increasing only beyond ${N({\rm CH})\gtrsim\!10^{15}~{\rm cm}~^{-2}}$ with a more pronounced increase seen in the 701~MHz line. 
As discussed above, the low excitation temperatures estimated for the HFS transitions of the first rotationally excited state can be attributed to collisional (de-)excitation rates prevailing over the rates of radiative (de-)excitation. This is because excitation through collisions becomes more efficient in denser environments, resulting in a reduced degree of inversion observed in the 3.264~GHz line for these velocity components.\\

\begin{figure*}
 \centering   \includegraphics[width=0.95\textwidth]{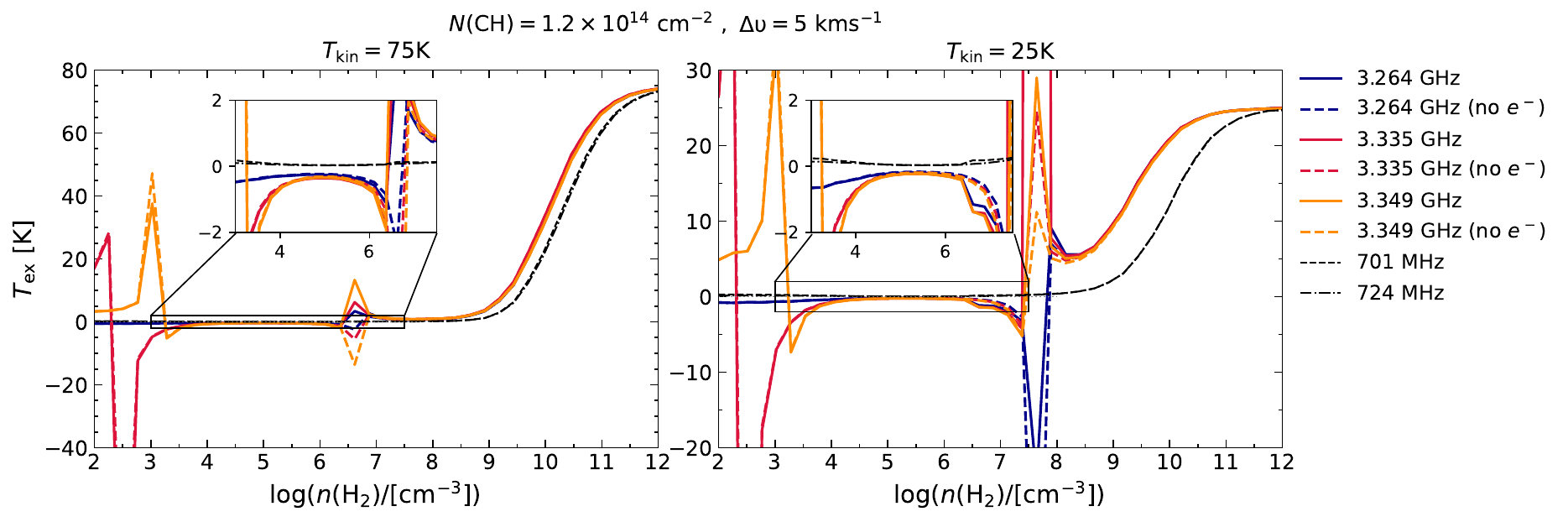}
    \caption{Variations in the MOLPOP-CEP modelled excitation temperatures of the ground-state and first rotationally excited transitions of CH as a function of gas densities for fixed values of $N(\text{CH})=1.2\times10^{14}~$cm$^{-2}$, $\Delta\upsilon = 5$~km~s$^{-1}$ and $T_{\rm kin} =75~$K (left) and $T_{\rm kin} =25~$K (right). The insets zoom-in on $T_{ex}$  between gas densities of $10^{3.5}$~cm$^{-3}$ and 10$^{7.5}$~cm$^{-3}$.}
    \label{fig:Tex-dens}
\end{figure*}
We visualise the excitation curve as a function of gas densities, while keeping the values of the other physical parameters constant (see Fig.~\ref{fig:Tex-dens}). Specifically, this analysis is carried out for fixed values of $N({\rm CH}) = 1.2\times10^{14}~$cm$^{-2}$, $T_{\rm kin} = 75~$K and $\Delta\upsilon=5~$km~s$^{-1}$, as these conditions account for the average gas properties of the different cloud components. The density parameter space probed is also extended to much higher values of $n_{\rm H} = 10^{12}~$cm$^{-3}$. Under these conditions and at both low (${n_{\rm H} < 320 ~}$cm$^{-3}$) and high gas densities (${n_{\rm H} > 3.5 \times10^{6}~}$cm$^{-3}$), we find the excitation curves for the 3.335~GHz and 3.349~GHz lines to show positive values of $T_{\rm ex}$, displaying inversion (${T_{\rm ex}<0}$~K) only within in a small range of intermediate gas densities. In contrast, the excitation curve for the 3.264~GHz line reveals that the models reproduce masing in this line up to gas densities of a few ${\times10^{6}~}$cm$^{-3}$, coincident with the critical density of the 532/536~GHz lines. Beyond this gas density, the inversion in the 3.264~GHz line is suppressed and the excitation temperature of the line becomes positive, after which all three of the 3.3~GHz lines are characterised by the same value of $T_{\rm ex}$. However, these lines only become thermalised at higher gas densities beyond 10$^{10}~$cm$^{-3}$. The slow thermalisation between gas densities of $10^{6}~$cm$^{-3}$ and $10^{10}~$cm$^{-3}$ is characterised by $T_{\rm kin}$ > $T_{\rm ex}$ > 0, 
leading to an enhanced absorption in the line. The modelled excitation for the main lines of the first rotationally excited state of CH is characterised by low excitation temperatures that remain almost constant up to gas densities of $\sim\!10^{8}~$cm$^{-3}$ beyond which they increase and tend toward thermalisation, but reaching it only at higher gas densities, as in the case of the 3.3~GHz lines. For the excited lines, we note that it is more difficult to discuss thermalisation, given the simplistic two-level definition, which particularly in the context of masers presents an incomplete picture, since masers by definition are multi-level systems. The right hand panel of Fig.~\ref{fig:Tex-dens} presents the modelled excitation temperatures as a function of gas densities, but for a gas temperature of 25~K, while keeping the other physical parameters fixed. The resulting excitation curve displays trends in $T_{\rm ex}$ that closely resemble the results of the higher gas temperature model but with shifts in the gas densities at which masing turns off, typically towards higher densities.

When compared to Paper~I, it is unclear if the inclusion of excitation by electrons has played a role in bringing the CH lines to thermalisation. To evaluate the impact of collisions with electrons on the excitation of the CH lines, we ran a grid of models for the same physical conditions but now excluding collisions of CH with electrons. Differences between both sets of models are most visible at densities where the masing action switches on and off and where the fractional ionisation, $x_{\rm e}$ is greater than $10^{-4}$. Since collisions with electrons are only efficient in this regime, it is unlikely that they play a significant role in thermalising the CH lines. Nonetheless, the current analysis, with the inclusion of collisions with electrons reproduces the observed excitation schemes in the HFS lines within both the ground and first excited states of CH.

\subsection{Non-detection of the CH 700~MHz lines towards other sight lines}\label{subsec:nondetection}
The fact that to date, the 700~MHz lines of CH are unambiguously detected towards a single molecular cloud, W51~E, is puzzling. While  the non-detections may be attributed to a number of reasons, a primary reason could be because of sensitivity issues arising from the weaker background continuum emission at 700~MHz. Although this clarifies the non-detection of these lines towards Sgr~B2(M), W43, and M8, the reason for the lack of detection of the 700~MHz lines in absorption towards sources with background continuum fluxes similar to that of W51~E, like DR21~Main and M17, remains unclear. The non-detection of the 700~MHz lines of CH towards these sources could be attributed to, at least in part, to the higher gas densities ($n_{\rm H}\approx 10^{6}$~cm$^{-3}$) prevailing in these regions. \\

\begin{figure}
    \centering
    \includegraphics[width=0.5\textwidth]{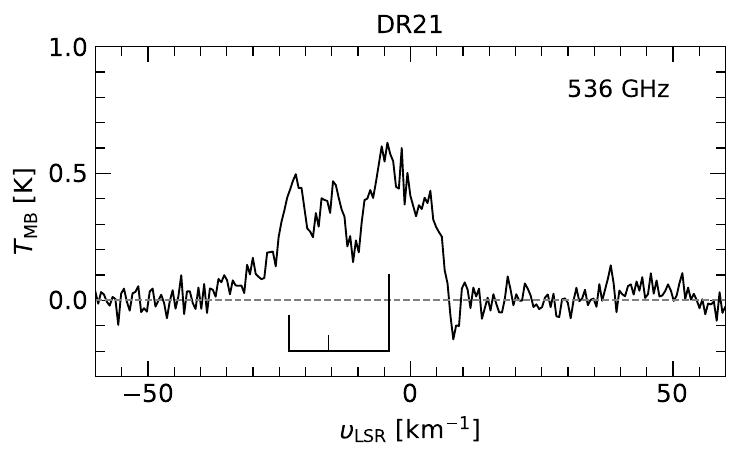}
    \caption{SOFIA/4GREAT spectrum of the $N, J=1,1/2 \rightarrow 2,3/2$ transitions of CH at 536~GHz, observed towards DR21~Main where the velocity scale is set by the strongest HFS component. The positions and relative intensities of the HFS lines are also marked. }
    \label{fig:DR21_536GHz}
\end{figure}

To verify this hypothesis, we further analyse the CH 536~GHz spectrum taken towards DR21 Main.  
A corresponding spectrum for M17 does not exist. Figure~\ref{fig:DR21_536GHz}  presents the spectrum of the CH 536~GHz line\footnote{Given that the 4G1 receiver provides two 4~GHz wide bands separated by 8~GHz only the 536~GHz HFS lines of the $N, J=1,1/2 \rightarrow 2,3/2$ were observed.} procured using channel 1 of the 4GREAT receiver \cite[4G1;][]{Duran2021}, which was on board SOFIA, under the program 83\_0802 (PI: Jacob). Strikingly, this  spectrum shows an asymmetric profile in emission, with weak absorption near 8~km~s$^{-1}$ arising from the more extended W75~N region.

As discussed in \citet{Koley2021} for OH, the one-sided geometry of the DR21 Main PDR suggests that CH traces only a part of the compact H~{\small II} region. In their subsequent analysis, these authors argued for a volume density of $n_{\rm H} = 10^6~{\rm cm}^{-3}$ for the OH-bearing gas
which lies at the interface between the dense molecular cloud and the observer, based on previous non-LTE calculations modelling radio wavelength OH lines.
 While a detailed radiative transfer analysis is challenging for the 536~GHz line of CH towards DR21~Main, the observed emission suggests that the densities associated with the CH-bearing gas should be roughly equivalent to the critical density for this transition, a few times $n_{\rm H} = 10^{6}~$cm$^{-3}$ (computed for collisions with H$_2$, at a gas temperature of 30~K). Hence, the gas densities within the DR21~Main PDR are notably (at least an order magnitude) greater than those required for the excitation of the 700~MHz CH lines (see Sects.~\ref{subsec:phyconditions} and \ref{subsec:excitation}). Qualitatively, this supports the non-detection of the 700~MHz lines of CH towards the DR21~Main PDR. 

\subsection{Other excited rotational transitions of CH}
To date, numerous transitions between the HFS split sublevels of rotationally excited levels of OH have been detected \citep[see, e.g.][]{Cesaroni1991}, with the highest being the $^{2}\Pi_{3/2}, J=9/2$ near 23~GHz reported by \citet{Winnberg1978}. In contrast, the only known detections of HFS lines from an excited rotational level of CH are  from its first excited level ($^{2}\Pi_{3/2}, J=3/2$) at 700~MHz, despite several searches. These included the search for allowed transitions between the $\Lambda$-doublet levels of the $^{2}\Pi_{3/2}, J=5/2$, $^{2}\Pi_{1/2}, J=3/2$, $^{2}\Pi_{3/2}, J=7/2$, $^{2}\Pi_{1/2}, J=5/2$, $^{2}\Pi_{3/2}, J=9/2$, $^{2}\Pi_{1/2}, J=7/2$ at 4.8~GHz, 7.3~GHz, 11.2~GHz, 14.7~GHz, 19.9~GHz and 24.4~GHz, respectively, by \citet{Sume1976} and \citet{Matthews1986}. While these early observations were hindered by significant uncertainties in the rest frequencies of the CH lines, coarse spectral resolution and modest sensitivity, more recent attempts to detect these lines aided by better spectroscopic constraints and  advanced instrumentation were also unsuccessful \citep{Tan2020}. 
Why have HFS transitions of CH in higher rotationally excited states  not been detected toward regions in which the analogous OH lines, many of which are masing, can be easily observed?
Due to the photodissociation of H$_2$O, OH attains a substantial abundance ($10^{-7}$ relative to H$_2$) in the hot ($\sim\,150$~K) dense ($10^6$~cm$^{-3}$) molecular envelopes of ultra-compact and compact H~{\small II} regions, for which W3(OH) and DR~21, respectively, are the archetypal examples \citep{Elitzur1978, Koley2021}. In contrast, no chemical pathway exists that would increase the CH abundance (which is at most a few times 10$^{-8}$ with respect to H$_2$, see \citealt{Sheffer2008}) under the above conditions.

\begin{figure}
    \centering
    \includegraphics[width=0.94\linewidth]{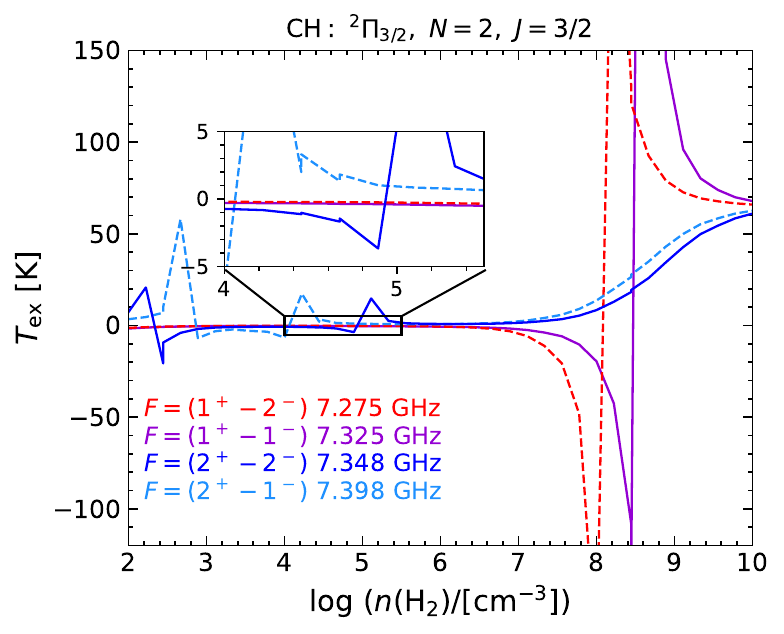}
    \\
    \includegraphics[width=0.94\linewidth]{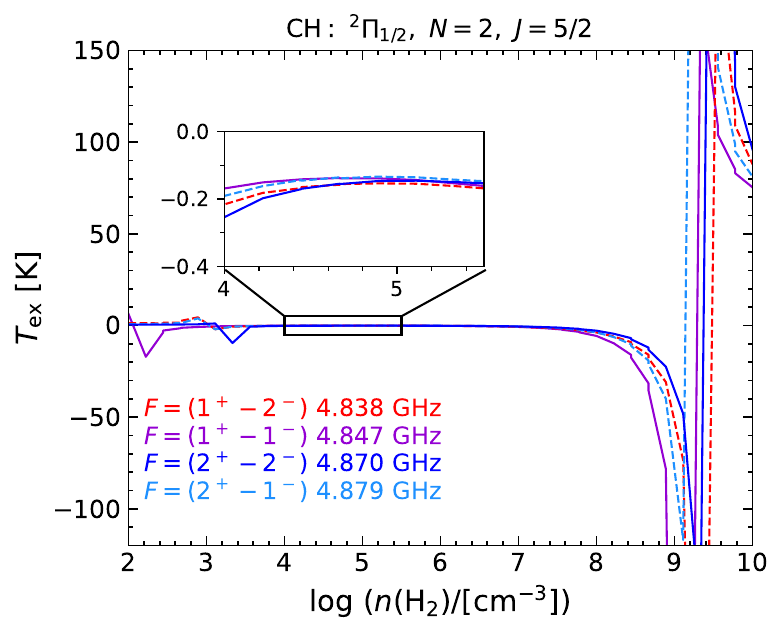}
    \caption{MOLPOP-CEP model predictions for the excitation temperature ($T_{\rm ex}$) for the HFS-split rotationally excited transitions between the $^{2}\Pi_{1/2}, N=2, J=3/2$ (top) and $^{2}\Pi_{3/2}, N=2, J=5/2$ (bottom) near 7.3~GHz and 4.8~GHz, respectively. The inset panels expand on the $T_{\rm ex}$ values for gas densities between 10$^{4}$~cm$^{-3}$ and 10$^{5.5}$~cm$^{-3}$.}
    \label{fig:higher_excited_lines_predictions}
\end{figure}

Using the non-LTE analysis discussed above, we evaluate the excitation conditions for excited rotational transitions that lie above the $^{2}\Pi_{3/2}, J=3/2$ state. Figure~\ref{fig:higher_excited_lines_predictions} illustrates the predicted excitation temperatures for the next lowest-lying rotationally excited CH lines associated with the $^{2}\Pi_{3/2}, J=5/2$ and $^{2}\Pi_{1/2}, J=3/2$ levels, for model parameters constrained based on considerations for the 51 km/s cloud component ($N$(CH) = $1.2\times10^{14}~$cm$^{-2}$, $\Delta\upsilon=5.7~$km~s$^{-1}$ and $T_{\rm kin}=63~$K). In general, for gas densities, $n_{\rm H}$ up to 10$^{7}$~cm$^{-3}$, the models predict both sets of transitions to exhibit excited state maser emission. 
We find that the excited states of CH are only weakly populated: The corresponding line optical depths of the main lines in each of the excited states modelled, lie close to $\tau\!\sim -10^{-5}$, which results in low maser gains and line intensities that are unlikely to be observable. It is possible that, similar to the first rotationally excited level of CH discussed in this work, the excitation of higher rotational levels of CH may also be dominated by collisional excitation. In the regime where collisions dominate, the excited lines appear in absorption against the bright background envelope. However, this absorption occurs only in unrealistically  dense regions with gas densities of $n_{\rm H}>10^{7}~$cm$^{-3}$ for the 7.3~GHz lines and $n_{\rm H}>10^{9}~$cm$^{-3}$ for the 4.8~GHz lines. Therefore, in contrast to OH, the possibility of detecting lines from  higher rotationally excited states of CH is highly unlikely.

\section{Conclusions} \label{sec:conclusions}
Revisiting the search for the spectral lines from the first rotationally excited state of CH, this paper presents the first interferometric detection of the elusive CH lines at 700~MHz in absorption towards W51\,E using the uGMRT. The absorption features of the main lines at 701~MHz and 724~MHz are detected near $\upsilon_{\rm LSR}\!\sim\!55~$km~s$^{-1}$ at significance levels of 3$\,\sigma$ and $2.6\,\sigma$, respectively, while the satellite lines show no clear detections. 

In contrast to the non-LTE radiative transfer analysis presented in Paper~I, also carried out using the MOLPOP-CEP code, this work incorporates the influence of collisions with electrons in exciting CH lines. The HFS lines within the first excited state of CH, around 700~MHz, are found to be excited in regions of high gas densities  ({$n_{\rm H} \sim$ a few ${\times10^{5}~}$cm$^{-3}$}), where collisional processes dominate the excitation of this line. These gas densities are close to the critical density of the 532/536~GHz lines, linking the first excited level to the ground state. While the observed anomalous excitation in the ground-state HFS lines of CH, particularly the enhancement of its lowest satellite line at 3.264~GHz, is best reproduced in low-density regions (${n_{\rm H} \leq 1800~}$cm$^{-3}$), the effects of IR trapping and line overlap are most dominant. It is for this reason that the ground-state masing is most effectively observed in velocity components near 65~km~s$^{-1}$ corresponding to the sub-mm/FIR background continuum while the absorption dip observed in the 700~MHz lines peaks closer to velocities of 55~km~s$^{-1}$, corresponding to the dense star forming clump. Exploring the  excitation conditions of these lines reveals that the 700~MHz lines are characterised by low and positive values of excitation temperature, and achieve thermalisation only as the masing in the 3.264~GHz is suppressed. Furthermore, from our analysis it is unclear if the inclusion of collisions between CH and electrons plays a significant role in thermalising the HFS lines of both the ground and first excited states, particularly at lower temperatures. It is worth noting that further exploration, employing perhaps a two-temperature model to separately characterise neutrals and electrons is necessary to accurately benchmark contributions from different collision partners.\\

Moreover, modelling the physical conditions conducive for the excitation of the 700~MHz lines has allowed us to constrain the conditions that give rise to the 532/536~GHz lines. This has proven to be instrumental in separating background emission features from absorption, a common occurrence in the spectra of these lines that complicates their analysis. This extends the use of the 532/536~GHz lines in determining CH column densities, which was previously limited due to the necessity of ad-hoc assumptions in modelling the background emission. While the models described in Paper I underscore the significance of the sub-mm/FIR rotational lines of CH in amplifying the widespread ground state maser of CH, the analysis presented in this work has delved into the nature of the HFS lines within the $\Lambda$-doublet levels of its first rotationally excited state. This exploration is critical for providing a comprehensive understanding of CH's ground-state maser.
The non-detection of the 700~MHz lines of CH towards the other sources in this study may be due to sensitivity issues, that make it challenging to discern the faint signal from the low background continua at this frequency. Alternatively, for a subset of these sources, the absence of the 700~MHz lines may also be attributed to higher gas densities within the CH-bearing sources. For now, this places W51~E, both with its strong background continuum emission and densities in some layers that aren't too high, in a Goldilocks paradigm.   

\section{Outlook} \label{sec:outlook}
The HFS lines from the first rotationally excited state of CH at 700~MHz display Zeeman splitting owing to their large Land\'{e}-g factors.  
In particular, excited in gas densities of a few times $10^{5}~$cm$^{-3}$, these transitions have been suggested to have the potential to probe the magnetic field strength in massive star-forming regions. As, for example, our study shows, their use as a Zeeman effect probe however, is greatly limited by the difficulties to detect these lines. Given that they probe regions with densities of around $10^5$\,cm$^{-3}$, following \citet{Crutcher2012} and \cite{Koley2021}, we empirically expect B-field strengths of a few hundred $\mu$G, which will be extremely challenging to measure. 
Based on the results of this work, focused searches for these lines are recommended towards bright continuum sources with gas densities of approximately $\sim\!10^{5}$cm$^{-3}$. If successful, any new detection of the 700~MHz lines would not only shed light on the excitation of the CH ground state levels but, if their Zeeman splitting could be detected, would provide a direct method for exploring magnetic fields in massive star-forming regions in a density regime that has been poorly addressed by previous efforts. Nonetheless, detecting the first rotationally excited lines of CH at 700~MHz with interferometry using the uGMRT marks progress for future studies on the Zeeman effect, which currently are sensitivity-limited. With an anticipated increase in sensitivity of more than an order of magnitude, the Square Kilometre Array \citep[see,][for more information]{Robishaw2015}, and the Deep Synoptic Array (DSA)-2000 \citep{Hallinan2019}, in addition to the Canadian Hydrogen Observatory and Radio-transient Detector \citep[CHORD;][]{Vanderlinde2019} promise to open new avenues, capable of achieving the requisite signal-to-noise ratios essential for detecting Zeeman splitting in these spectral lines.\\

The analysis presented in this work and Paper~I can be extended to examine the excitation in other similar systems, including the ground state transitions of OH between 1.6 and 1.7~GHz (18~cm), where a difference lies in the number of HFS components with the ground state $\Lambda$-doublet: while the CH ground state has three, the ground state of OH, lying in the $^{2}\Pi_{3/2}, N=1, J =3/2$ state, splits into four HFS components.

As for CH, up to date excitation studies of OH would profit from the new hyperfine-resolved collisional rate coefficients that were recently calculated by \citet{Klos2020} and \citet{Dagdigian2023}.

\section{Data availability}
Appendices A and B are available at

\begin{acknowledgement}
We thank the staff of the GMRT that made these observations possible. GMRT is run by the National Centre for Radio Astrophysics of the Tata Institute of Fundamental Research. We are grateful to the referee, for a careful review of the manuscript of this article and their insightful comments which have helped to improve its clarity. The authors extend their gratitude to Ayan Acharyya for valuable help in automating routines and Rohit Dokara for his valuable comments on data reduction. A.~M.~J. and D.~A.~N gratefully acknowledge the support of grant SOF 08-0038 from USRA. A.~M.~J. would also like to acknowledge the support of the Max Planck Society. M.~N. acknowledges postdoctoral fellowship support from a Max
Planck–India Partner Group Grant and support from the Indian Institute of Science, Bangalore.
\end{acknowledgement}

\bibliographystyle{aa} 
\bibliography{refs}

\begin{appendix}
\onecolumn 
\section{Examining the robustness of the spectral line fits}\label{appendix:fit_checks}
\begin{figure*}[h]
    \centering
    \includegraphics[width=0.31\textwidth]{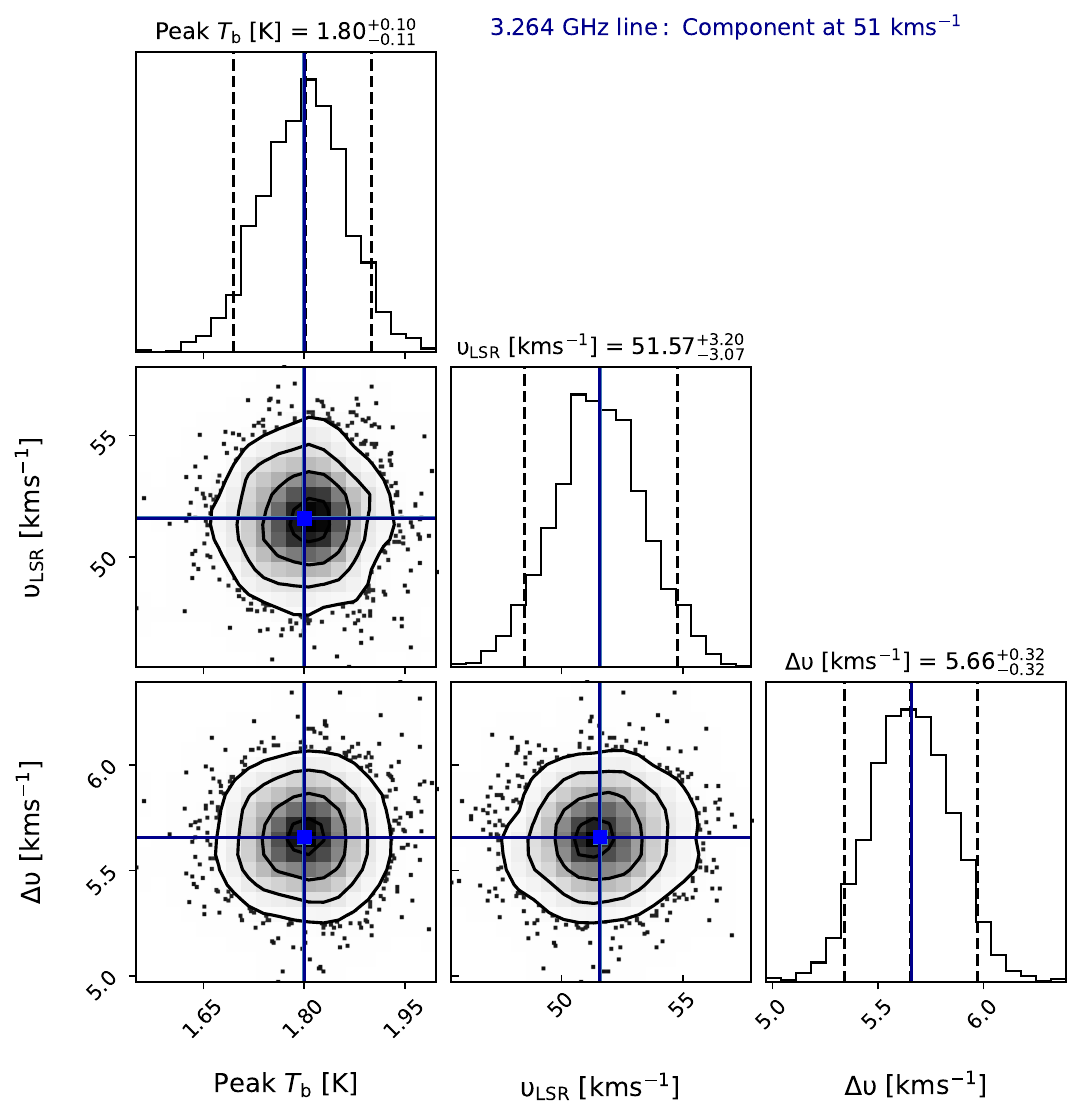} \quad
    \includegraphics[width=0.31\textwidth]{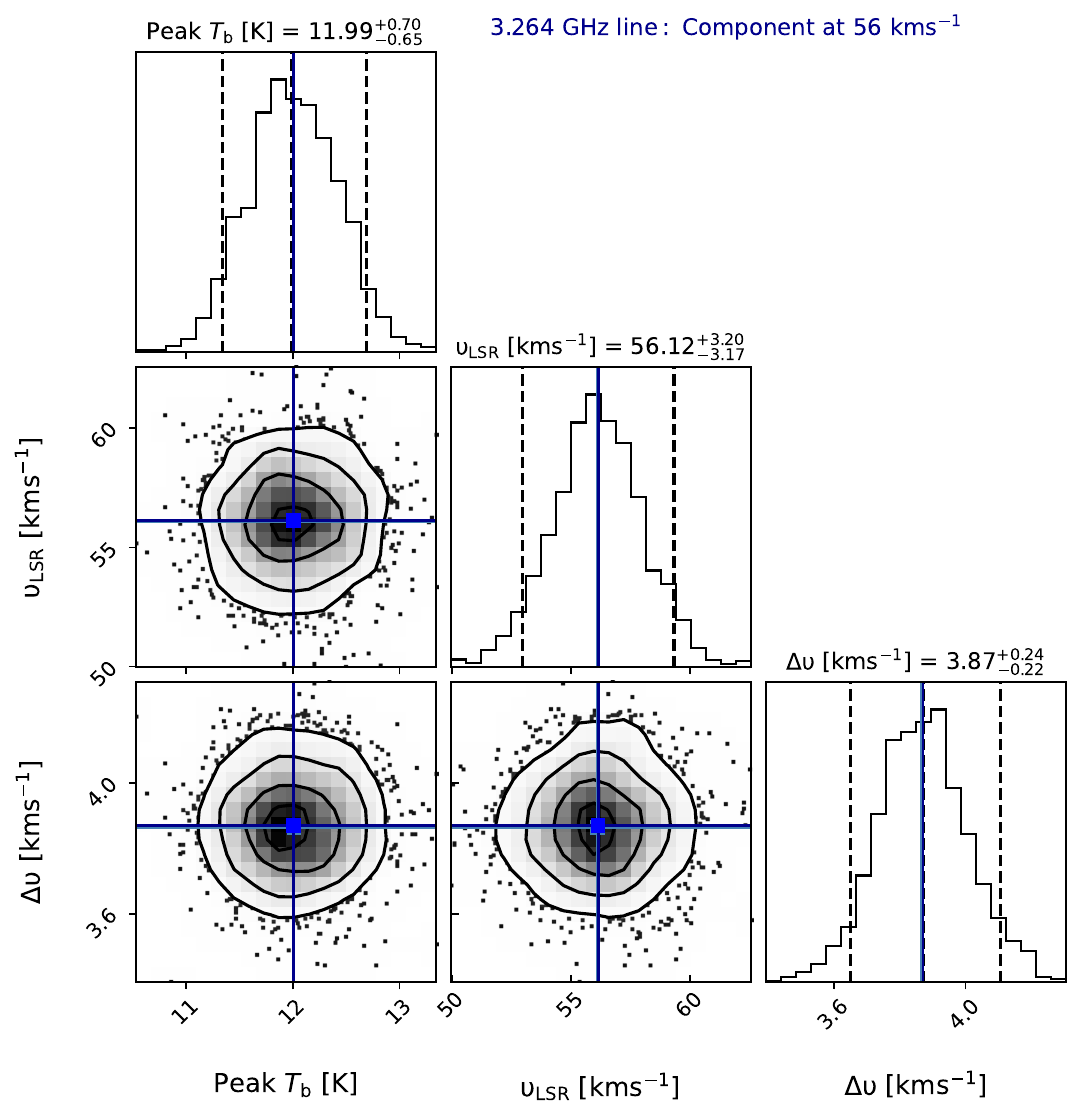} \quad
        \includegraphics[width=0.31\textwidth]{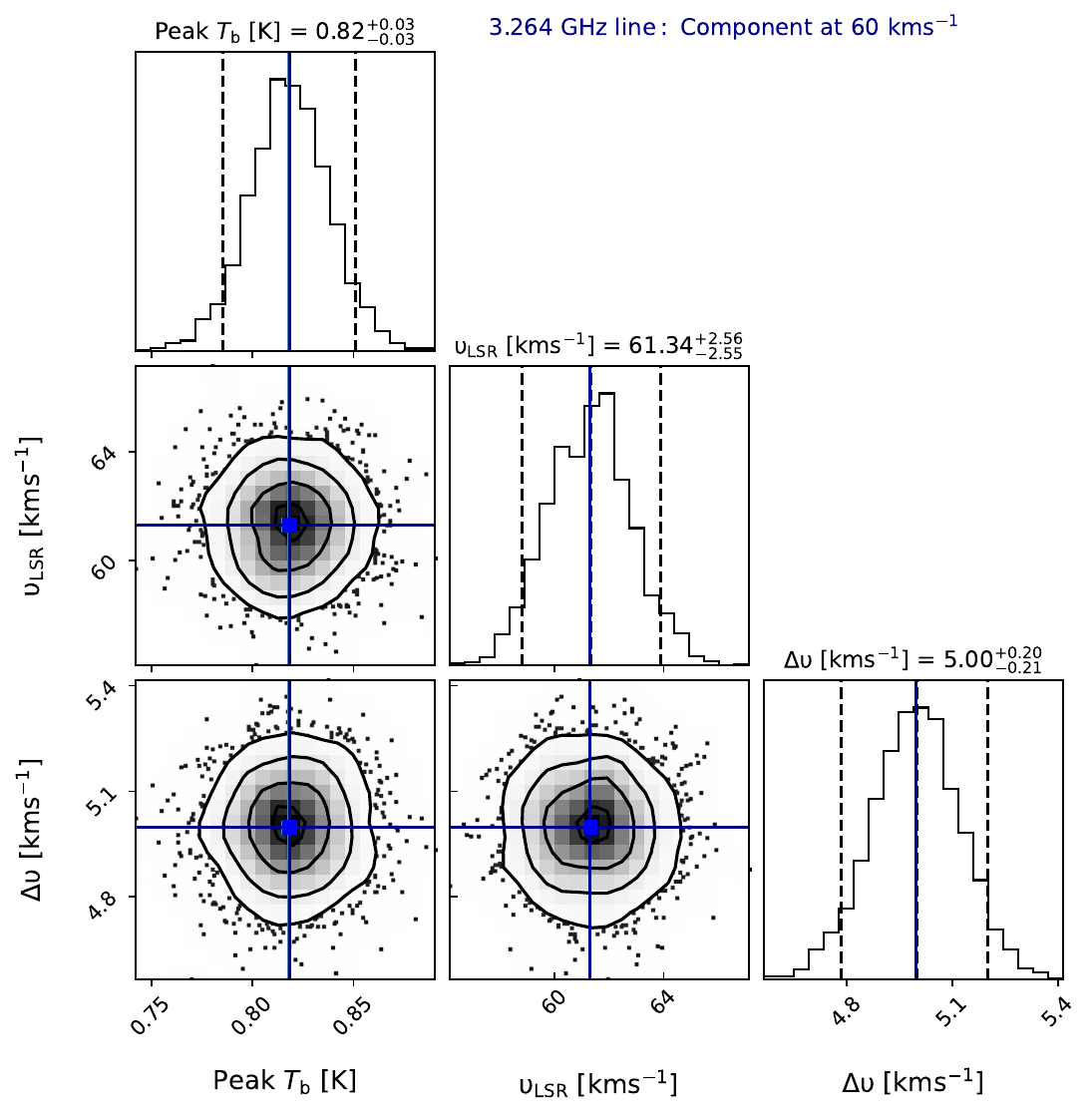} \\
    \includegraphics[width=0.31\textwidth]{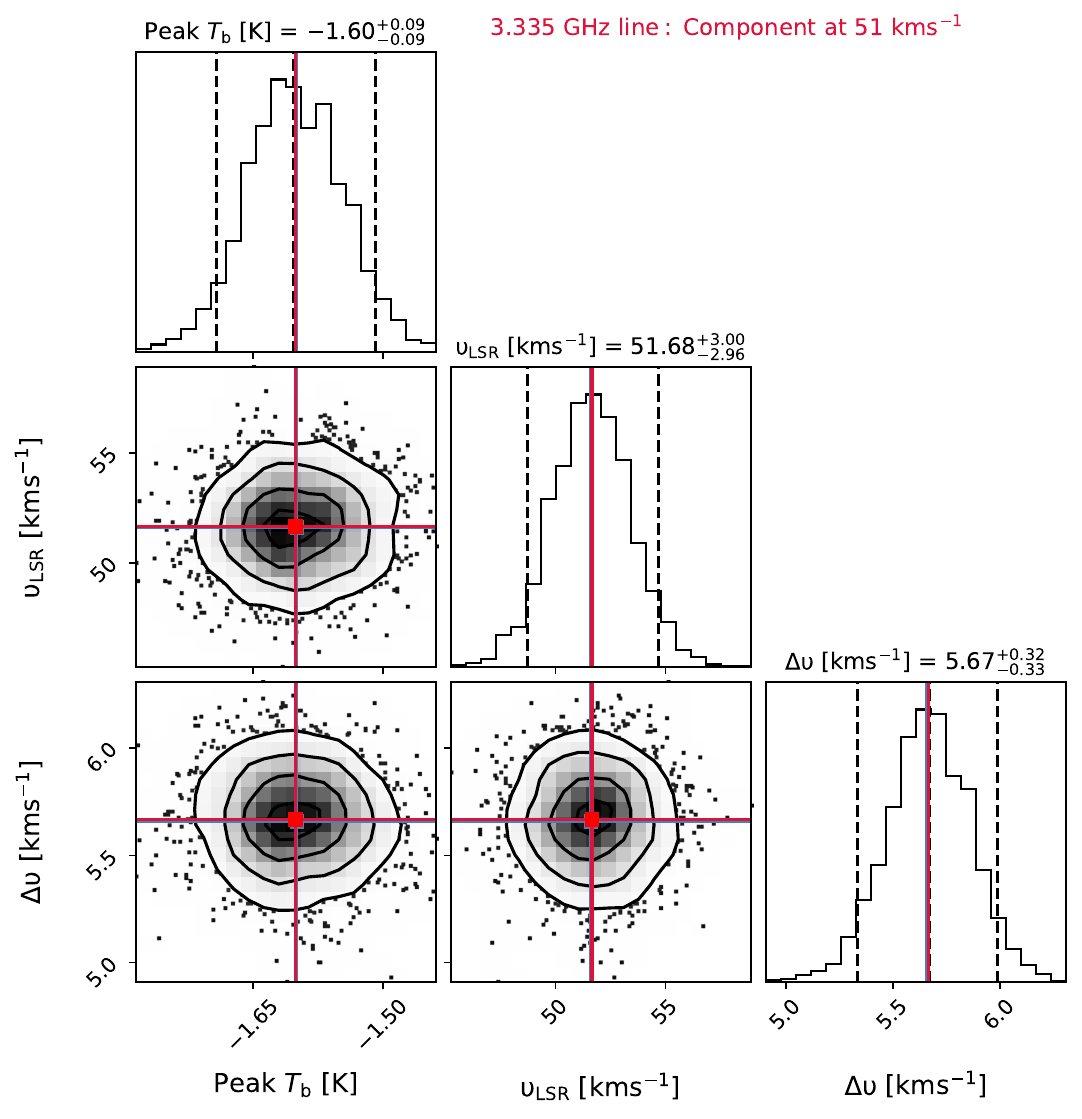} \quad
    \includegraphics[width=0.31\textwidth]{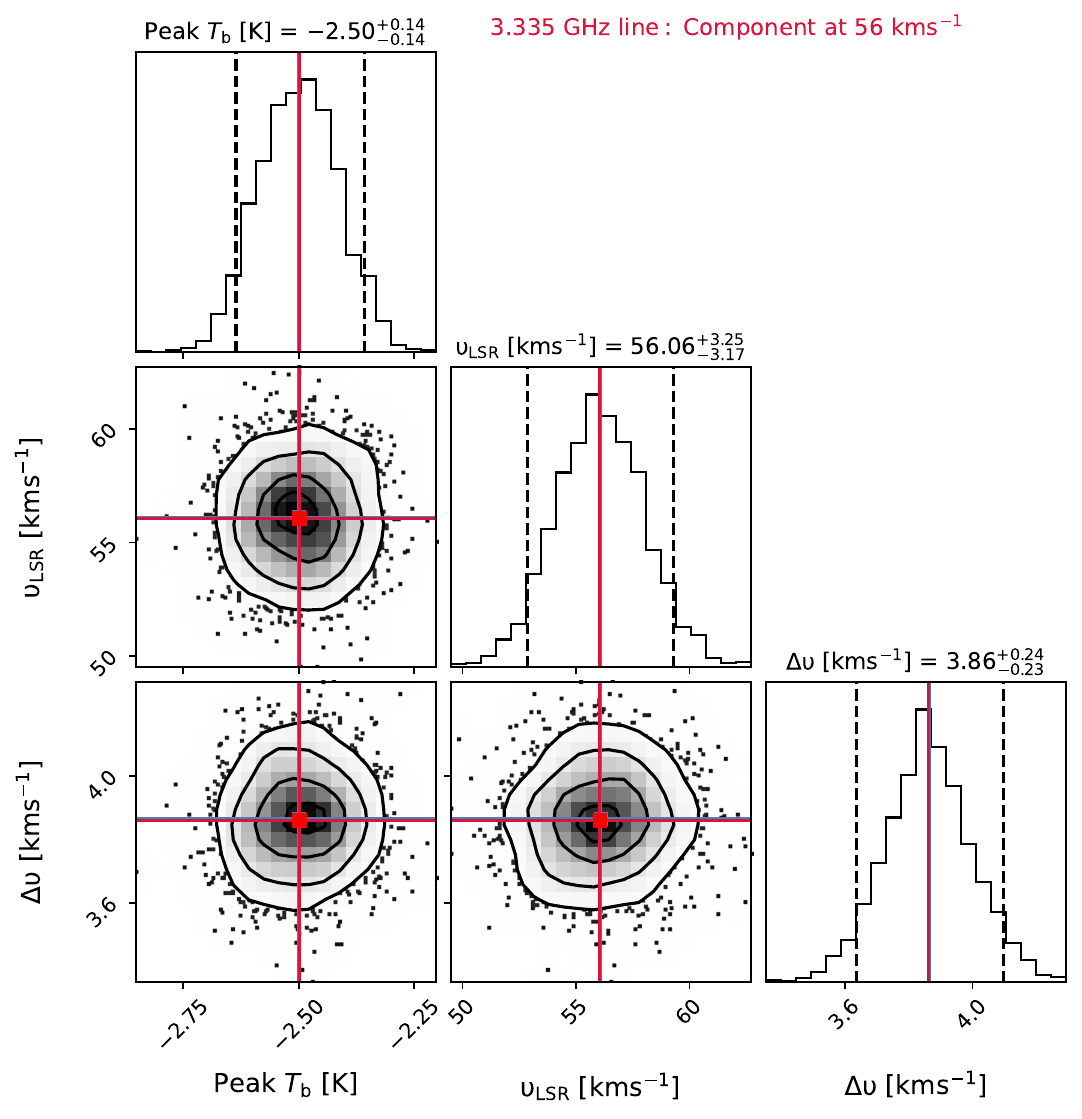}\quad
    \includegraphics[width=0.31\textwidth]{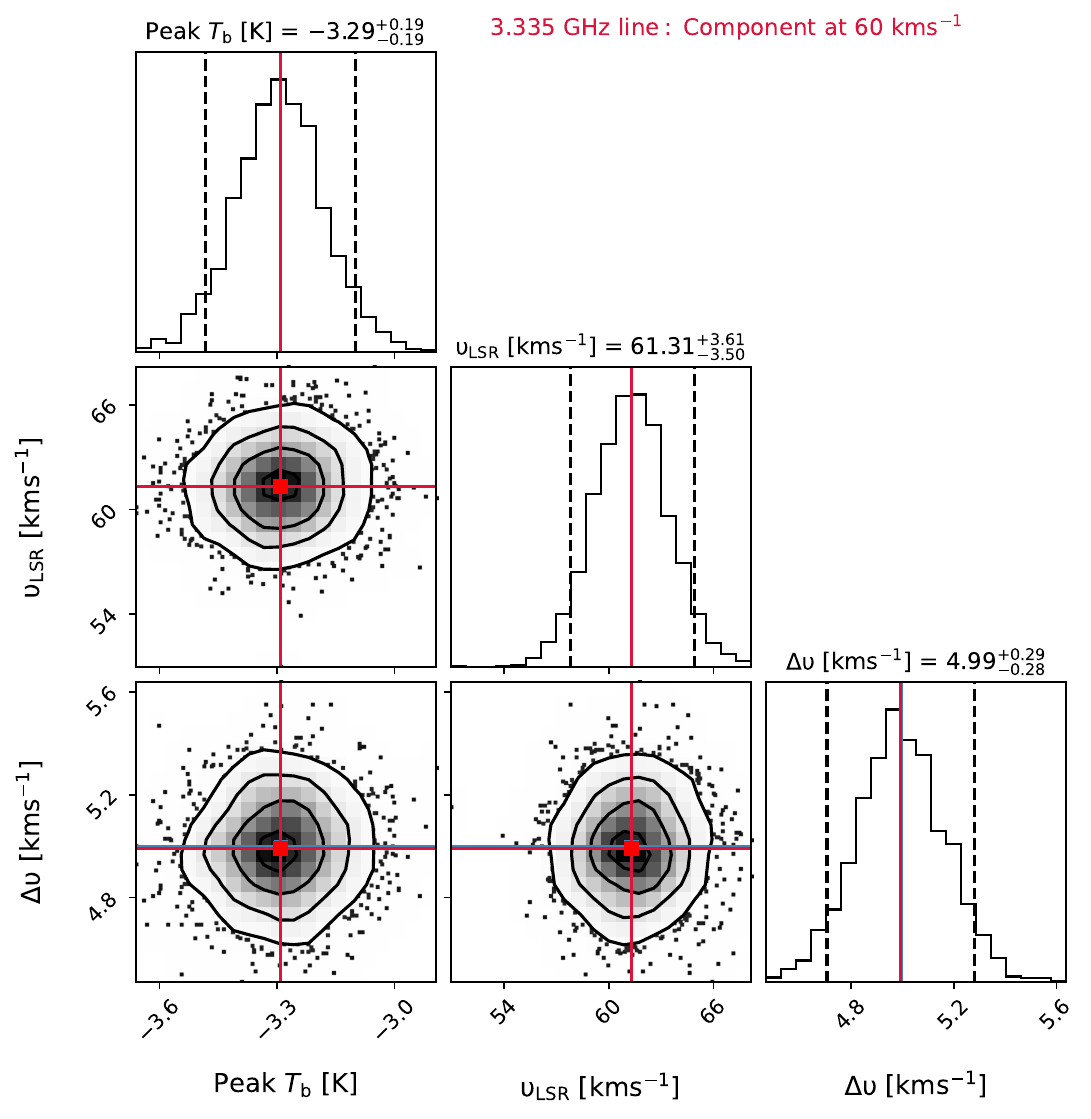}\\
    \includegraphics[width=0.31\textwidth]{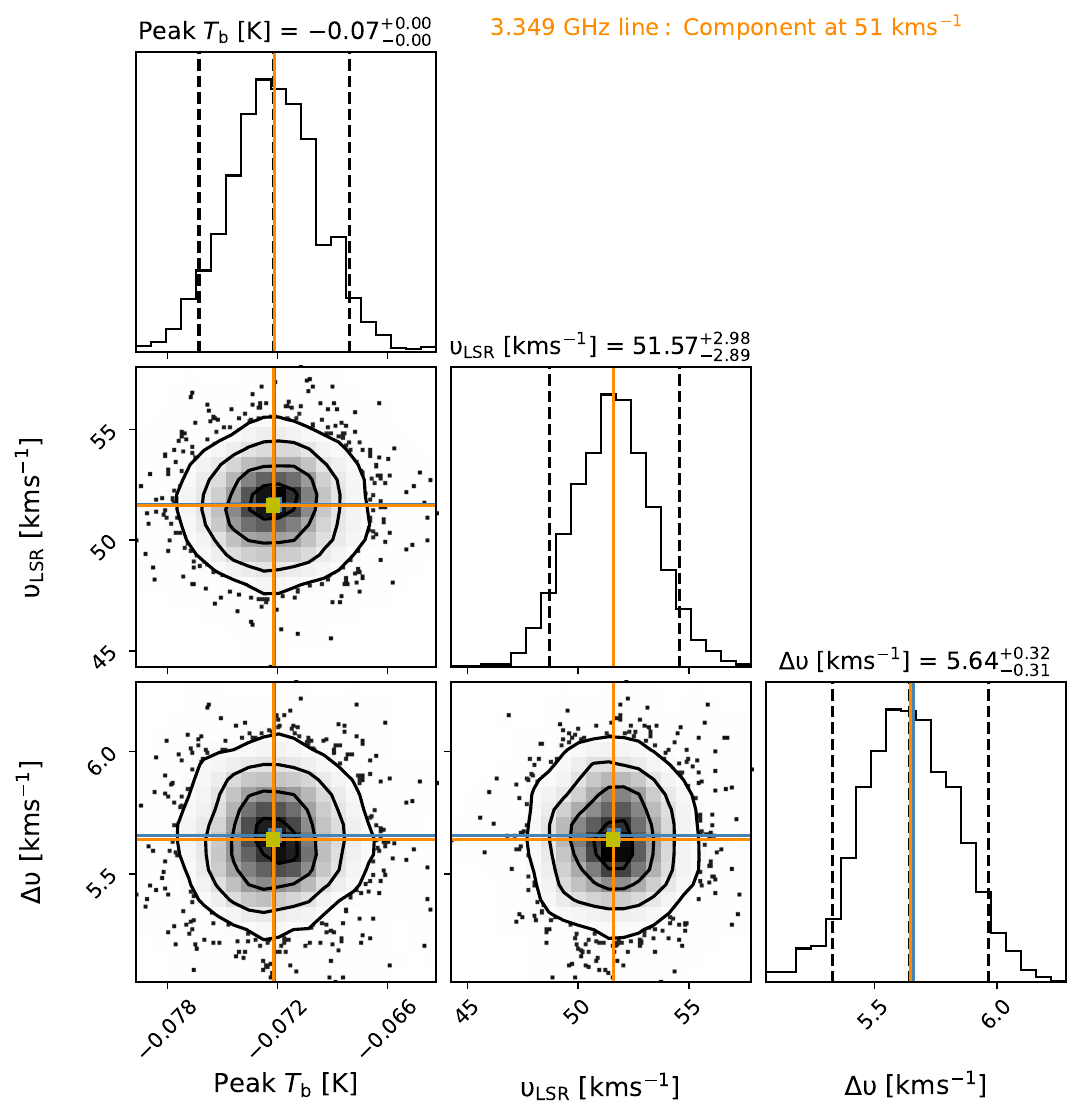} \quad 
    \includegraphics[width=0.31\textwidth]{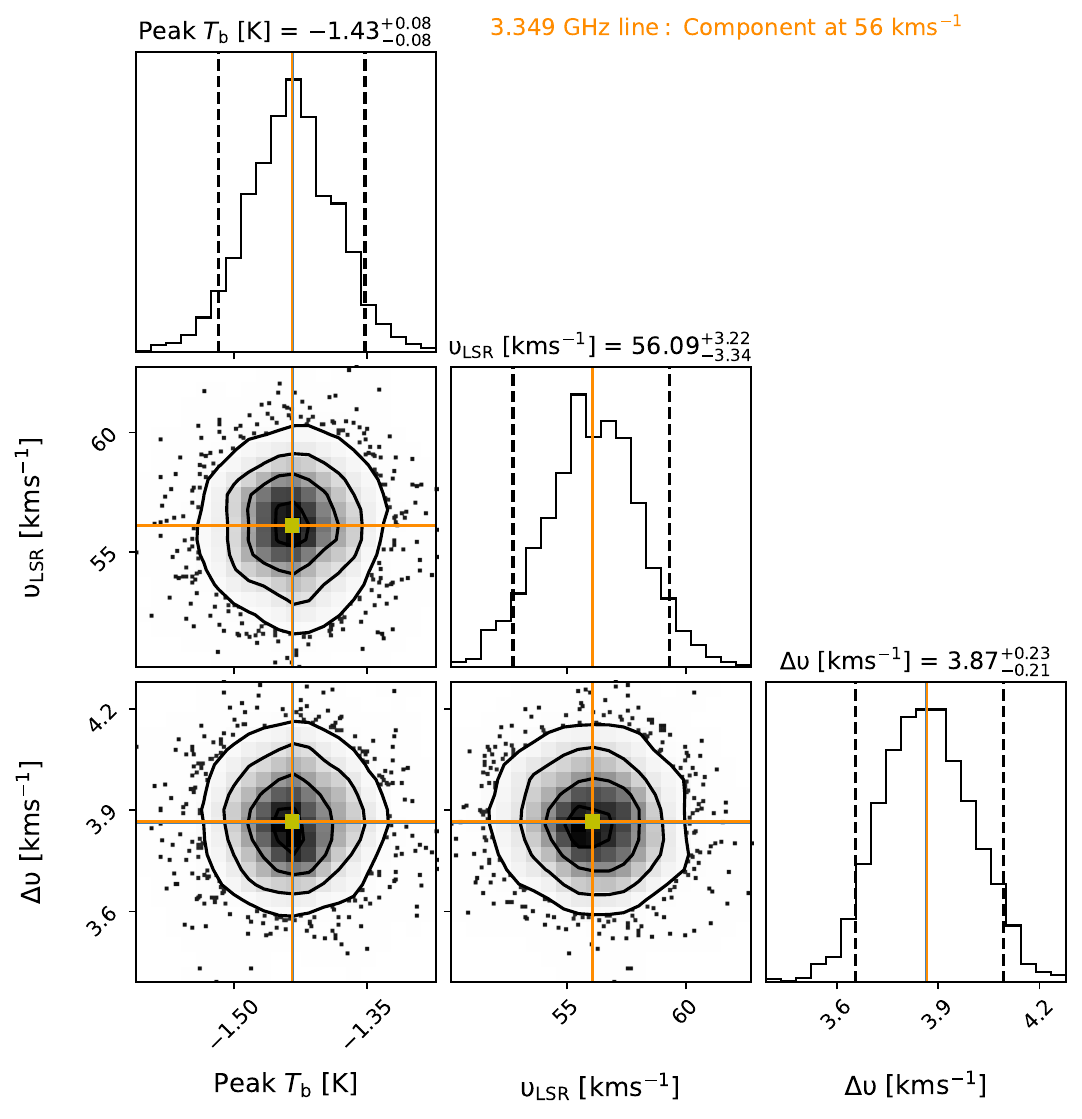} \quad
        \includegraphics[width=0.31\textwidth]{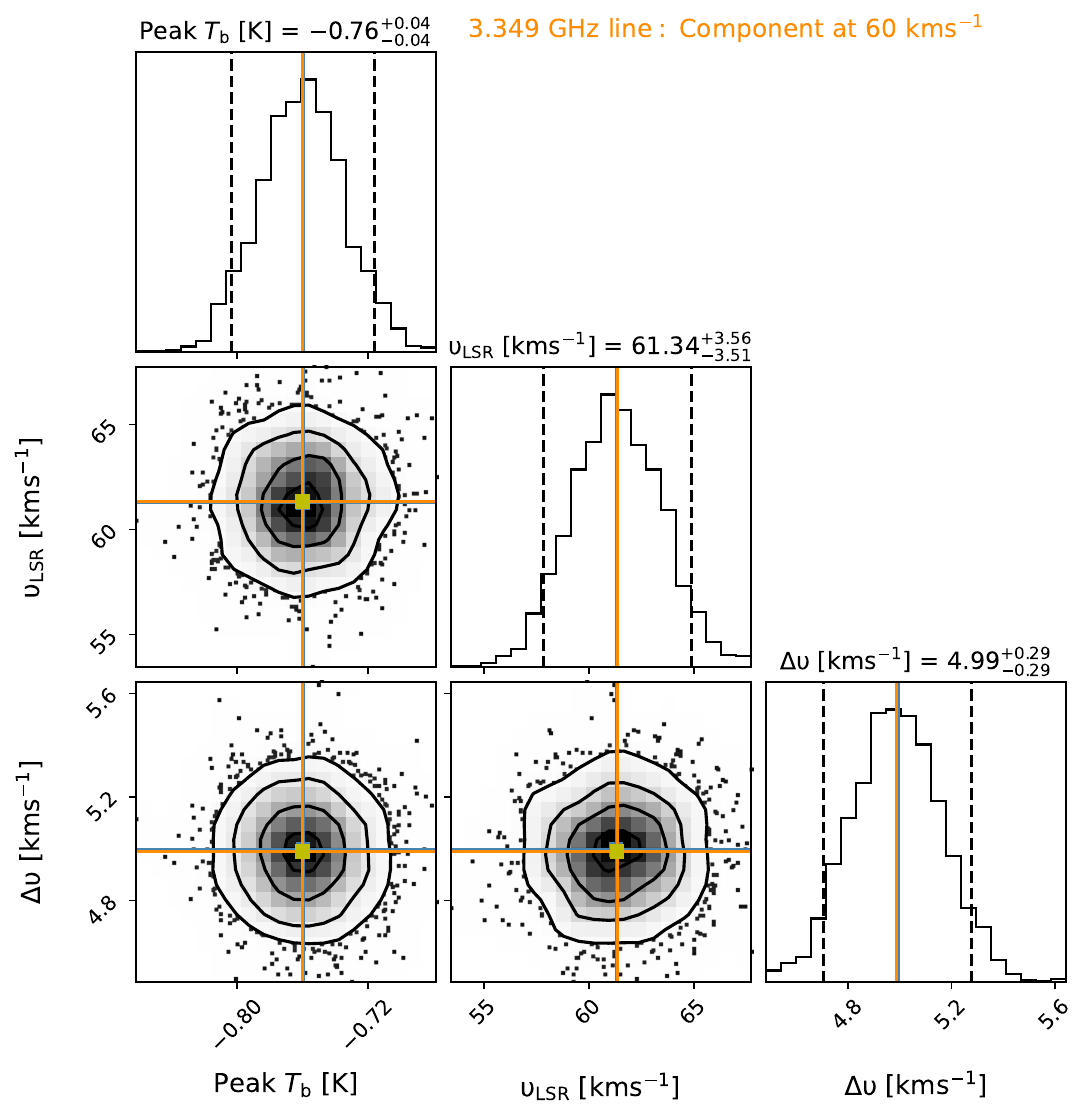} \\
    \caption{ Clockwise from the top to bottom: Corner plots presenting the 2D histograms and probability distribution functions of the Gaussian line fitting parameters (Peak $T_{\rm b}$, $\upsilon_{\rm LSR}$ and $\Delta \upsilon$) for the 3.264~GHz (dark blue), 3.335~GHz (red), and 3.349~GHz (dark orange) spectra, respectively, for the cloud component at 51~km~s$^{-1}$ (left), 57~km~s$^{-1}$ (centre) and 60~km~s$^{-1}$ (right). The solid coloured and dashed black lines indicate the best fit parameter values and the $2\sigma$ confidence intervals, respectively.}
    \label{fig:fit_checks1}
\end{figure*}

\begin{figure*}[h]
    \centering
    \includegraphics[width=0.31\textwidth]{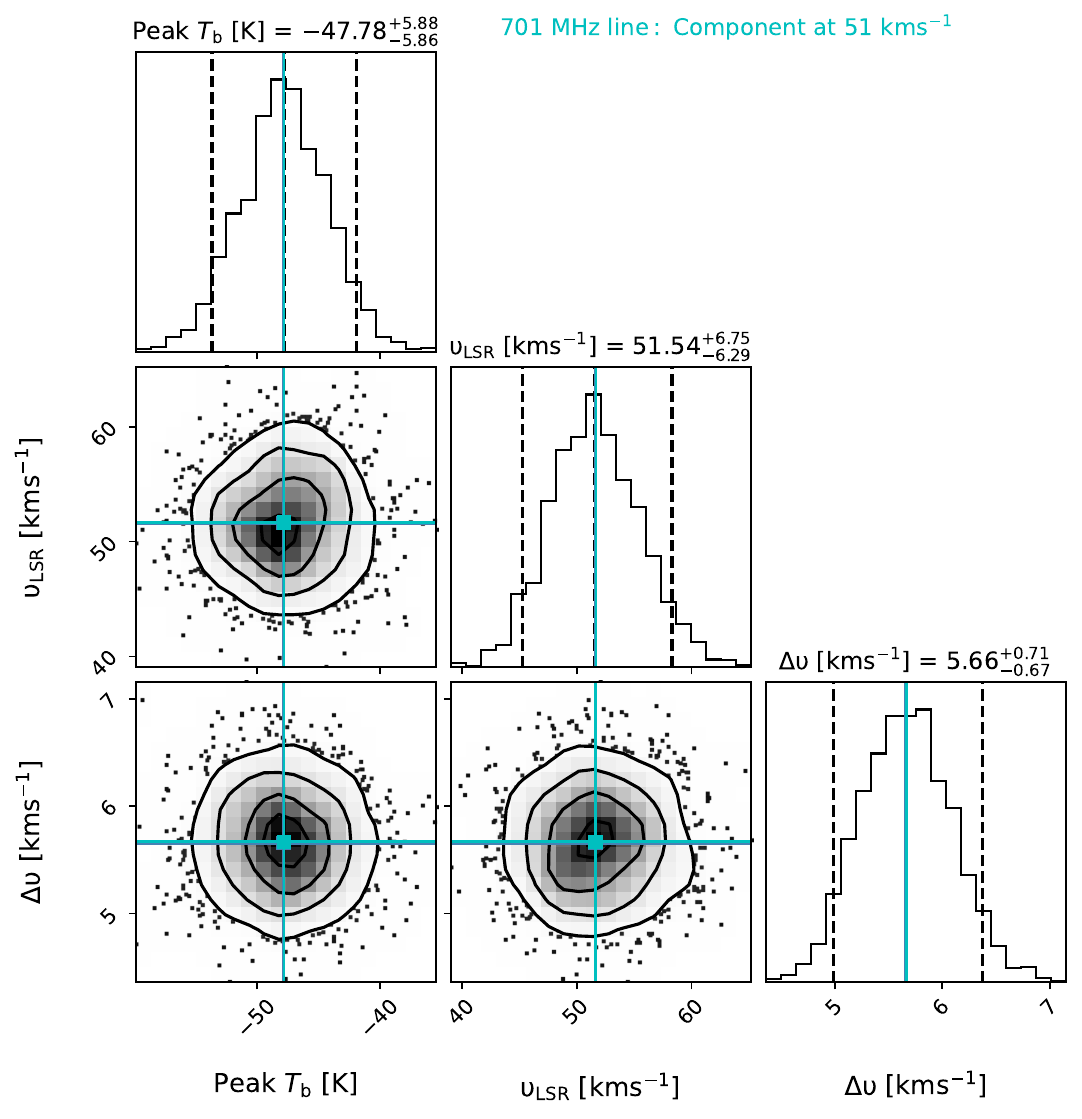} \quad
    \includegraphics[width=0.31\textwidth]{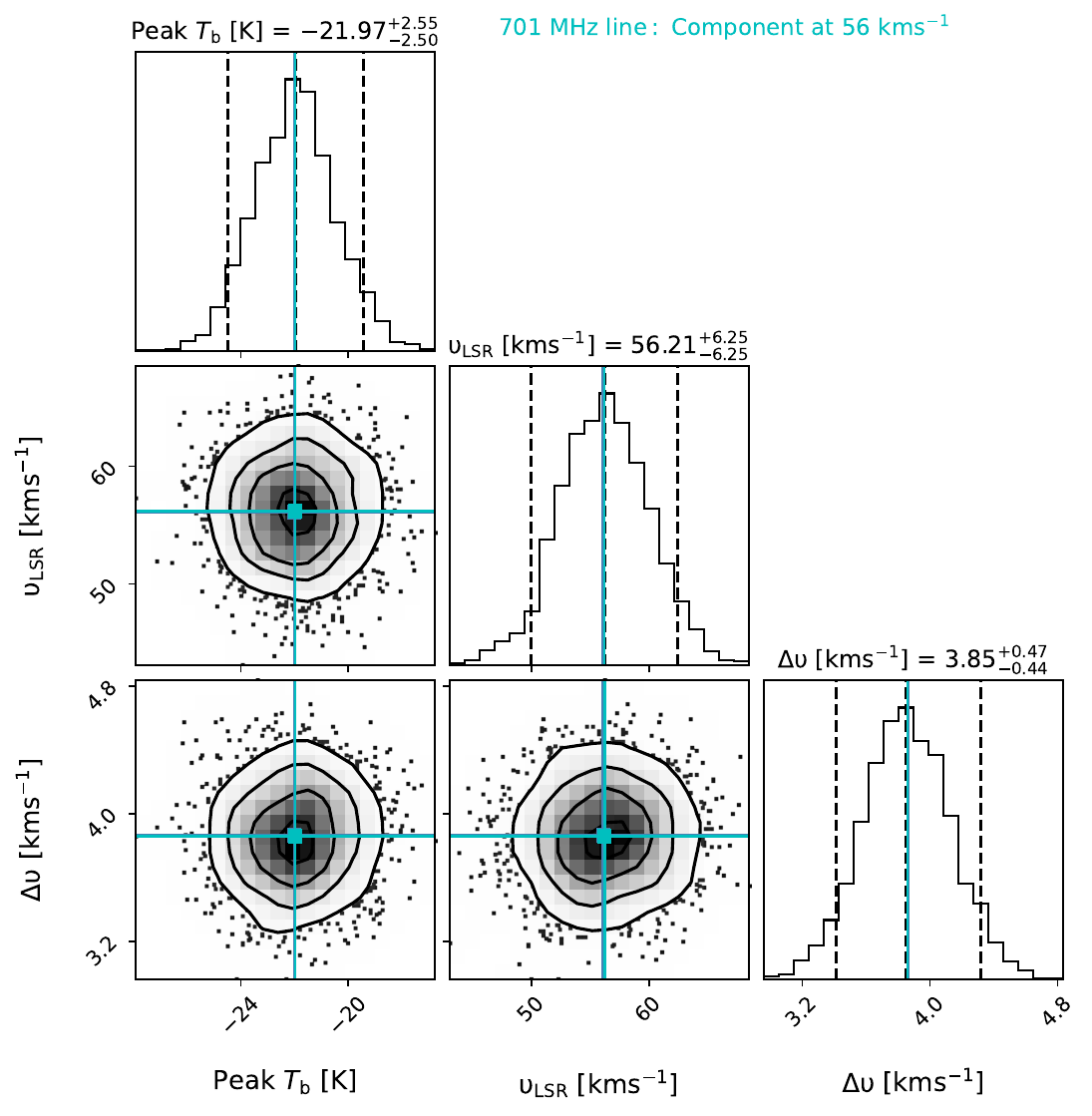} \quad
    \includegraphics[width=0.31\textwidth]{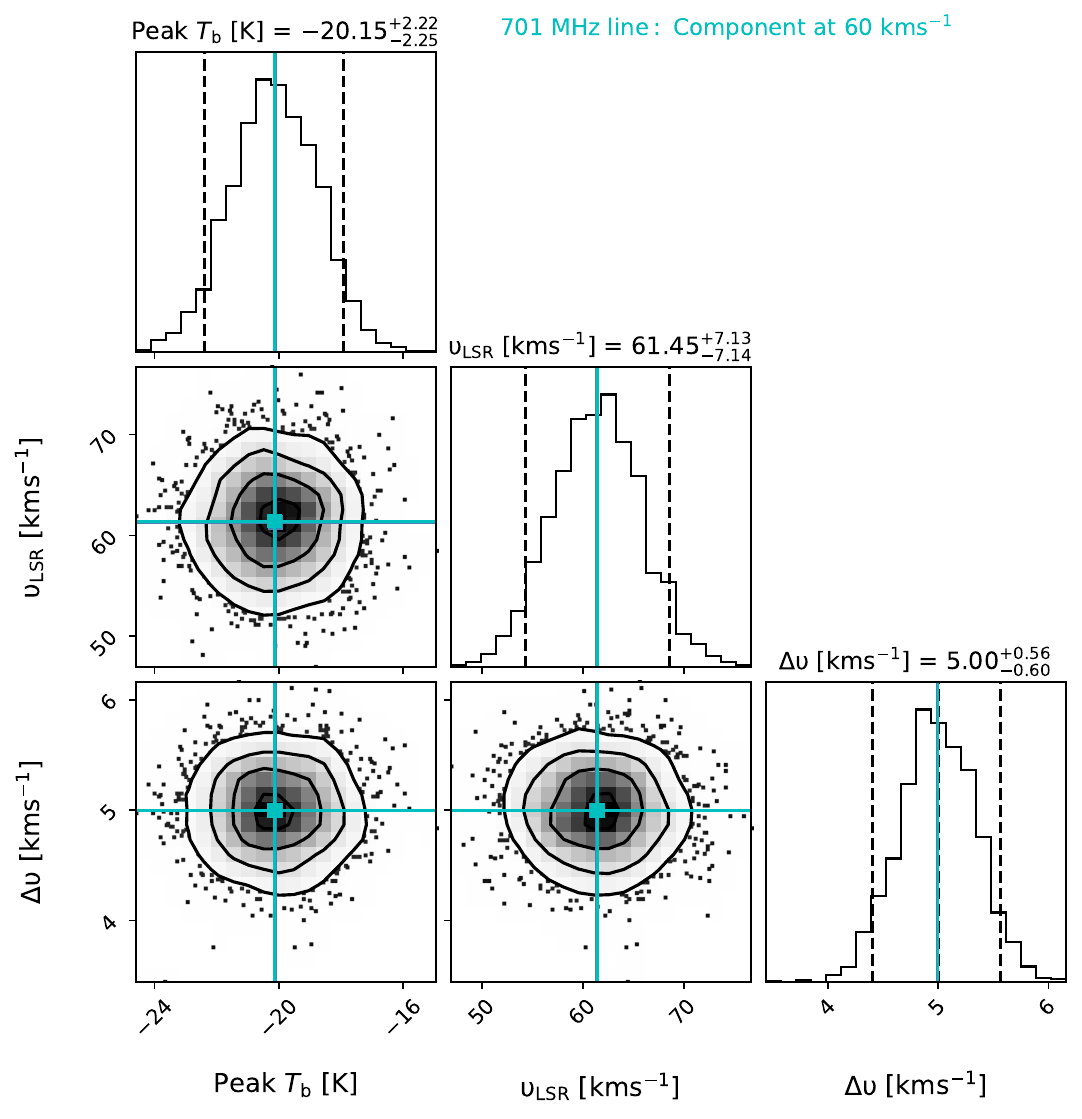}\\
    \includegraphics[width=0.31\textwidth]{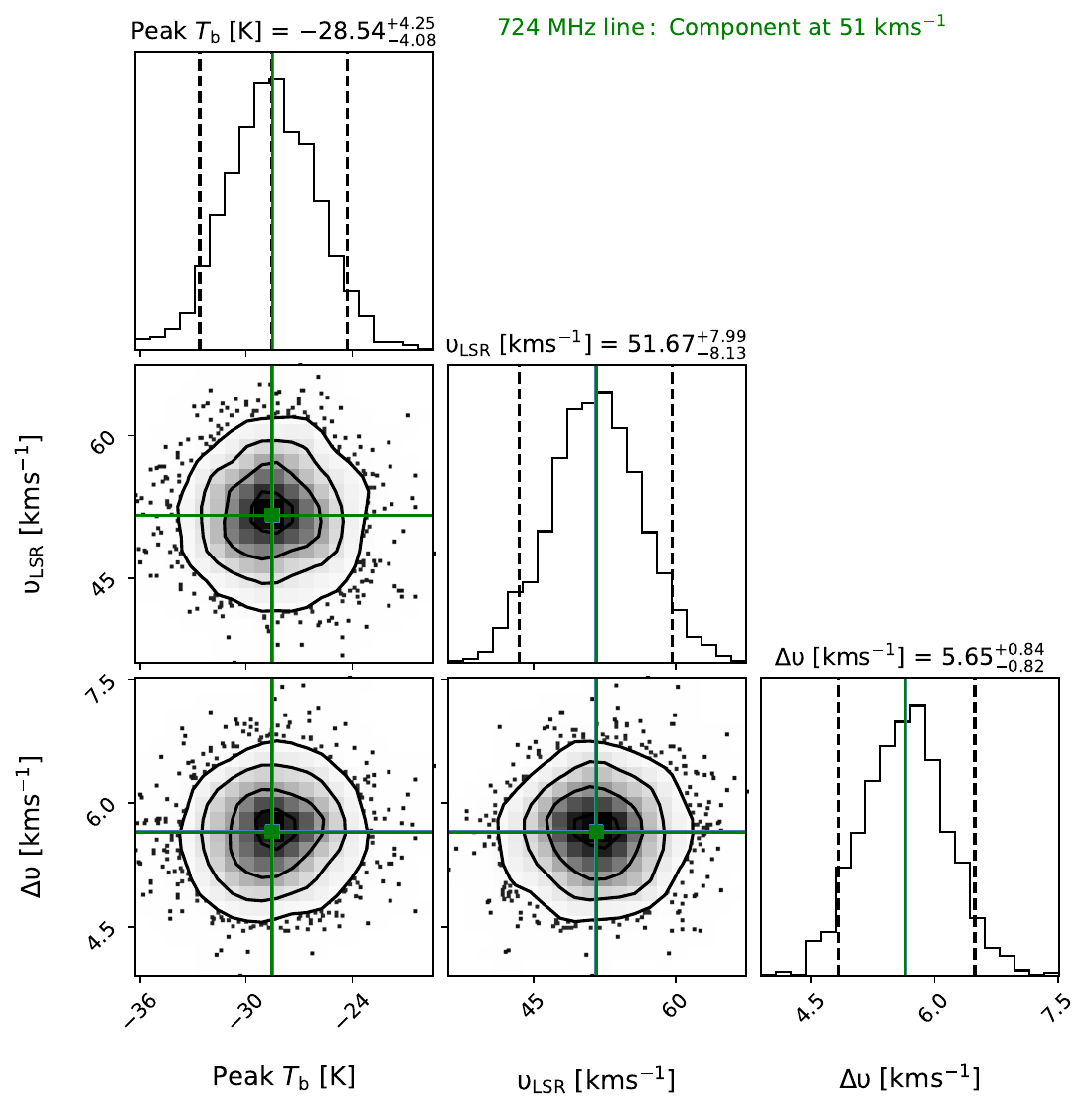} \quad
    \includegraphics[width=0.31\textwidth]{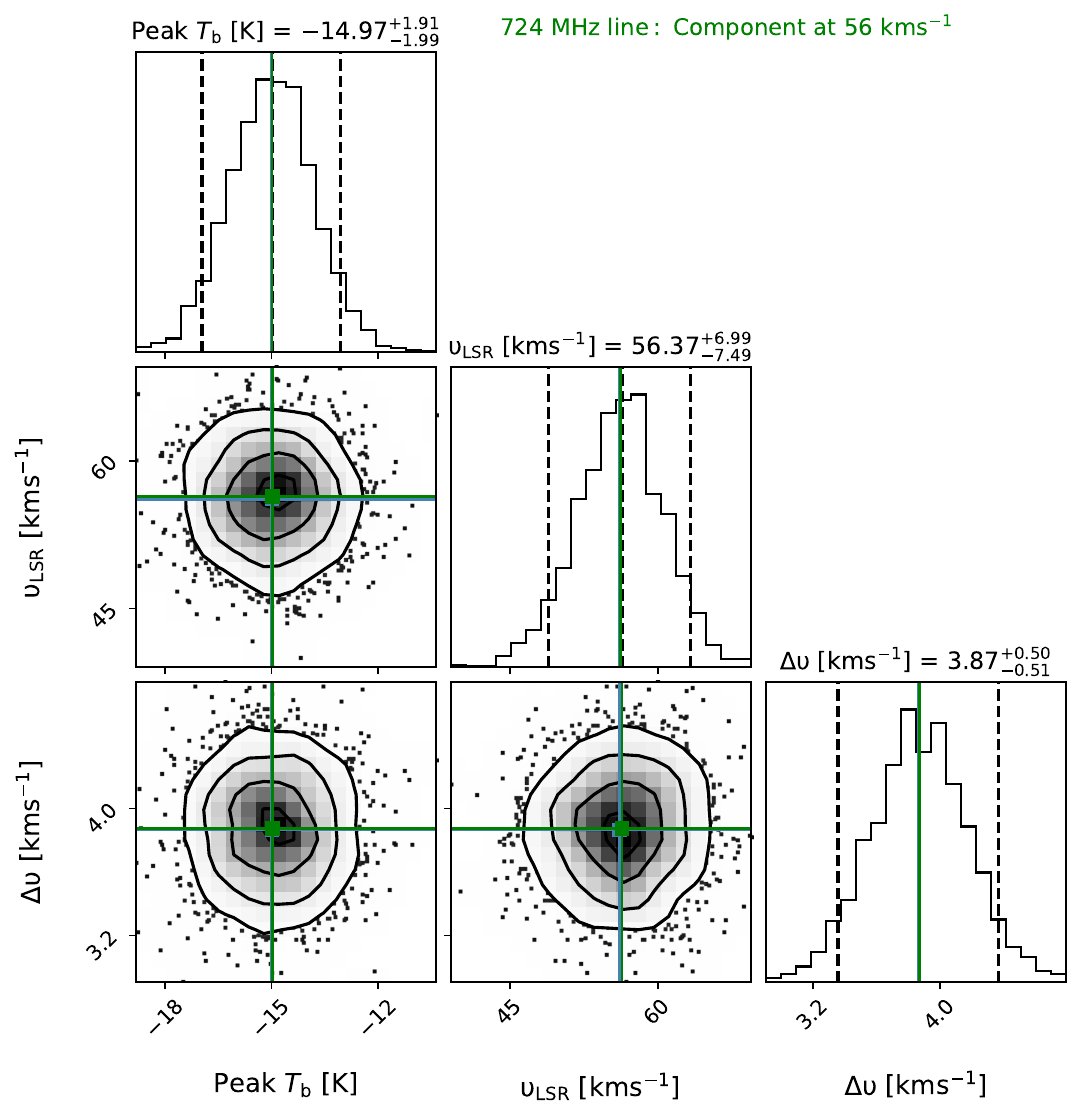} \quad\includegraphics[width=0.31\textwidth]{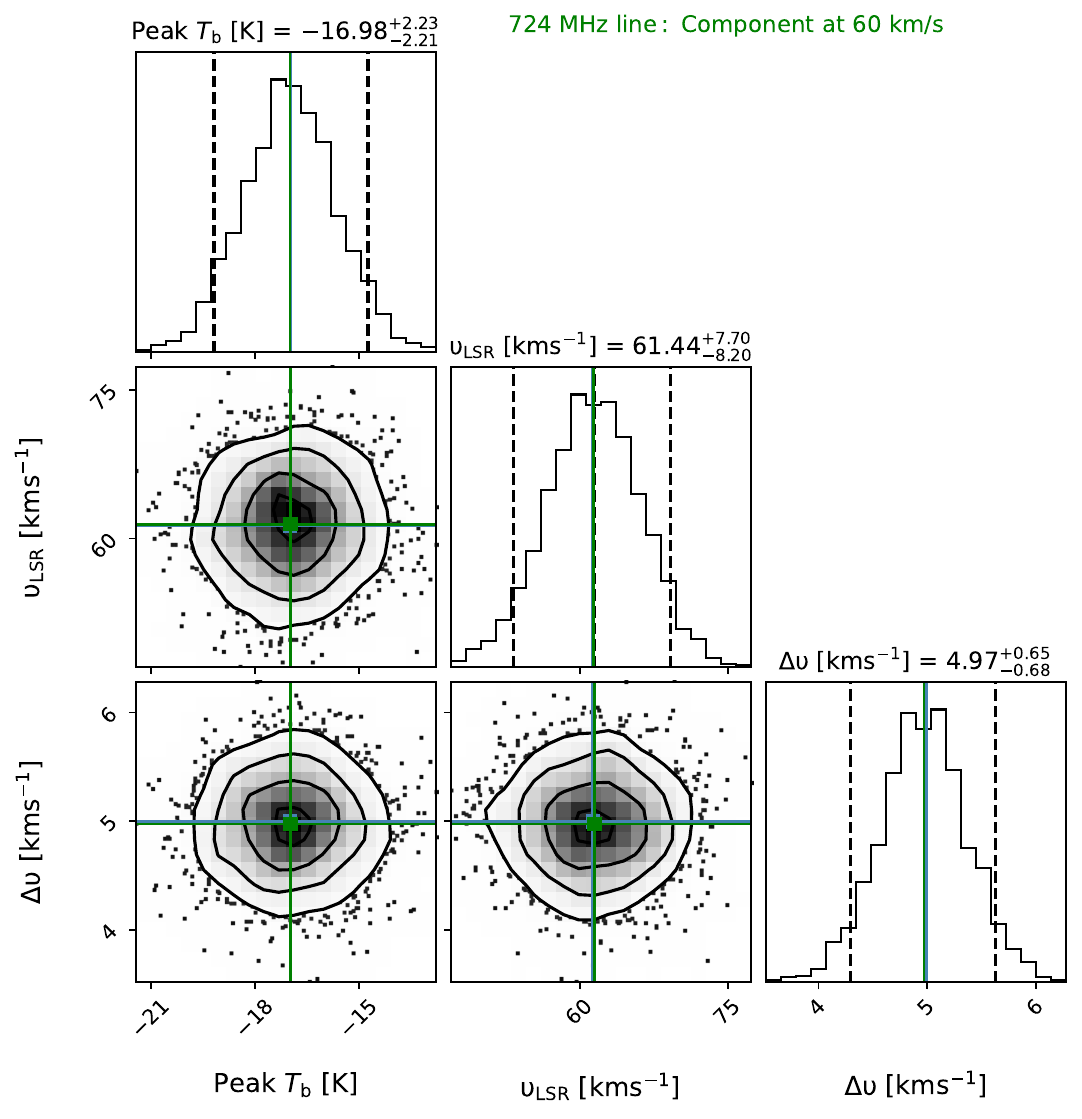}\\
    \caption{Same as Fig.~\ref{fig:fit_checks1} but for the 701~MHz (top) and 724~MHz (bottom) spectra in cyan and green, respectively.}
    \label{fig:fit_checks2}
\end{figure*}

\begin{figure*}[h]
    \centering
    \includegraphics[width=0.31\textwidth]{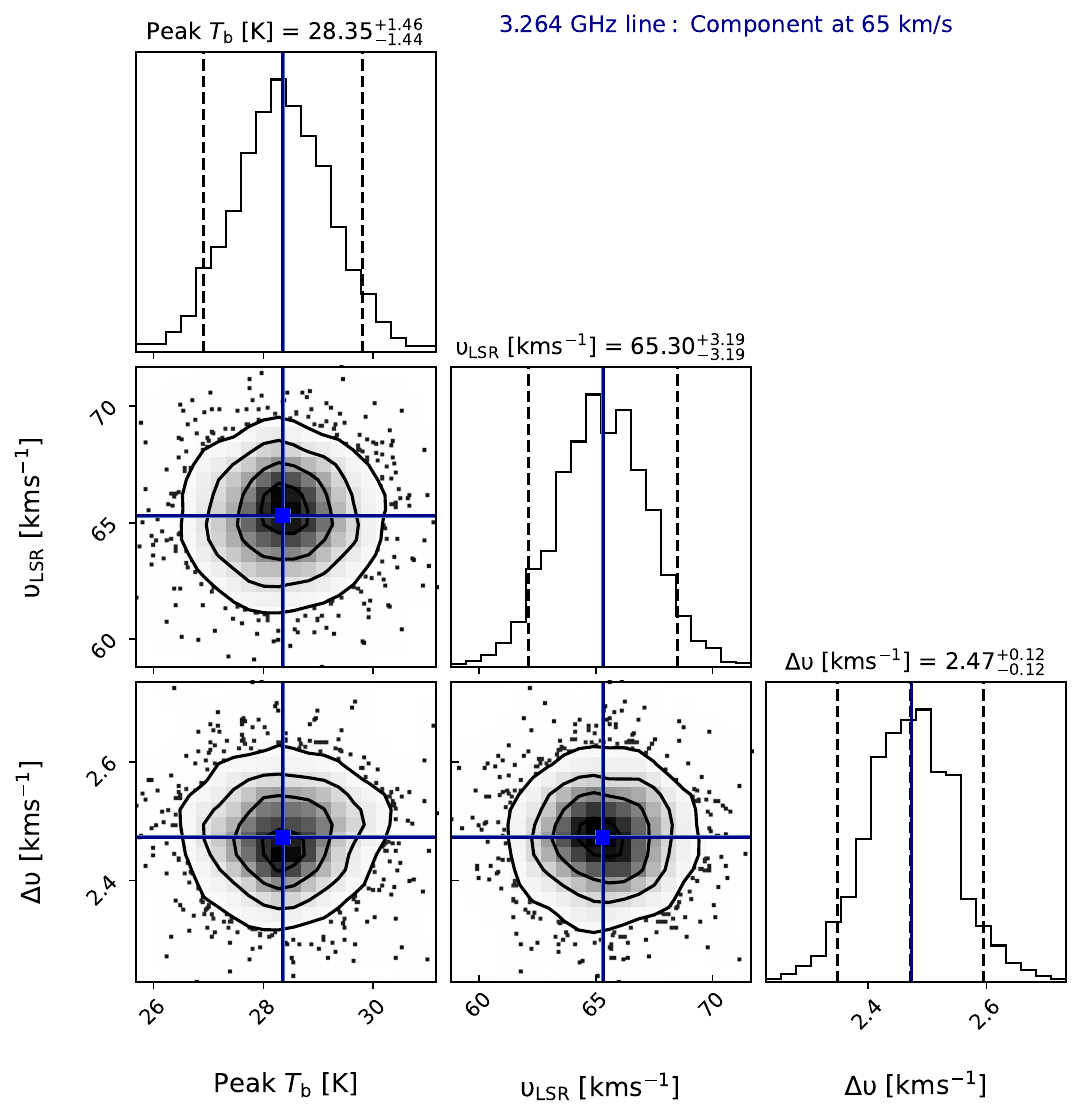} \quad
    \includegraphics[width=0.31\textwidth]{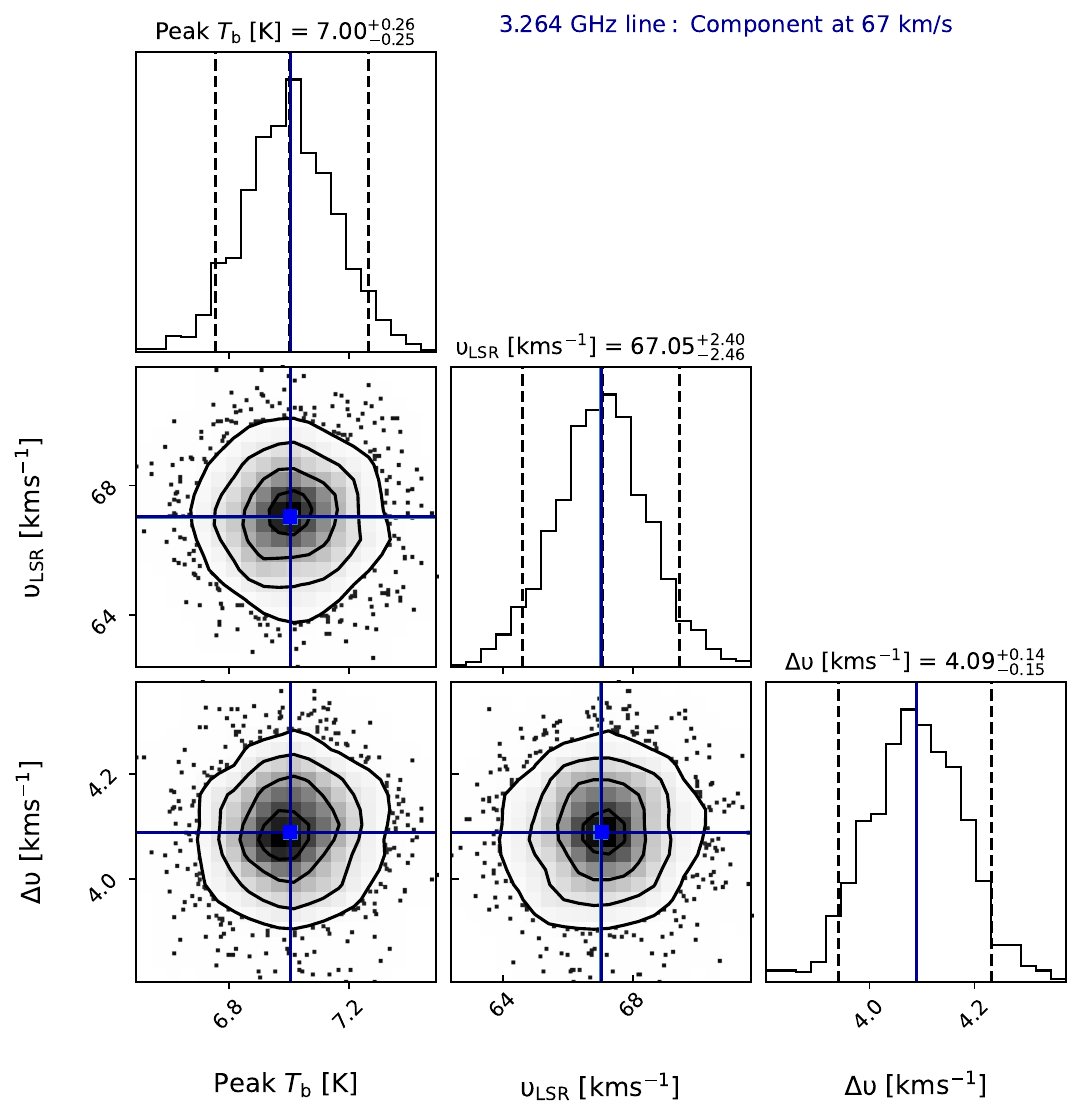} \\
    \includegraphics[width=0.31\textwidth]{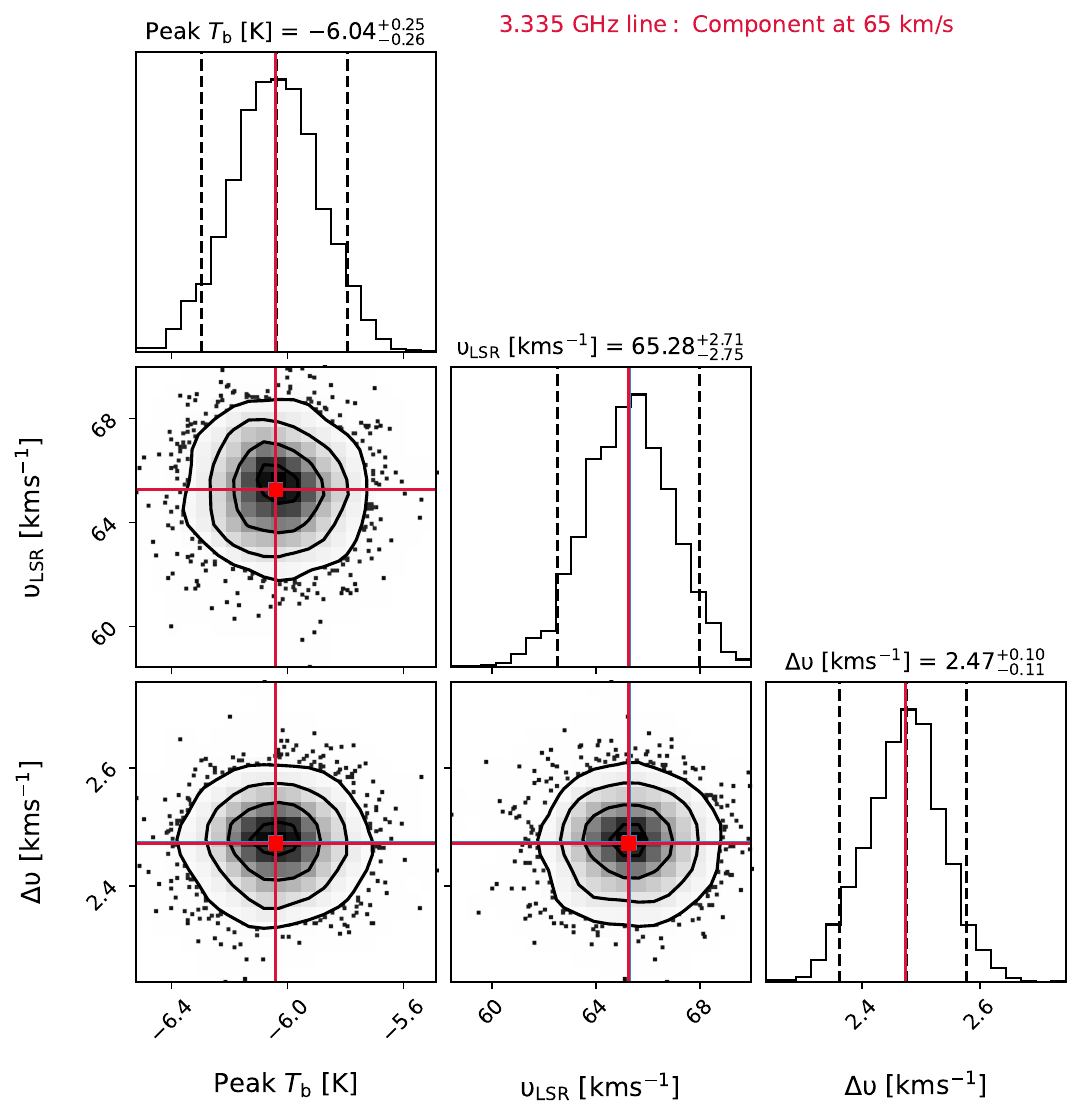} \quad
    \includegraphics[width=0.31\textwidth]{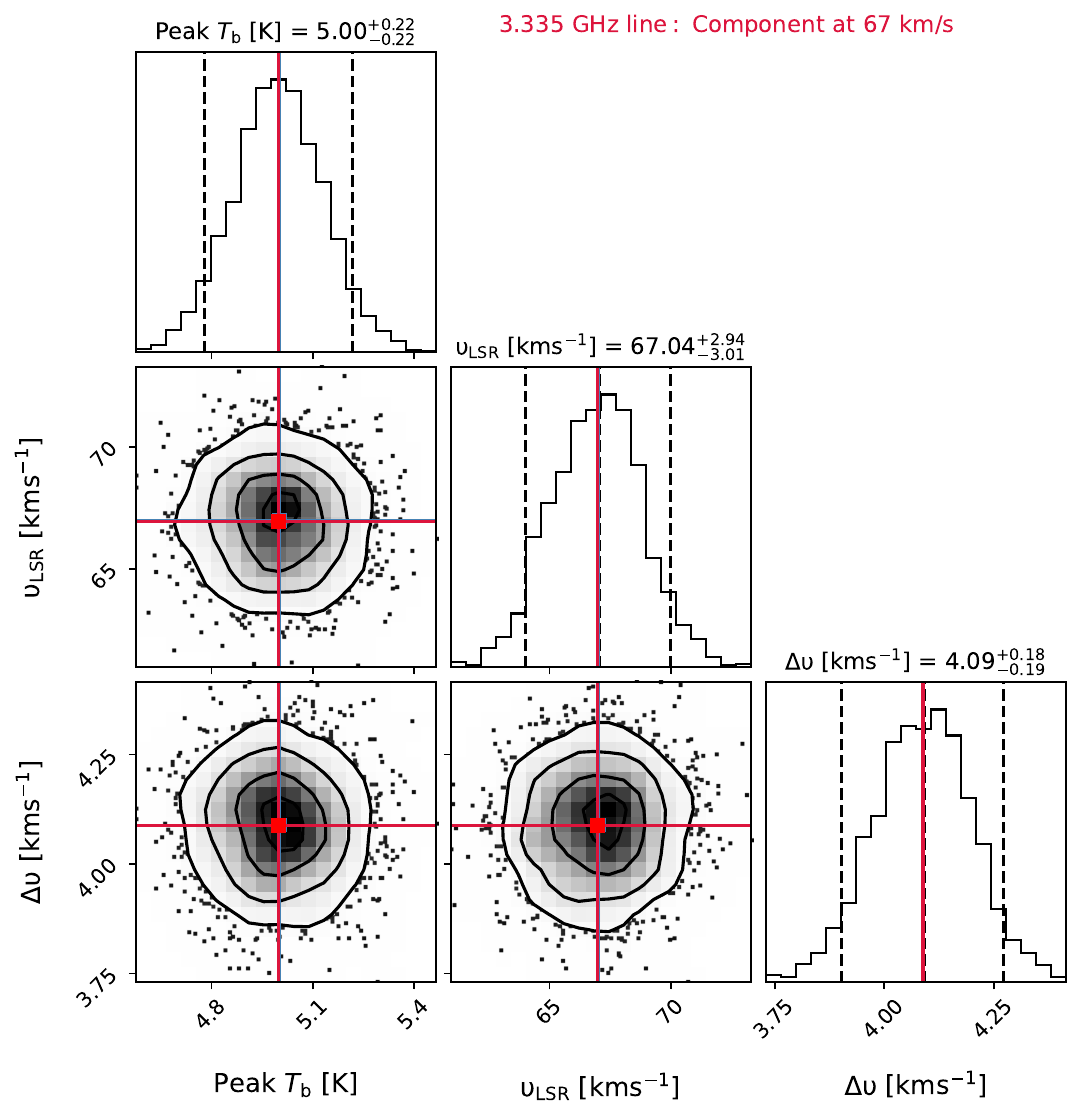}\\
    \includegraphics[width=0.31\textwidth]{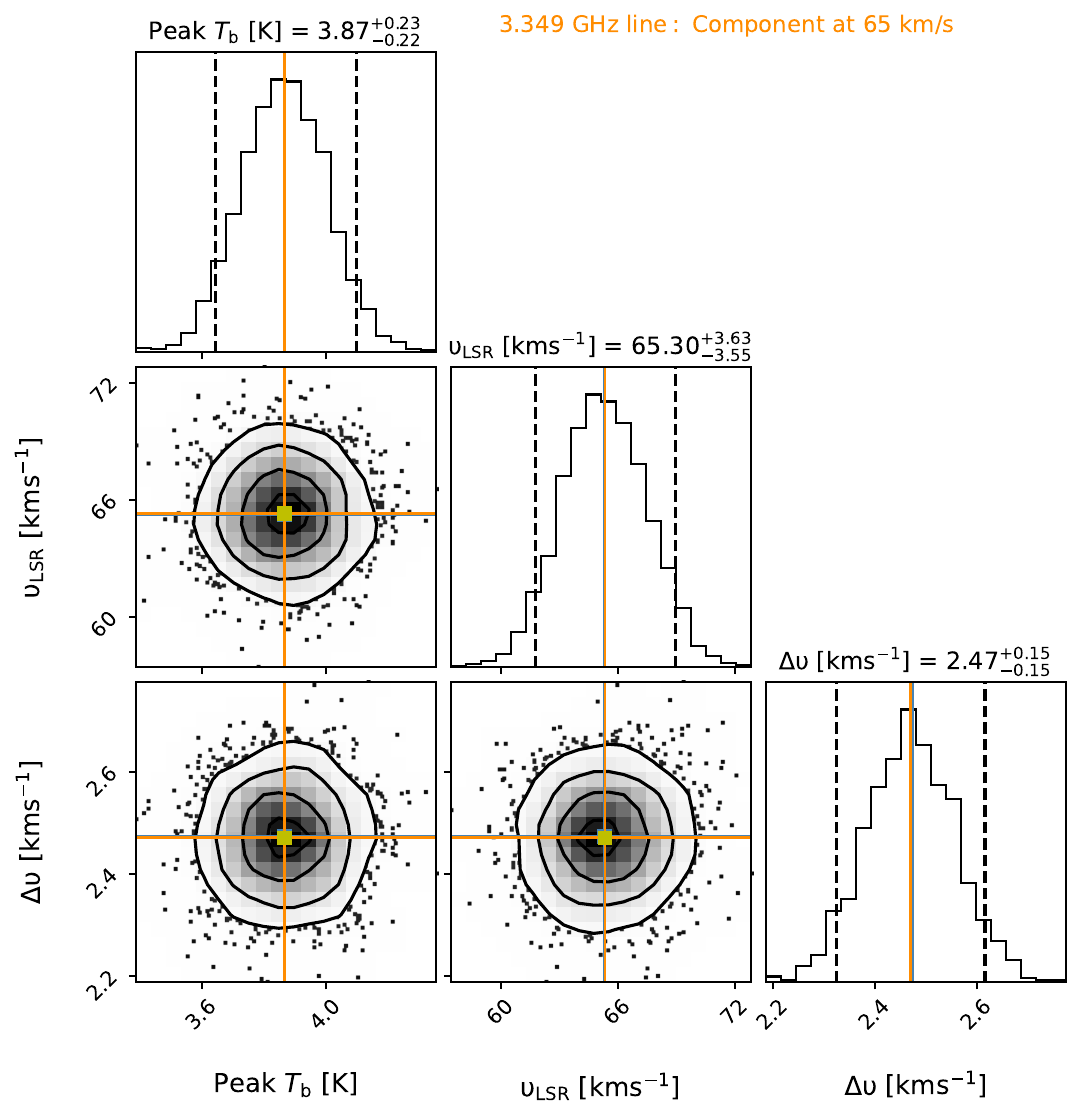} \quad
    \includegraphics[width=0.31\textwidth]{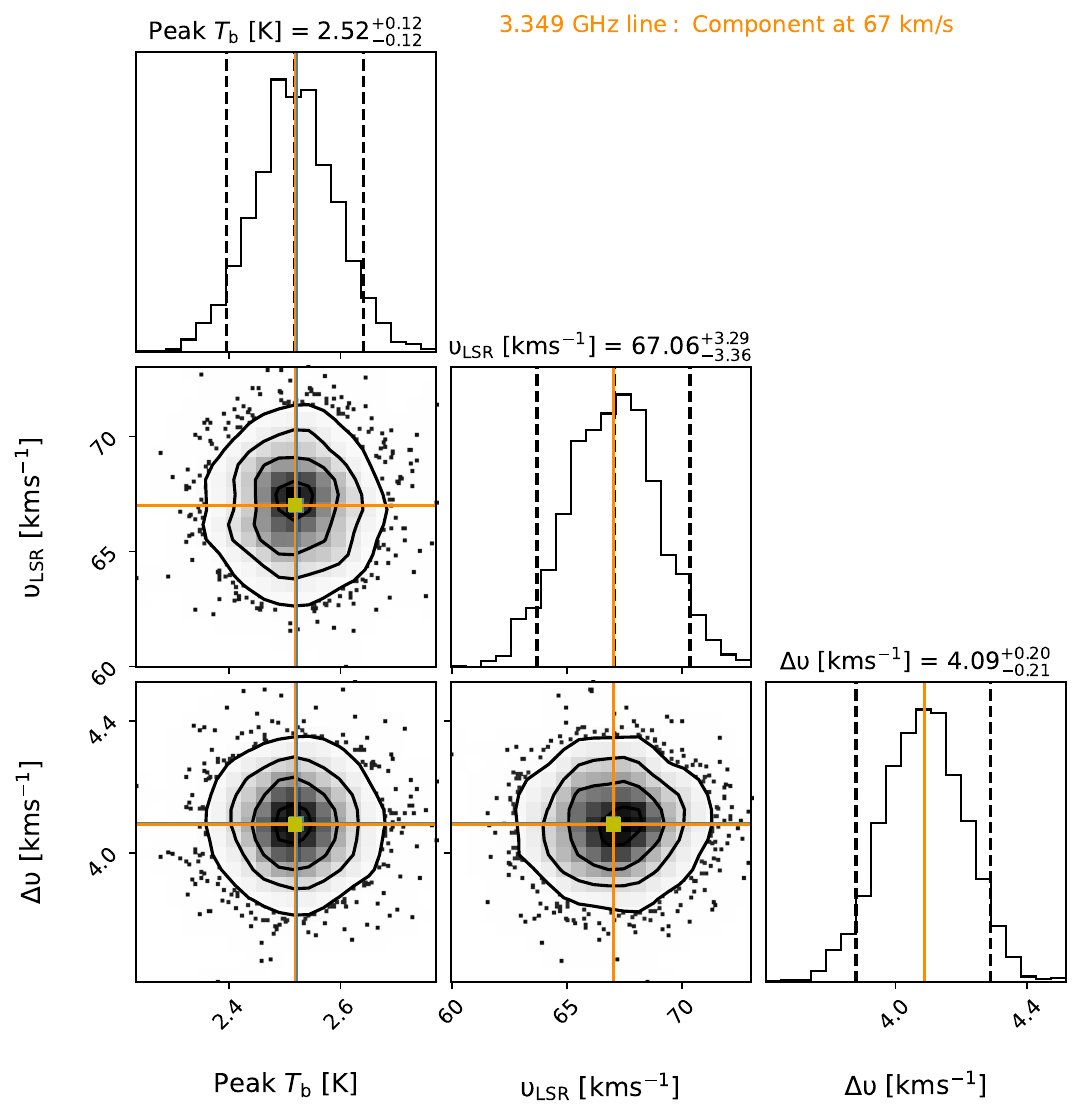}
    \caption{Same as Fig.~\ref{fig:fit_checks1} but for the fit components toward the 65~km~s$^{-1}$ (left) and 67~km~s$^{-1}$ (right) clouds.}
    \label{fig:fit_checks3}
\end{figure*}

\FloatBarrier
Due to the low signal-to-noise ratio present in the detected 700~MHz lines, as discussed in Sect.~\ref{subsec:line_profile}, it becomes imperative to evaluate the robustness of the Gaussian fits, particularly because the fitted parameters constitute key input parameters for the non-LTE analysis that follows. Figures~\ref{fig:fit_checks1} to \ref{fig:fit_checks3} display corner plots \citep{corner2016} illustrating the probability density distribution functions for all the fitted components across the whole parameter range, for each of the CH lines. These figures display the probability density distributions of each free parameter alongside the projected 2D histogram for a given pair of parameters. The solid lines mark the best-fit of the distribution while the dashed lines mark the bounds of the $2\sigma$ confidence interval.

\FloatBarrier
\section{Non-detections of the CH 700~MHz lines}\label{appendix:non_detections}
\begin{figure*}[h]
    \centering
    \includegraphics[width=0.9\textwidth]{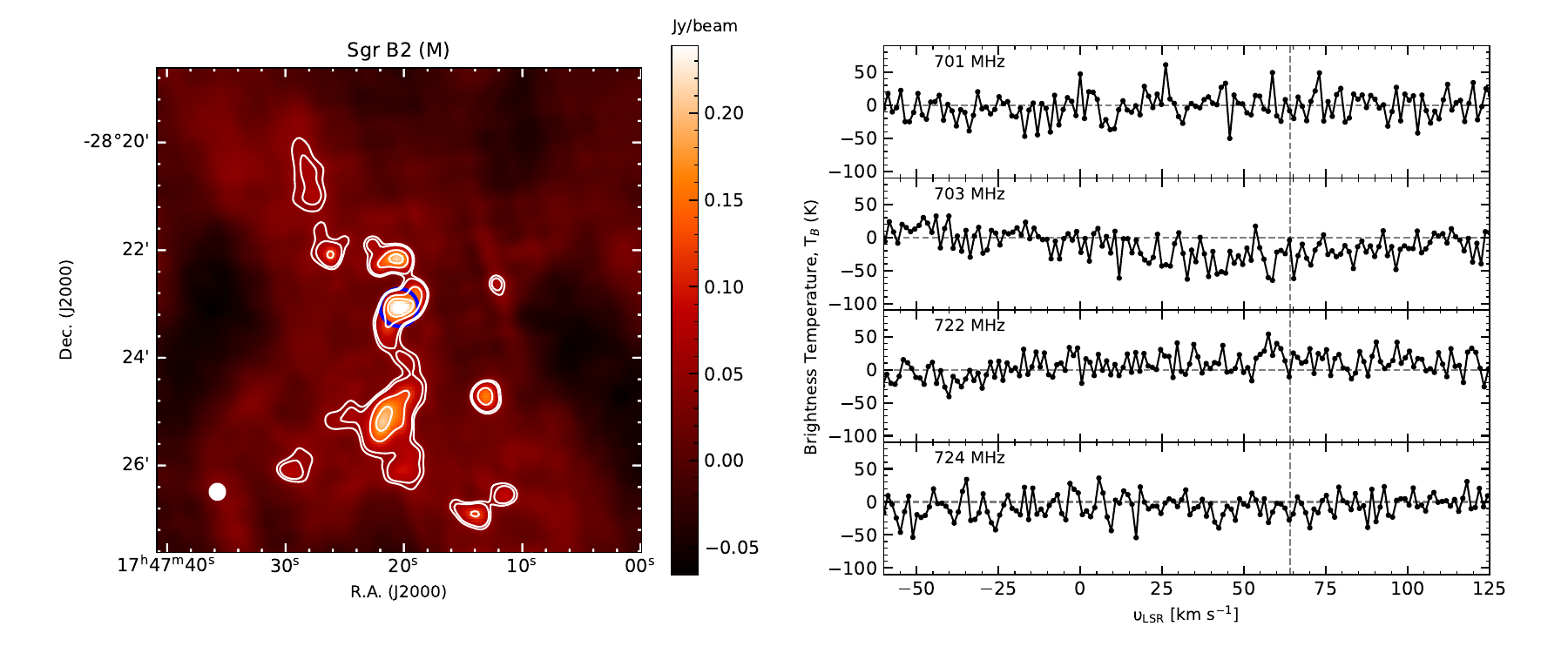}
    \caption{Left: Colour map displaying the background continuum emission at 700~MHz towards Sgr~B2~(M) alongside white contours marking continuum emission levels from $1.8\times,2\times, 4\times, 7\times$ and $10\times1\sigma$ where  $1\sigma = 2.7\times10^{-2}~$Jy/beam. The filled white ellipses on the bottom left-hand corner displays the synthesised beam and the blue circle marks the position towards which the spectra of the first rotationally excited lines of CH are extracted from. Right: From top-to-bottom the resulting baseline subtracted spectra of the 701~MHz, 703~MHz, 722~MHz and 724~MHz lines of CH.}
    \label{fig:Spec_sgr}
\end{figure*}

\begin{figure*}[h]
    \centering
    \includegraphics[width=0.9\textwidth]{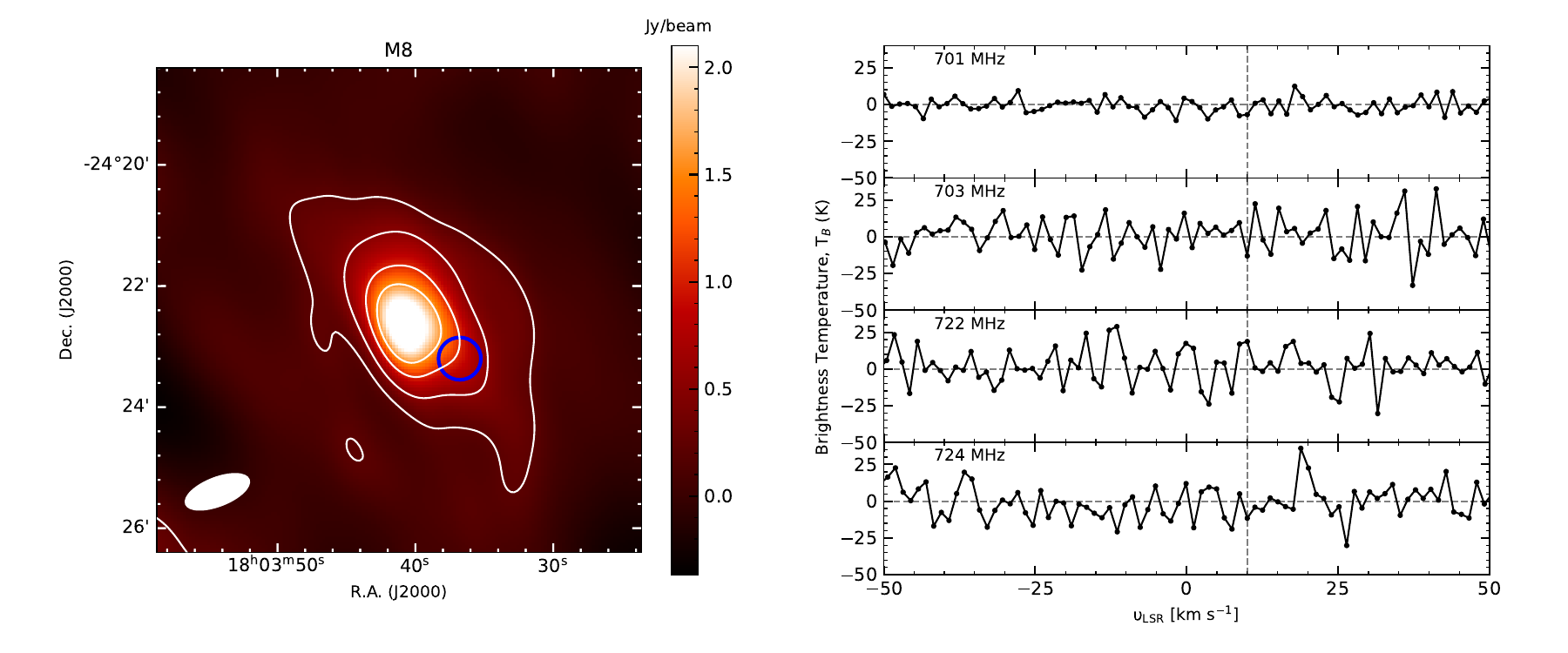}
    \caption{Same as Fig.~\ref{fig:Spec_sgr} but towards M8, where the contours are plotted for continuum emission levels from $1\times,2\times, 4\times,7\times$ and $10\times1\sigma$ where  $1\sigma = 0.24~$Jy/beam.}
    \label{fig:Spec_M8}
\end{figure*}

\begin{figure*}[h]
    \centering
    \includegraphics[width=0.9\textwidth]{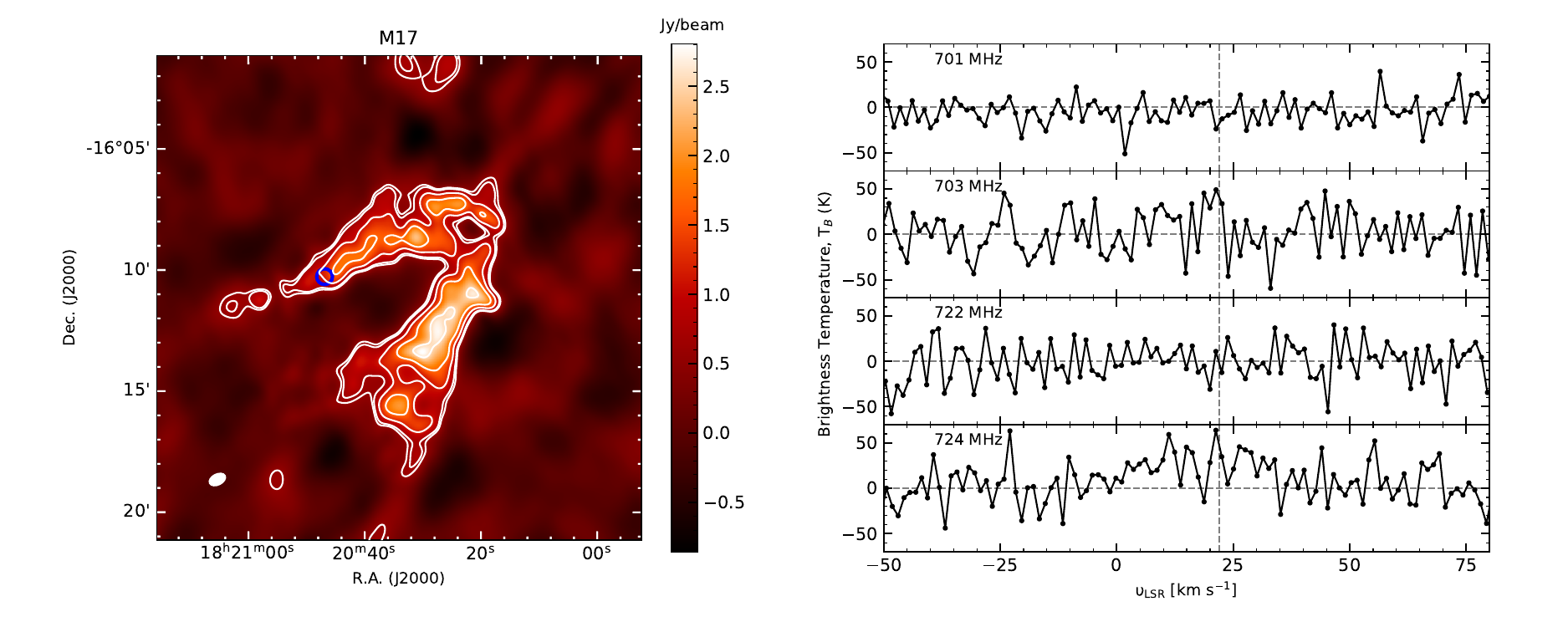}
    \caption{Same as Fig.~\ref{fig:Spec_sgr} but towards M17, where the contours are plotted for continuum emission levels from $2\times,3\times, 4\times,7\times$ and $10\times1\sigma$ where  $1\sigma = 0.25~$Jy/beam.}
    \label{fig:Spec_M17}
\end{figure*}

\begin{figure*}[h]
    \centering
    \includegraphics[width=0.9\textwidth]{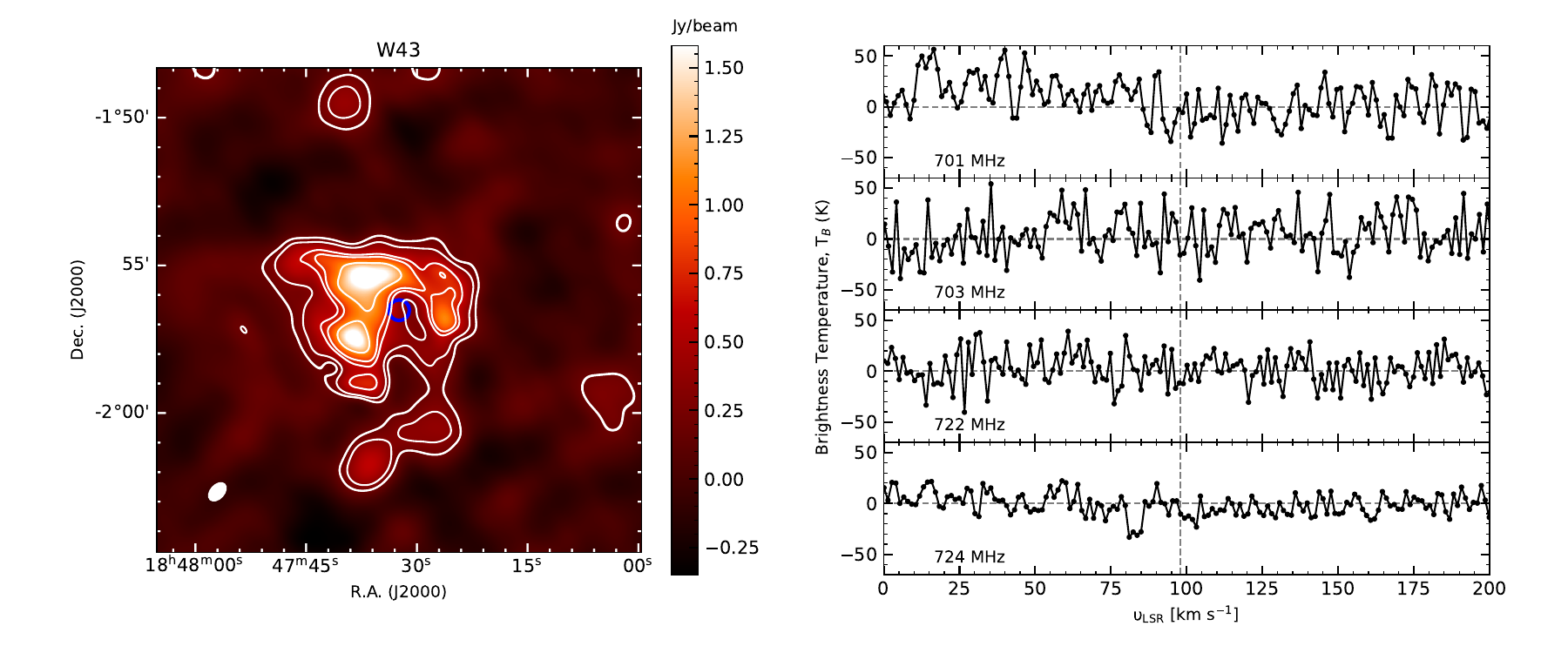}
    \caption{Same as Fig.~\ref{fig:Spec_sgr} but towards W43, where the contours are plotted for continuum emission levels from $1.08\times,1.12\times,2\times,3\times, 4\times,7\times$ and $10\times1\sigma$ where  $1\sigma = 0.17~$Jy/beam.}
    \label{fig:Spec_W43}
\end{figure*}

\begin{figure*}[h]
    \centering
    \includegraphics[width=0.9\textwidth]{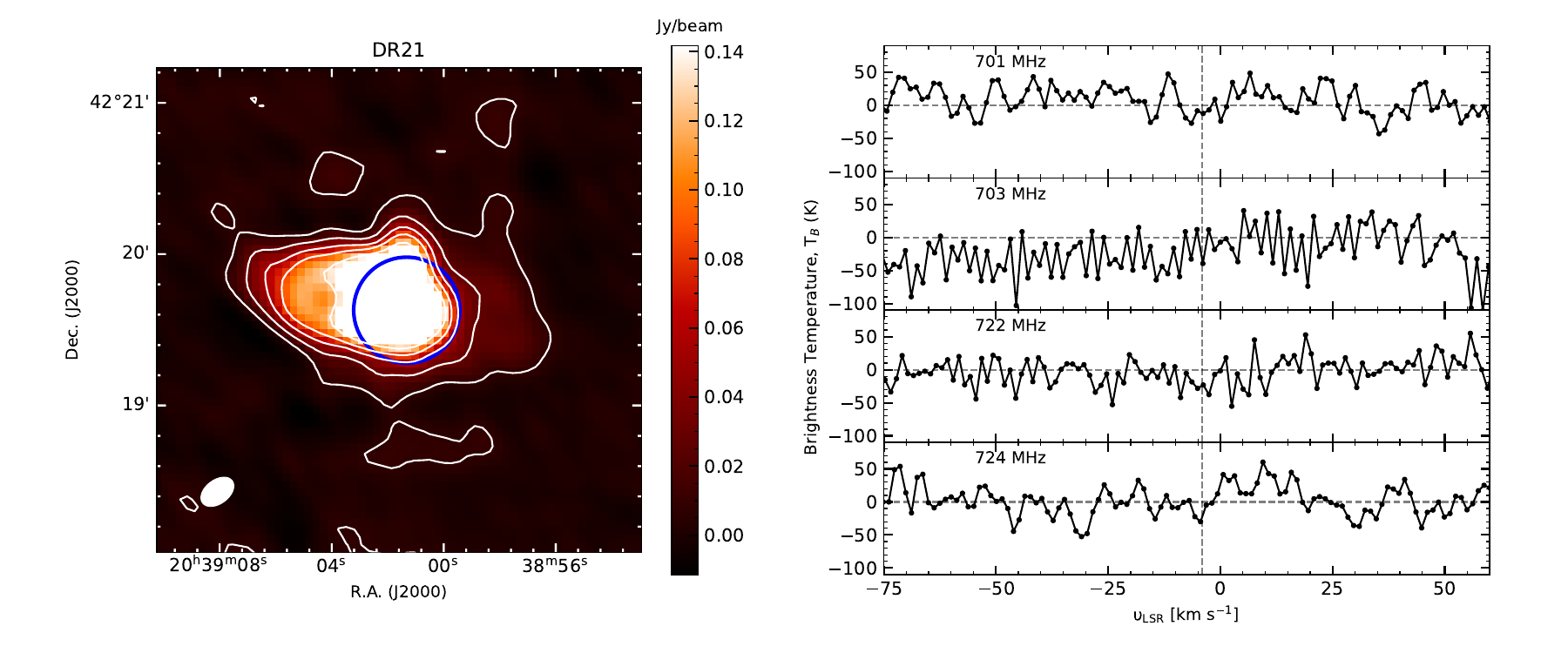}
    \caption{Same as Fig.~\ref{fig:Spec_sgr} but towards DR21~Main, where the contours are plotted for continuum emission levels from $0.5\times,2\times, 4\times,7\times$ and $10\times1\sigma$ where $1\sigma = 1.3\times10^{-2}~$Jy/beam.}
    \label{fig:Spec_DR21}
\end{figure*}

\FloatBarrier
This Appendix presents the non-detections (down to the rms noise levels quoted in Table~\ref{tab:source_parameters}) of the rotationally excited lines of CH near 700~MHz  towards Sgr~B2~(M), M8, M17, W43, and DR21~Main, respectively. Figures~\ref{fig:Spec_sgr} to \ref{fig:Spec_DR21} present the average continuum map at 700~MHz towards each source alongside their corresponding CH spectra extracted from a region enclosed in a $40\rlap{.}^{\prime\prime}4$ beam. This beam is centred at the position where the sub-mm/FIR continuum emission is the strongest and for which there exists previous observations of the sub-mm and FIR lines of CH.

\FloatBarrier 
\section{Scattering calculations for collisions with electrons }\label{appendix:collisions-e}

\FloatBarrier
In the Born approximation, the cross-section, $\sigma_{ij}$, for the electron-impact de-excitation from the upper energy state of a molecule, $i$, to a lower state, $j$, is related to the spontaneous radiative decay rate, $A_{ij}$, by the relation \citep{Takayanagi1966, Loreau2022}
\begin{equation*}
	\sigma_{ij}={e^2 c^3 A_{ij} \ln r \over 4 \pi^2 \hbar v_i^2 \nu^3}  = {\alpha  A_{ij} c \lambda^3 \ln r\over 4 \pi^2 v_i ^2 } ,
 \end{equation*}
	where $e$: elementary charge, $c$: speed of light, $h$: Planck constant, $\hbar$: reduced Planck constant ($=h/2\pi$), $m_e$: mass of an electron, $r=(v_f + v_i)/(v_f - v_i)$, $v_i$ and $v_f = (v_i^2 + v_t^2)^{1/2}$ are the initial and final speeds
	of the electron, $v_t=(2h \nu/m_e)^{1/2}$ is the threshold speed for exciting an upwards transition from $j$ to $i$, 
	$\nu$ is the transition frequency, $\lambda = c/\nu$ is the transition wavelength, and $\alpha = e^2/(\hbar c) \sim 1/137$ is the fine-structure constant. 
    Here, Gaussian units have been adopted
	and the right-hand-side of \citet{Loreau2022} Eqn.~(5) has been divided by $e^2 a_0^2$, where $a_0 = \hbar^2/(m_e e^2)$ is the Bohr radius. \\
	\noindent
	The quantity $r$ may be rewritten as
    \begin{equation*}
        r={(v_f + v_i) \over (v_f - v_i)} = {(v_f + v_i)^2 \over (v_f^2 - v_i^2)} = {(v_f + v_i)^2 \over v_t^2}
	= (z + (z + 1)^{1/2})^2,
    \end{equation*}
	
	where $z=v_i/v_t$.
	Averaging  over a Maxwell-Boltzmann distribution of electron velocities,
	we obtain the collisional de-excitation rate coefficient:

\begin{equation}
 q_{ij} = \langle v_i \sigma_{ij} \rangle = {\alpha  A_{ij} c \lambda^3 \over 4 \pi^2 v_t}  \langle \ln (z + (z + 1)^{1/2})^2 / z \rangle,
 \label{eqn:1}
 \end{equation}

 where $\langle Q \rangle$ denotes the average of quantity Q over a Maxwell-Boltzmann 
	distribution of electron velocities,
    \begin{equation*}
 	\langle Q \rangle = {1 \over \pi^{1/2} t^{3/2}} \int{z^2 Q \exp(-z^2/t) dz},       
    \end{equation*}

	and $t = 2kT/(m_e v_t^2) = kT/h\nu$
	
	\subsection*{Behaviour in the limits of small and large $T$}
	
	In the limit of small velocity, $z \ll 1$, a Taylor expansion yields
	\begin{equation*}
	    \ln r \sim 2z - z^3/3 + O(z^5)
	\end{equation*}
	and $Q = \ln r / z = 2(1-z^2/6+ O(z^4))$.
	In the limit $t \ll 1$, we then obtain  
	$\langle Q \rangle = 2 (1 - t/4)$ and finally
 \begin{equation}
	q_{ij}={\alpha  A_{ij} c \lambda^3 \over 2 \pi^2 v_t} (1 - t/4) 
 \label{eqn:2}
 \end{equation}
	In the limit  $t \rightarrow 0$, $q_{ij}$ tends to a maximum value that depends only on
	$A_{ij}$ and $\lambda$:
	\begin{equation}
 q_{0}={\alpha  A_{ij} c \lambda^3 \over 2 \pi^2 v_t} = {A_{ij} \over 2 \pi^2} 
	\biggl({R_\infty \over h\nu}\biggr)^{1/2} \lambda^3 = {A_{ij} \over 2 \pi^2} 
	\biggl({\lambda \over \lambda_0}\biggr)^{1/2} \lambda^3
 \label{eqn:3}
 \end{equation}
	where $R_\infty = 1/2 \alpha m_e c^2 = 13.6 \rm \,eV$ is the Rydberg constant for hydrogen and
	$\lambda_0 = hc/R_\infty = 91.2\,\rm nm$ is the wavelength at the Lyman limit.
	
	\noindent In the limit of large velocity, $z \gg 1$, $\ln r = \ln (4z^2) + O(1/z^2)$.
	In the limit $t \gg 1$, with the approximation $\ln r$ = $\ln (4z^2)$, we find that the
	integral needed to determine $\langle Q \rangle$ 
	can be solved analytically, yielding
	\begin{equation}
    q_{ij}= {q_0 (\ln (4t) - \gamma) \over (\pi t)^{0.5}}, 
    \label{eqn:4}
    \end{equation}
	where $\gamma=0.5772...$ is Euler's constant.
	\FloatBarrier
	\subsection*{Summary and analytic fit in two parts}
	
\begin{figure}[h]
    \includegraphics[width=0.5\textwidth]{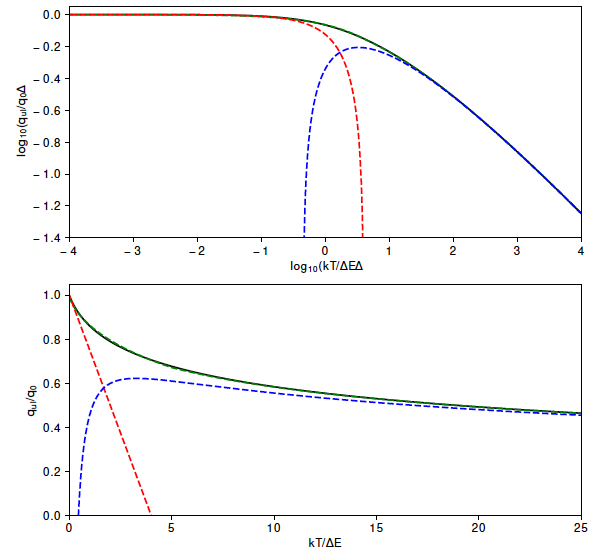}
		\caption{The ratio $q_{ij}/q_{0}$ as a function of $t=kT/h\nu$.  The quantity $q_{0}$ is given by Eqn.~\ref{eqn:3} in the text.
			Black curve: exact result. Red and blue dashed curves: behaviour (Eqns.~\ref{eqn:2} and \ref{eqn:4}) in the limits 
			$t\ll 1$ and $t\gg 1$, respectively. Green curves: analytic fit in two parts (see text).
		}
  \label{fig:ba}
	\end{figure}

	Figure~\ref{fig:ba} shows $q_{ij}/q_{0}$ as a function of $t=kT/h\nu$.  Here the black curve shows the 
	exact result (Eqn.~\ref{eqn:1}) obtained by numerical integration, and the red and blue dashed
	curves show the behaviour (Eqns.~\ref{eqn:2} and \ref{eqn:4}) in the limits $t\ll 1$ and $t\gg 1$, respectively.
	The green dashed curve shows the following analytic fit to the entire function, $q_{ij}(t)/q_{0}$,
	which is accurate to better than 1$\%$ for all t.
	\begin{align*}
	    {q_{ij} \over q_{0}} &= {1 \over 1 + 0.1587 \, t^{0.7}} \phantom{0000000} t < 5 \\
	 {q_{ij} \over q_{0}} &= {\ln (4t) - \gamma \over (\pi t)^{0.5}} \biggl(1 + {1 \over 2t} \biggr) \qquad t > 5
	\end{align*}
    
\end{appendix}

\end{document}